\documentclass[useAMS,usenatbib]{mn2e}
\usepackage{epsfig}
\title[Study of two binary clusters]{Multi-colour
photometry and Gaia EDR3 astrometry of two couples of binary clusters (NGC 5617 and Trumpler 22) and (NGC 3293 and NGC 3324).}  

\author[Bisht et al.]
   {D. Bisht$^{1}$,\thanks{E-mail: dbisht@ustc.edu.cn; zhuqf@ustc.edu.cn; shashikiran.ganesh@gmail.com; rkant@aries.res.in;
   alokdurgapal@gmail.com; geetarangwal91@gmail.com; deveshpathsariya@gmail.com; jiang@phys.nthu.edu.tw}
Qingfeng Zhu$^{1}$,
R.\ K.\ S.\ Yadav$^{2}$,
Shashikiran Ganesh$^{3}$,
Geeta Rangwal$^{4}$,
\and
Alok Durgapal$^{4}$,
Devesh P. Sariya$^{5}$,
Ing-Guey Jiang$^{5}$
\\\\
    $^{1}$ Key Laboratory for Researches in Galaxies and Cosmology, University of Science and
           Technology of China, Chinese Academy of Sciences,\\
           Hefei, Anhui, 230026, China\\
    $^{2}$Aryabhatta Research Institute of Observational Sciences,
           Manora Peak, Nainital 263129, India\\
    $^{3}$Physical Research Laboratory, Ahmedabad 380009, India\\
    $^{4}$Center of Advanced Study, Department of physics, D.S.B. campus, Kumaun University, Nainital, 263002, India\\
    $^{5}$Department of Physics and Institute of Astronomy, National Tsing-Hua University, Hsin-Chu, Taiwan\\
}

\begin{document}

\date{Accepted ....... Received  ......; in original form ......}

\pagerange{\pageref{firstpage}--\pageref{lastpage}} \pubyear{2010}

\maketitle

\label{firstpage}

\begin{abstract}

This paper presents a comprehensive analysis of two pairs of binary clusters (NGC 5617 and Trumpler 22)
and (NGC 3293 and NGC 3324) located in the fourth quadrant of our Galaxy. For this purpose we use different data taken from
VVV survey, WISE, VPHAS, APASS, GLIMPSE along with Gaia~EDR3 astrometric data. We identified 584, 429, 692 and 273
most probable cluster members with membership probability higher than $80 \%$ towards the region of clusters
NGC 5617, Trumpler 22, NGC 3293 and NGC 3324. We estimated the value of $R=\frac{A_{V}}{E(B-V)}$ as $\sim$ 3.1
for clusters NGC 5617 and Trumpler 22, which indicates normal extinction law. The value of $R$ $\sim3.8$ and $\sim1.9$
represent the abnormal extinction law towards the clusters NGC 3293 and NGC 3324.
Our Kinematical analysis show that all these clusters have circular orbits.
Ages are found to be $90\pm10$ and $12\pm3$ Myr for the cluster
pairs (NGC 5617 and Trumpler 22) and (NGC 3293 and NGC 3324), respectively. The distances of $2.43\pm0.08$, $2.64\pm0.07$,
$2.59\pm0.1$ and $2.80\pm0.2$ kpc estimated using parallax are alike to the values calculated by using the distance modulus. We have also
identified 18 and 44 young stellar object candidates present in NGC 5617 and Trumpler 22, respectively. Mass function slopes are
found to be in fair agreement with the Salpeter's value. The dynamical study
of these objects shows a lack of faint stars in their inner regions, which leads to the mass-segregation effect. Our study
indicates that NGC 5617 and Trumpler 22 are dynamically relaxed but the other pair of clusters are not.
Orbital alongwith the physical parameters show that the clusters in both pairs are physically connected.

\end{abstract}


\begin{figure*}
\begin{center}
\hbox{
\includegraphics[width=9.5cm, height=9.0cm]{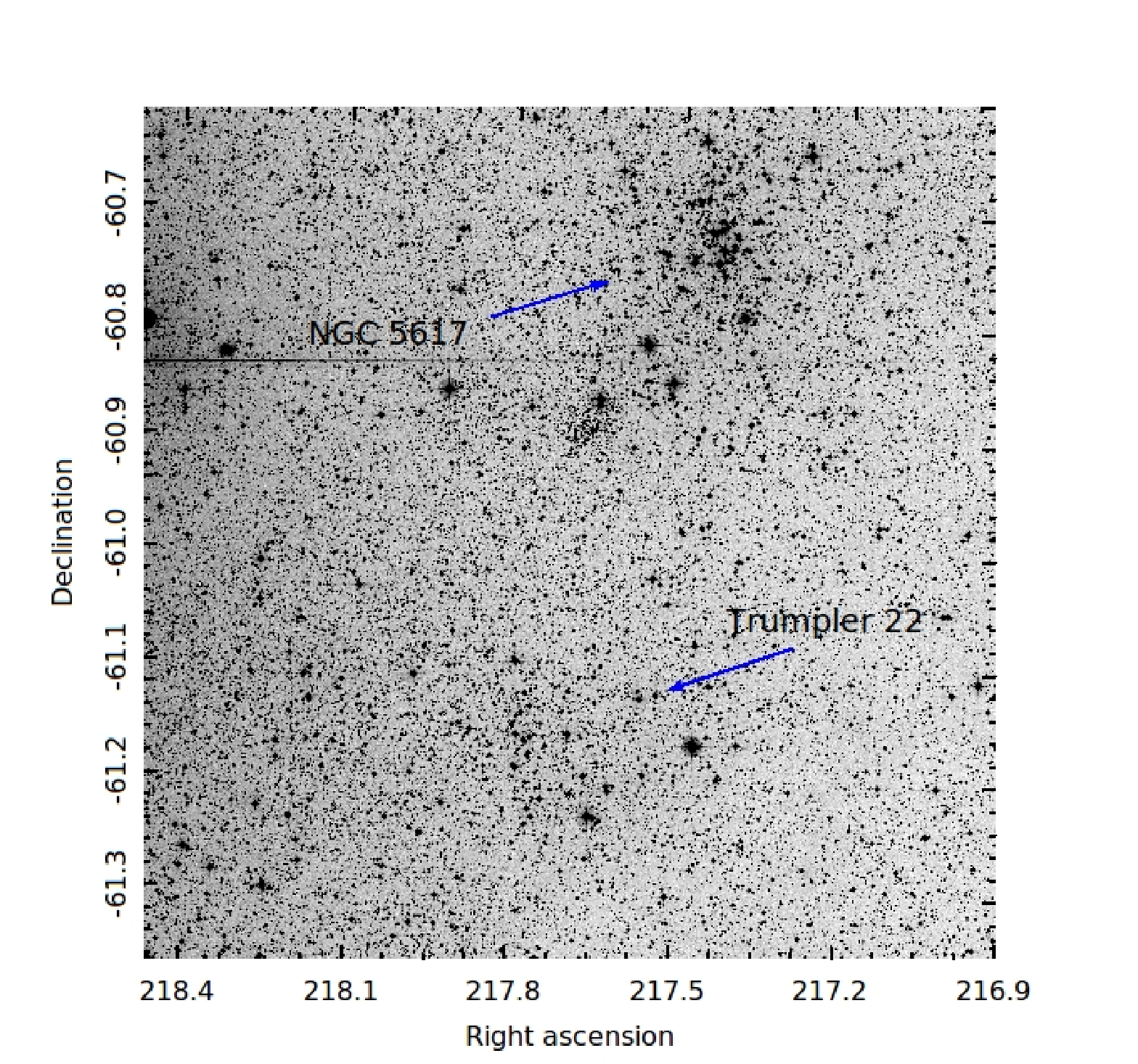}
\includegraphics[width=9.0cm, height=9.0cm]{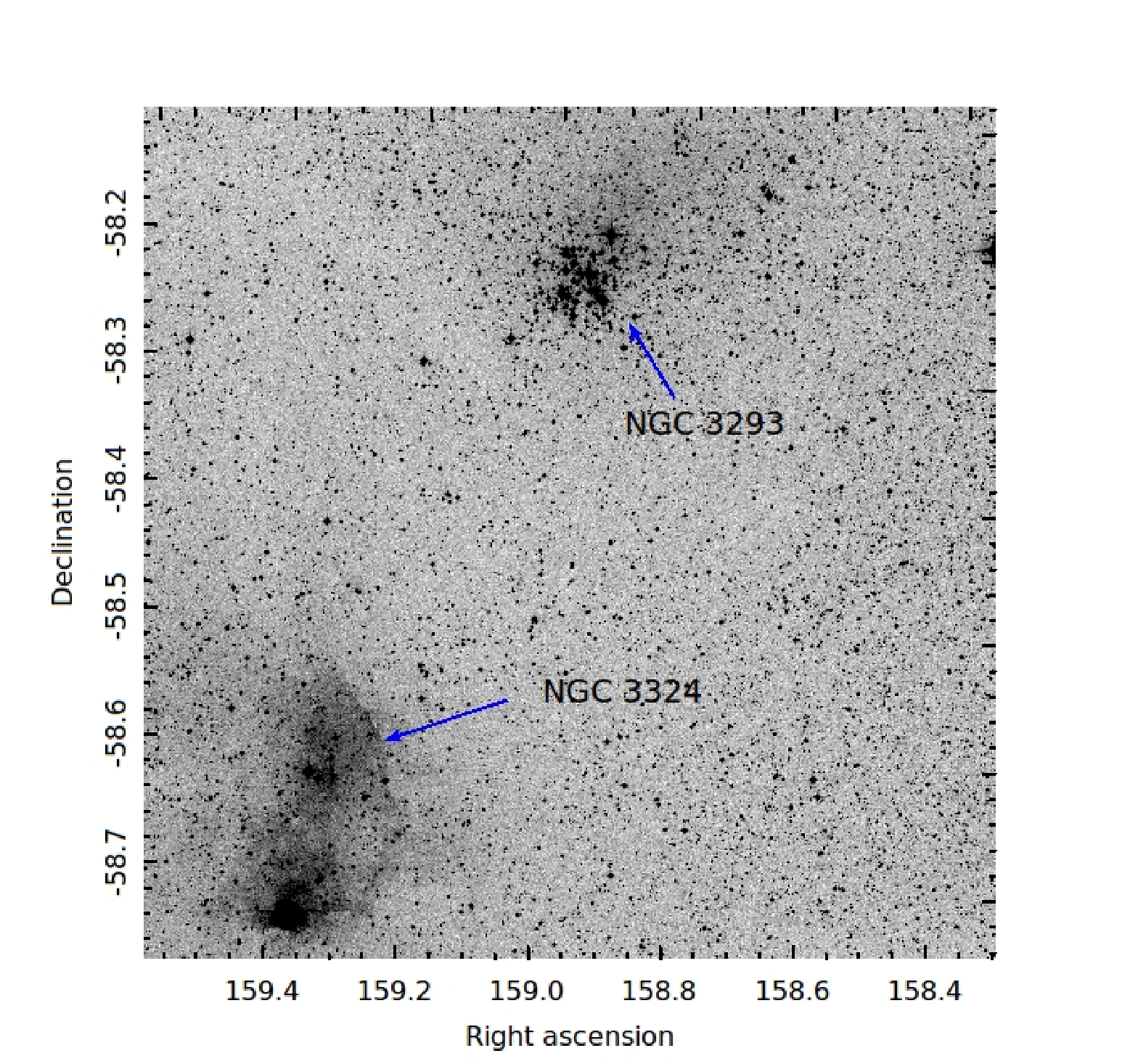}
}
\caption{Identification maps of two pair of clusters (NGC 5617 and Trumpler 22) and (NGC 3293 and NGC 3324) as taken
from the Digitized Sky Survey.} 
\label{id}
\end{center}
\end{figure*}
\section{Introduction}

The open clusters (OCs) are considered excellent laboratories for studies of stellar evolution and the dynamics of stellar
systems. The structure of the cluster is a result of its evolutionary processes such as initial physical conditions of 
the molecular clouds, external tidal perturbation, etc (Chen et al., 2004; Sharma et al., 2008). OCs
become beneficial objects for the stellar evolution because they are formed by the collapse and fragmentation of a turbulent
molecular cloud (Harris \& Pudritz 1994; Bate et al. 2003). OCs are influenced by the contamination of field stars. In recent years,
the detailed membership analysis of stars in the cluster field has become a subject of intense investigation, mainly in view to understand the
cluster properties (Carraro et al., 2008; Yadav et al., 2008; Joshi et al., 2014 ; Cantat-Gaudin et al. 2018). Recently many authors
have been estimated membership probability for clusters using Gaia DR2 kinematical data (Cantat-Gaudin et al. 2018, 2019;
Castro-Ginard et al. 2018, 2019; Bisht et al. 2019, 2020). The (early) Third Gaia Data Release (hereafter EDR3; Gaia
Collaboration et al. 2020) was made public on 3$^{rd}$ December 2020. EDR3 consists the central coordinates, proper motions in
right ascension and declination and parallax angles $(\alpha, \delta, \mu_{\alpha}cos\delta, \mu_{\delta}, \pi)$ for around 1.46
billion sources with a limiting magnitude of 3 to 21 mag in $G$ band. The Gaia EDR3 data are much accurate than
second data release of Gaia mission.

Bhatia (1990) has suggested that the lifetime of binary clusters depends on cluster separation, tidal force of the parental
Galaxy and encounters with giant molecular clouds. In the Large and Small Magellanic Clouds (LMC and SMC), $\sim$ $10\%$ of 
the well known OCs may be in pairs and around $50\%$ of them are primordial binary clusters (Bhatia \& Hatzidimitriou 1988;
Dieball \& Grebel 2000; Dieball, Muller \& Grebel 2002). In our Milky Way Galaxy around 10 $\%$ of total OCs have been
proposed to be in binary or multiple systems (Subramaniam et al. 1995; de la Fuente Marcos \& le la Fuente Marcos 2010).
The main aim of the paper is to study the properties of the binary open clusters NGC 5617, Trumpler 22 and NGC 3293,
 NGC 3324. The available information about these objects in the literature are as follows-

{\bf NGC 5617}: (Cl426-605) ($\alpha_{2000} = 14^{h}29^{m}48^{s}$, $\delta_{2000}$=-60\degr43\arcmin00\arcsec;
$l$=314\fdg67, $b$=-0\fdg11). Lindoff (1968) has estimated age of the cluster as $\sim$4.6$\times10^{7}$ yr using
photographic data. Based on photographic-photoelectric photometry Haug (1978) obtained parameters for this cluster as;
$E(B-V)=0.53$, $A_{V}$=1.69 and a distance of 1.8 Kpc. CCD $UBV$ photometry has been reported by Kjeldsen \& Frandsen
(1991, hereafter KF91), who got a smaller reddening $E(B-V)=0.48\pm0.02$, a larger distance of $2.05\pm0.2$ Kpc and an
age of 70 Myr. It is an intermediate age open cluster ($8.2\times10^{7}$ years) containing red giants and blue straggler
stars (Ahumada \& Lapasset 2007) in its surroundings, which membership of the cluster is still in doubt.

{\bf Trumpler 22}: ($\alpha_{2000} = 14^{h}31^{m}02^{s}$, $\delta_{2000}=-61\degr10\arcmin00\arcsec$;
$l$=314\fdg64, $b$=-0\fdg58). Haug (1978) studied this object using photographic data. De Silva et al. (2015) has done
photometric and spectroscopic analysis of both clusters NGC 5617 and Trumpler 22. He has obtained common age,
distance and radial velocity for both clusters as $70\pm10$ Myr, $2.1\pm0.3$ Kpc and $38.5\pm2.0$ Km/sec respectively.

{\bf NGC 3293}: ($\alpha_{2000} = 10^{h}35^{m}51^{s}$, $\delta_{2000}=-58\degr13\arcmin48\arcsec$;
$l$=285\fdg85, $b$=0\fdg07). This object is moderately younger and belongs to the rich Carina complex.
Preibisch et al. (2017) studied this object using Chandra X-ray observations. They found the
age of this object as 8-10 Myr. Delgado et al. (2011) have estimated parameters of pre-main sequence stars in
this cluster. They obtained flatter mass function slope than the Salpeter's value. 
Slawson et al. (2007) studied the stellar
mass spectrum of NGC 3293 using CCD $UBVRI$ images. They found significantly fewer lower mass stars towards the region of
NGC 3293. They confirmed the age of this cluster as 10 Myr on the basis of the presence of some intermediate-mass stars
near the main sequence in the HR diagram. Tuvikene \& Sterken (2006) checked the variability of stars in
NGC 3293. Out of 15 candidates they found 3 constant stars, 10 stars with significant variability while 2 of them were
considered as suspected variables. Photometric study has been done by Baume et al. (2003) using CCD photometric
observations at $UBVRI_CH_{\alpha}$. They found distance as $2750\pm250$ pc and age as $8\pm1$ Myr. The 
initial mass function slope was estimated as $1.2\pm0.2$, a bit flatter than the typical slope for field stars.

{\bf NGC 3324}: ($\alpha_{2000} = 10^{h}37^{m}20^{s}$, $\delta_{2000}=-58\degr38\arcmin30\arcsec$;
$l$=286\fdg23, $b$=-0\fdg18). This object is also situated in proximity with NGC 3293 in Carina complex. 
Carraro et al. (2001) reported the first CCD $UBVRI$ photometry of NGC 3324, 
and found that this cluster is very young and contains several
pre-main sequence stars. Claria (1977) presented wide band (UBV) and narrow band (H$_{\alpha}$) photometry of this object.
According to this study, NGC 3324 contains at least twenty O-and B-type members and it is located at 3.12 kpc in
the Carina spiral feature. A mean colour excess and age are found as 0.47 mag and $2.2 \times10^{6}$ years, respectively.

\begin{figure}
\centering
\includegraphics[width=8.5cm,height=8.5cm]{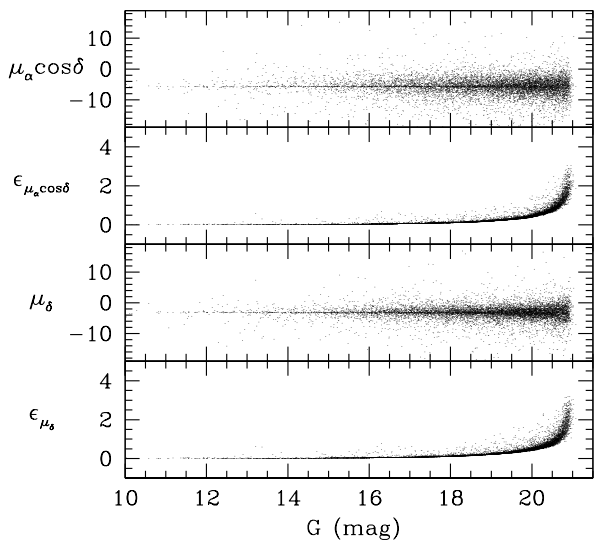}
\caption{Plot of proper motions and their errors versus $G$ magnitude for the cluster NGC 5617 is shown as an example.}
\label{error_pm} 
\end{figure}

\begin{figure*}
\centering
\includegraphics[width=10.5cm,height=10.5cm]{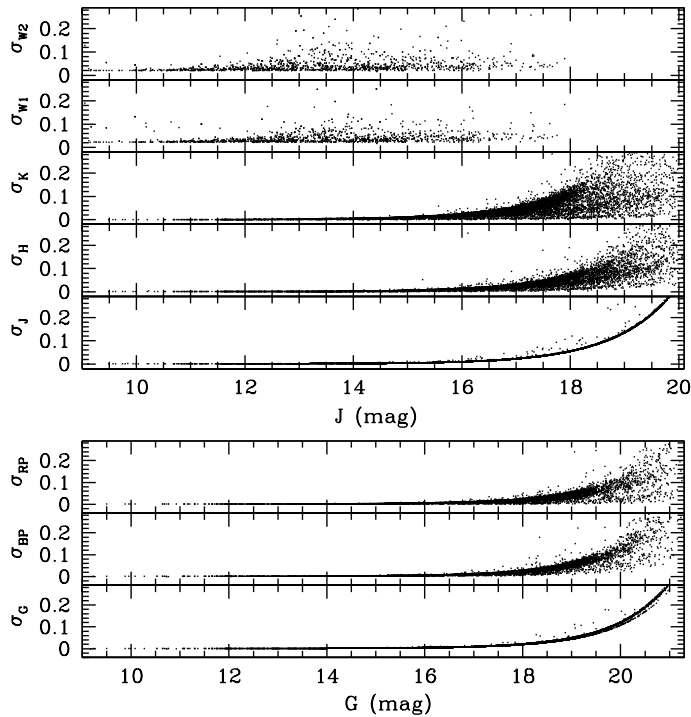}
\caption{Photometric errors in Gaia pass bands $G$, $G_{BP}$ and $G_{RP}$ against $G$ magnitude in three lower panels while photometric
errors in $J$, $H$, $K$, $W_{1}$ and $W_{2}$ magnitudes against $J$ magnitude in upper five panels.} 
\label{error_pbands}
\end{figure*} 

Apart from this available information for the clusters under study, membership is still a question of debate.
Our main goal is to estimate the membership probability for these objects and determine the more precise fundamental
parameters, Galactic orbits, luminosity function, mass function and dynamical state of the clusters NGC 5617,
Trumpler 22, NGC 3293 and NGC 3324 using multiwavelength photometric data along with high-precision astrometric
data from the Gaia EDR3 catalog. Gaia EDR3 involves photometric magnitudes in three bands ($G, G_{BP}, G_{RP}$),
astrometric data at the sub milliarcsecond level along with parallax values (Gaia Collaboration et al. 2020).

Proper motion is a very important parameter of open clusters. Another important implication of the cluster's proper
motion is the determination of membership probabilities for individual stars (Sanders 1971). The investigation of OCs
also offers to understand the mass function (MF) of stellar objects, which is an important tool to study the
star formation history (Sharma et al. 2017; Jose et al. 2017, and references therein). In recent years, many authors
have estimated the present day MF for plenty of OCs (Dib et al. 2017; Joshi et al. 2020). The spatial
distribution of massive and faint stars within the clusters provides important information to understand the mass-segregation
in OCs (Bisht et al. 2016).

The outline of the paper is as follows. The brief description of the data used has been described in Section 2. Section 3 is devoted to the
study of mean proper motion and estimation of membership probability of stars. In Section 4, orbits of the clusters are calculated.
The cluster structure has been explained in Section 5. The main fundamental parameters of the clusters are discussed in Section 6. 
The dynamical properties of the clusters are described in Section 7. Binarity of clusters have been discussed in Section 8.
The conclusion of this paper has been given in Section 9.


\section{Data Used}
\label{OBS}

We collected the astrometric and photometric data from Gaia EDR3 along with broad-band photometric data from VVV, WISE, APASS,
GLIMPSE, and VPHAS data for clusters NGC 5617, Trumpler 22, NGC 3293 and NGC 3324. The finding charts for the clusters are taken from
Digitized Sky Survey (DSS) and shown in Fig. \ref{id}. We cross-matched each catalog for the clusters under study.
The brief description has been given for each data sets as follows:
 
\subsection{The multi-dimensional Gaia EDR3 data}

We used Gaia~EDR3 (Gaia Collaboration et al. 2020) data for the astrometric analysis of the clusters NGC 5617, Trumpler 22,
NGC 3293 and NGC 3324. Data should be considered complete down to G=18-19 mag. The $G$, $G_{BP}$ and
$G_{RP}$ bands cover the wavelength range from 330 to 1050 nm, 330-680 nm and 630-1050 nm, respectively (Evans et al. 2018).
We have plotted the photometric errors in $G$, $G_{BP}$ and $G_{RP}$ versus $G$ band as shown in the bottom panels of
Fig. \ref{error_pbands}. The uncertainties in parallaxes have the range of $\sim$ 0.02-0.03 milliarcsecond (mas) for sources
at $G\le15$ mag and $\sim$ 0.07 mas for sources with $G\sim17$ mag. The uncertainties in the respective proper motion components
are up to 0.01-0.02 $mas~ yr^{-1}$ (for $G\le15$ mag), 0.05 $mas~ yr^{-1}$ (for $G\sim17$ mag) and 0.4 $mas~ yr^{-1}$ (for $G\sim20$ mag).
The proper motion and their corresponding errors are plotted against $G$ magnitude in Fig. \ref{error_pm}.

\subsection{WISE data}

This database is a NASA Medium Class Explorer mission that conducted a digital imaging survey of the entire
sky in the mid-IR bands. The effective wavelength of mid-IR bands are $3.35 \mu m$ (W1), $4.60 \mu m $(W2),
$11.56 \mu m$ (W3) and $22.09 \mu m$ (W4) (Wright et al. 2010). We have taken data for clusters NGC 5617, Trumpler 22,
NGC 3293 and NGC 3324 from the ALLWISE source catalog. This catalog has achieved $5\sigma$ point source sensitivities
better than 0.08, 0.11, 1 and 6 mJy at 3.35, 4.60, 11.56, and 22.09 $\mu m$, which is expected to be more than $99\%$ of
the sky (Bisht et al. 2020). These sensitivities are 16.5, 15.5, 11.2, and 7.9 for W1, W2, W3, and W4 bands correspond
to vega magnitudes.

\subsection{VVV data}

The VVV survey is an ESO infrared Large Public survey (Minniti et al. 2010; Saito et al. 2012b) which uses the 4-meter VISTA telescope located
at Cerro Paranal Observatory, Chile. The effective wavelength of near-infrared broadband filters are $0.87 \mu m$ (Z), $1.02 \mu m$ (Y),
$1.25 \mu m$ (J), $1.64 \mu m$ (H) and $2.14 \mu m$ (K). The telescope has a near-infrared camera, VIRCAM (Dalton et al. 2006), consisting
of an array of 16 detectors with $2048\times2048$ pixels.
The errors given in the VVV catalog for the ($J$, $H$, $K$) bands and $W1$, $W2$ bands from the WISE catalog are plotted against $J$ magnitudes
in the top panels of Fig. \ref{error_pbands}.

\begin{figure*}
\begin{center}
\hbox{
\includegraphics[width=8.5cm, height=8.5cm]{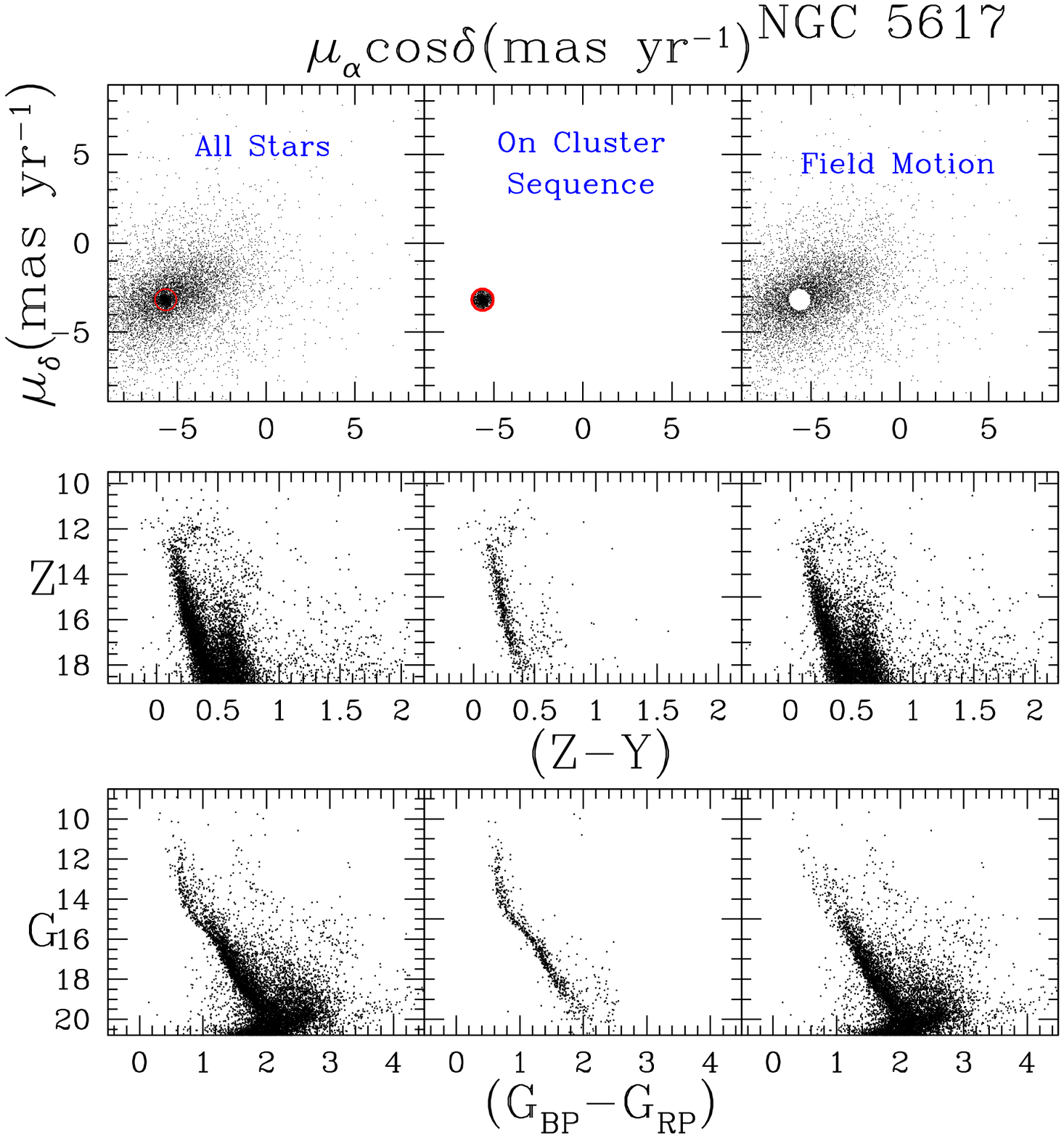}
 \includegraphics[width=8.5cm, height=8.5cm]{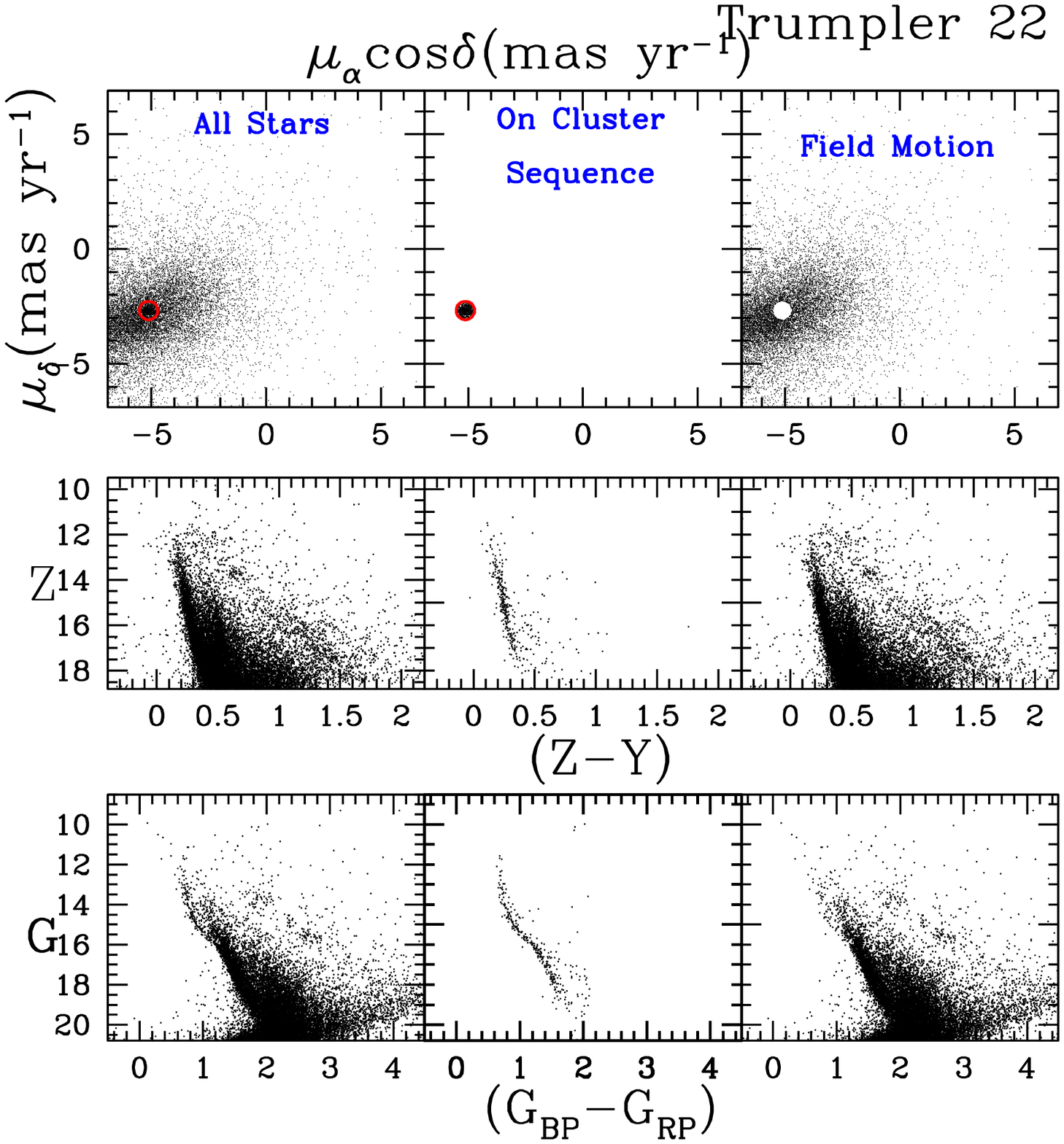}
}
\caption{(Top panels) VPDs for clusters NGC 5617 and Trumpler 22. (Middle panels)
$Z$ versus $Z-Y$ colour magnitude diagrams. (Bottom panels) $G$ versus $(G_{BP}-G_{RP})$ colour magnitude diagrams.
For each cluster CMDs, (Left panel) The entire sample. (Center) Stars within the circle of $0.6~ mas~ yr^{-1}$ and
$0.4~ mas~ yr^{-1}$ radius for clusters NGC 5617 and Trumpler 22 centered around the mean proper motion of the
clusters. (Right) Probable background/foreground field stars in the direction of these clusters. All plots show only
stars with PM error smaller than $0.4~ mas~ yr^{-1}$ in each coordinate.} 
\label{vpd}
\end{center}
\end{figure*}


 \begin{figure*}
 \begin{center}
 \hbox{ 
 \includegraphics[width=8.5cm, height=8.5cm]{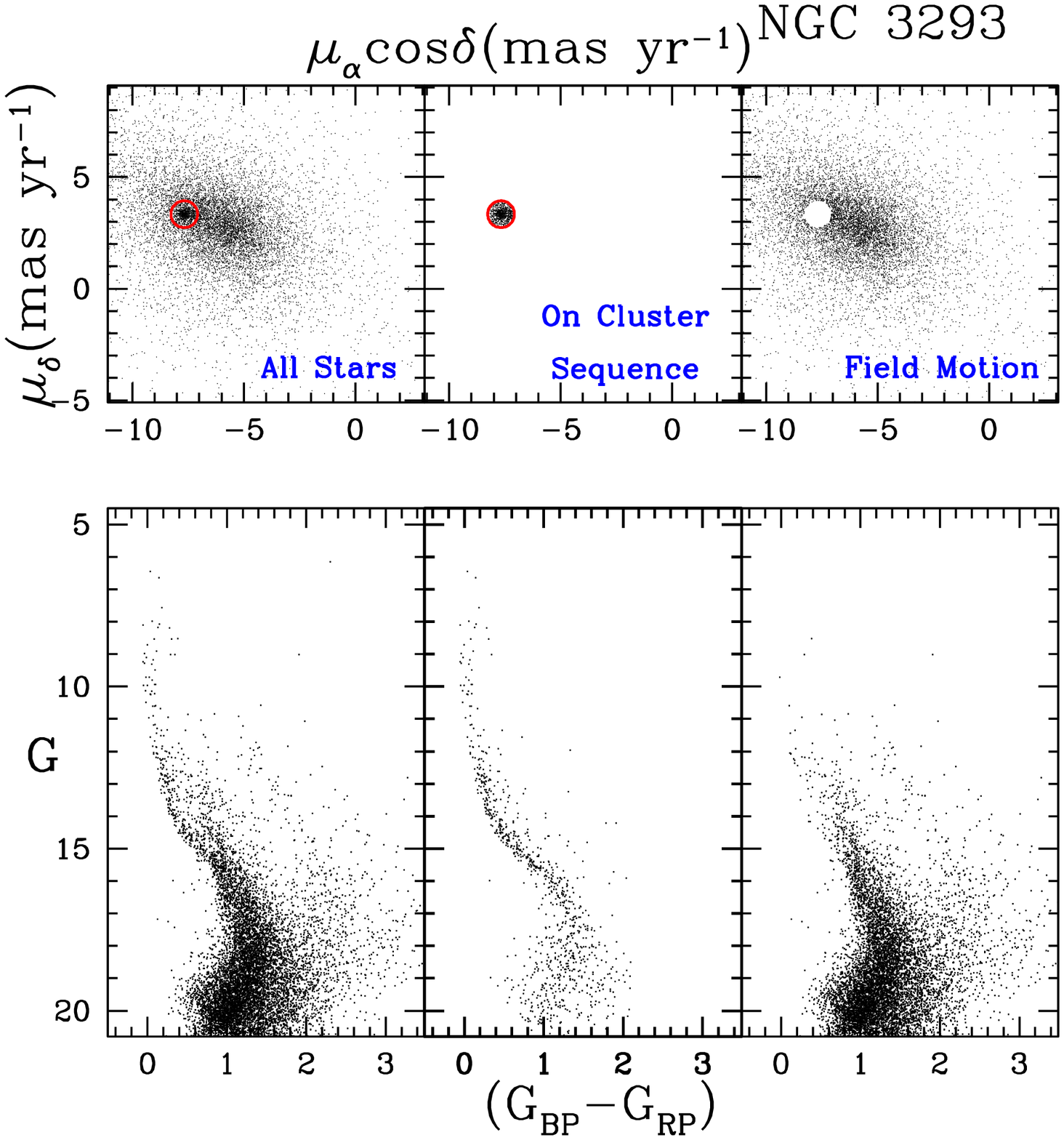}
 \includegraphics[width=8.5cm, height=8.5cm]{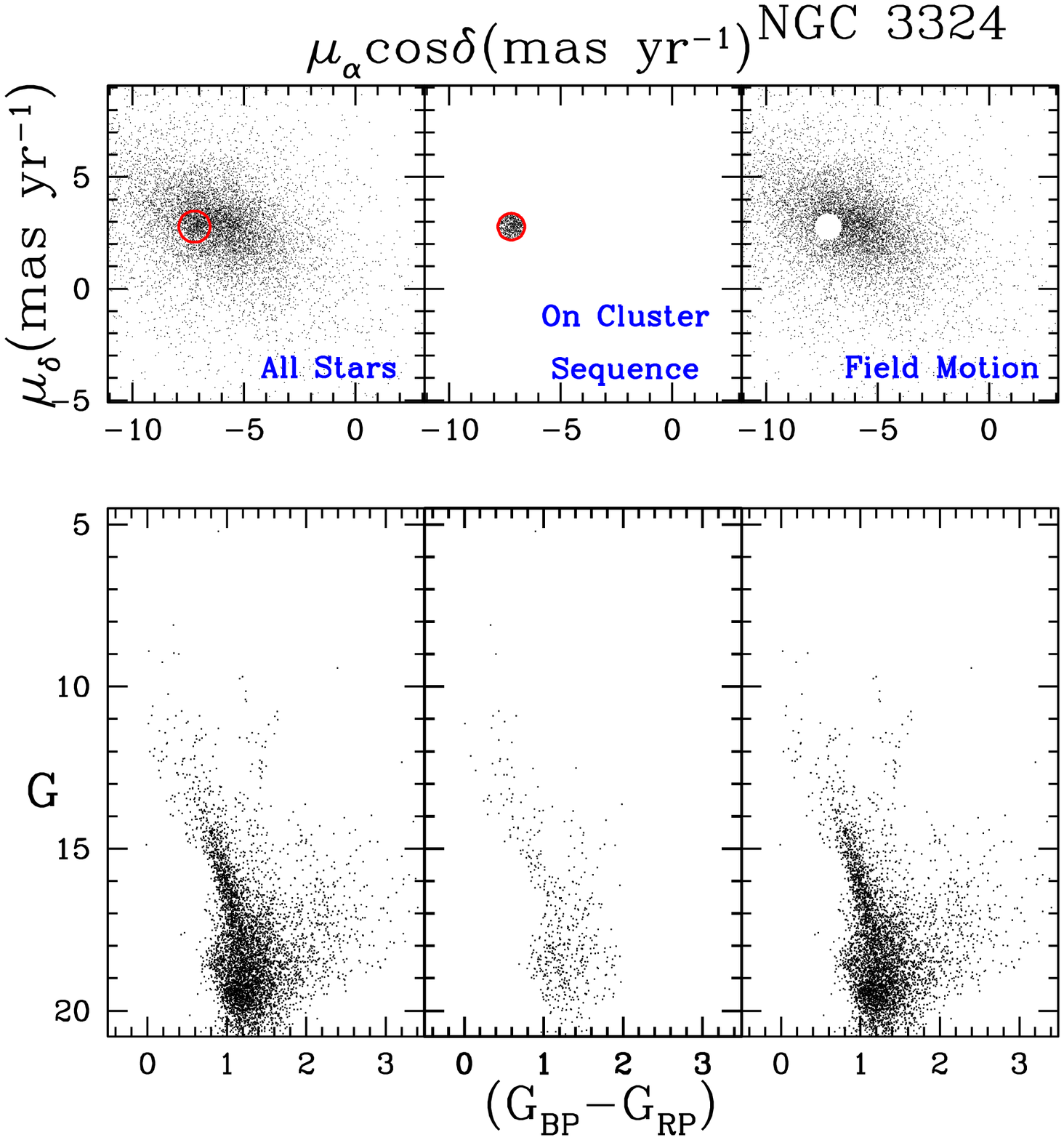}
 }
\caption{(Top panels) VPDs for clusters NGC 3293 and NGC 3324.
(Bottom panels) $G$ versus $(G_{BP}-G_{RP})$ colour magnitude diagrams. For each cluster CMDs, (Left panel)
The entire sample. (Center) Stars within the circle of $0.6~ mas~ yr^{-1}$ radius for clusters NGC 3293 and
NGC 3324 centered around the mean proper motion of the clusters. (Right) Probable background/foreground field
stars in the direction of these clusters. All plots show only stars with PM error smaller than
$0.4~ mas~ yr^{-1}$ in each coordinate.} 
 \label{vpd2}
 \end{center}
 \end{figure*}

\subsection{APASS data}

The American Association of Variable Star Observers (AAVSO) Photometric All-Sky Survey (APASS) is organized in five filters:
$B$, $V$ (Landolt) and $g^{\prime}$, $r^{\prime}$, $i^{\prime}$ proving stars with $V$ band magnitude range from 7 to 17 mag
(Heden \& Munari 2014). DR9 is the latest catalog and covers about $99\%$ sky (Heden et al. 2016). We have extracted this
data from $http://vizier.u-strasbg.fr/viz-bin/VizieR?-source=II/336$.

\subsection{GLIMPSE data}

The Galactic Legacy Infrared Mid-Plane Survey Extraordinaire (GLIMPSE; Benjamin et al. 2003; Churchwell et al. 2004) data
has been used for clusters NGC 5617 and Trumpler 22. The basic calibration of the GLIMPSE IRAC frames was performed by the
Spitzer Science Center Pipeline (Spitzer Observers Manual 2004). This database consists only of high-reliability sources with
each source must be detected twice in any of the four IRAC bands (3.6, 4.5 5.8, 8.0 $\mu m$).

\subsection{VPHAS data}

The VST/Omegacam Photometric $H_{\alpha}$ Survey (VPHAS) is imaging the entire Southern Milky Way in visible light at $\sim$ 1 arcsec
angular resolution down to $\ge$ 20 mag using the VLT Survey Telescope in Chile. We have extracted data from VPHAS catalog (Drew et al. 2014)
for the analysis of clusters under study. This catalog includes data in $u$, $g$, $r$, $i$ and $H_{\alpha}$ passbands.

\section{Mean Proper motion and cluster membership}

\begin{figure}
\begin{center}
\hbox{
\includegraphics[width=4.2cm, height=4.2cm]{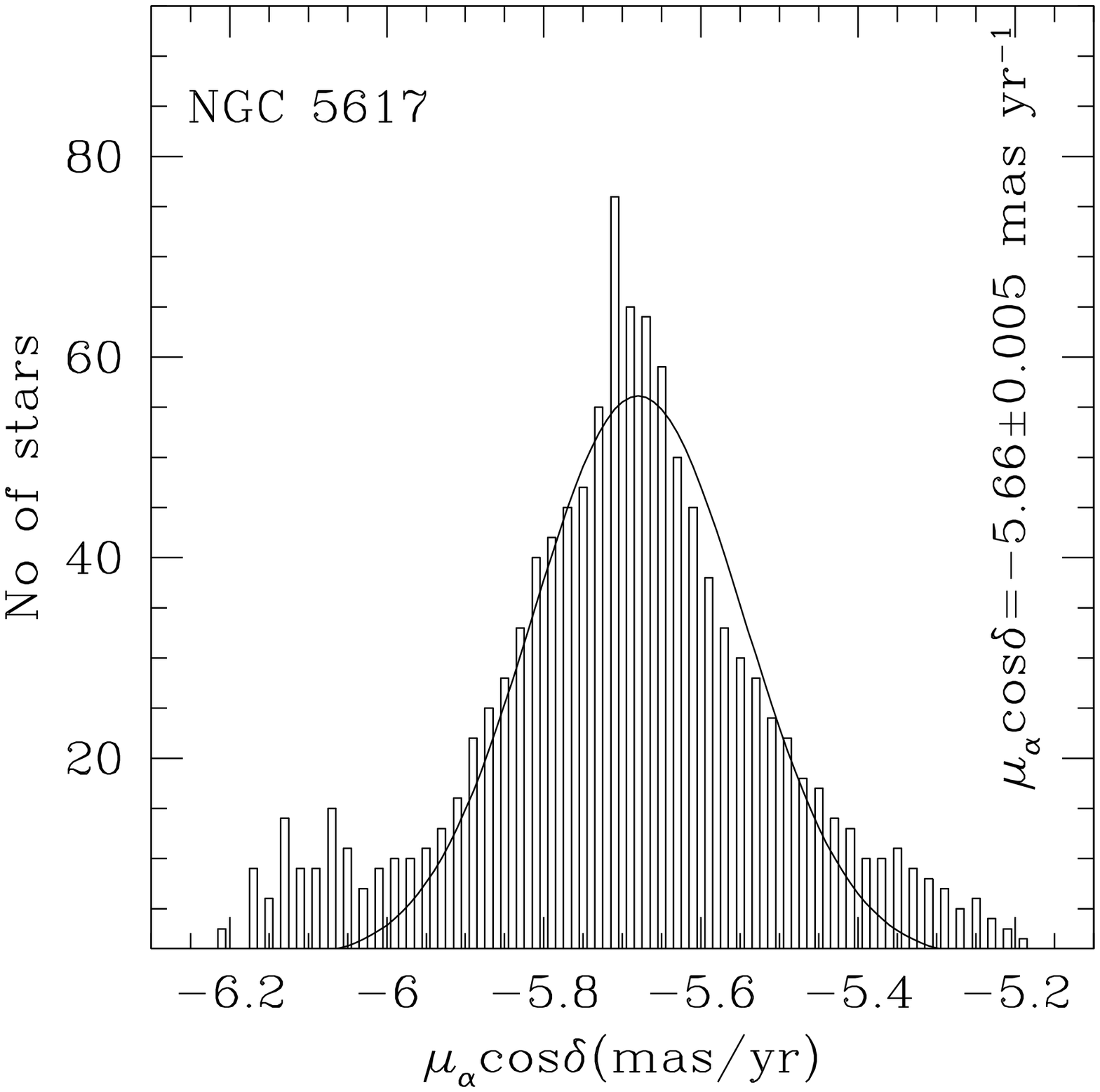}
\includegraphics[width=4.2cm, height=4.2cm]{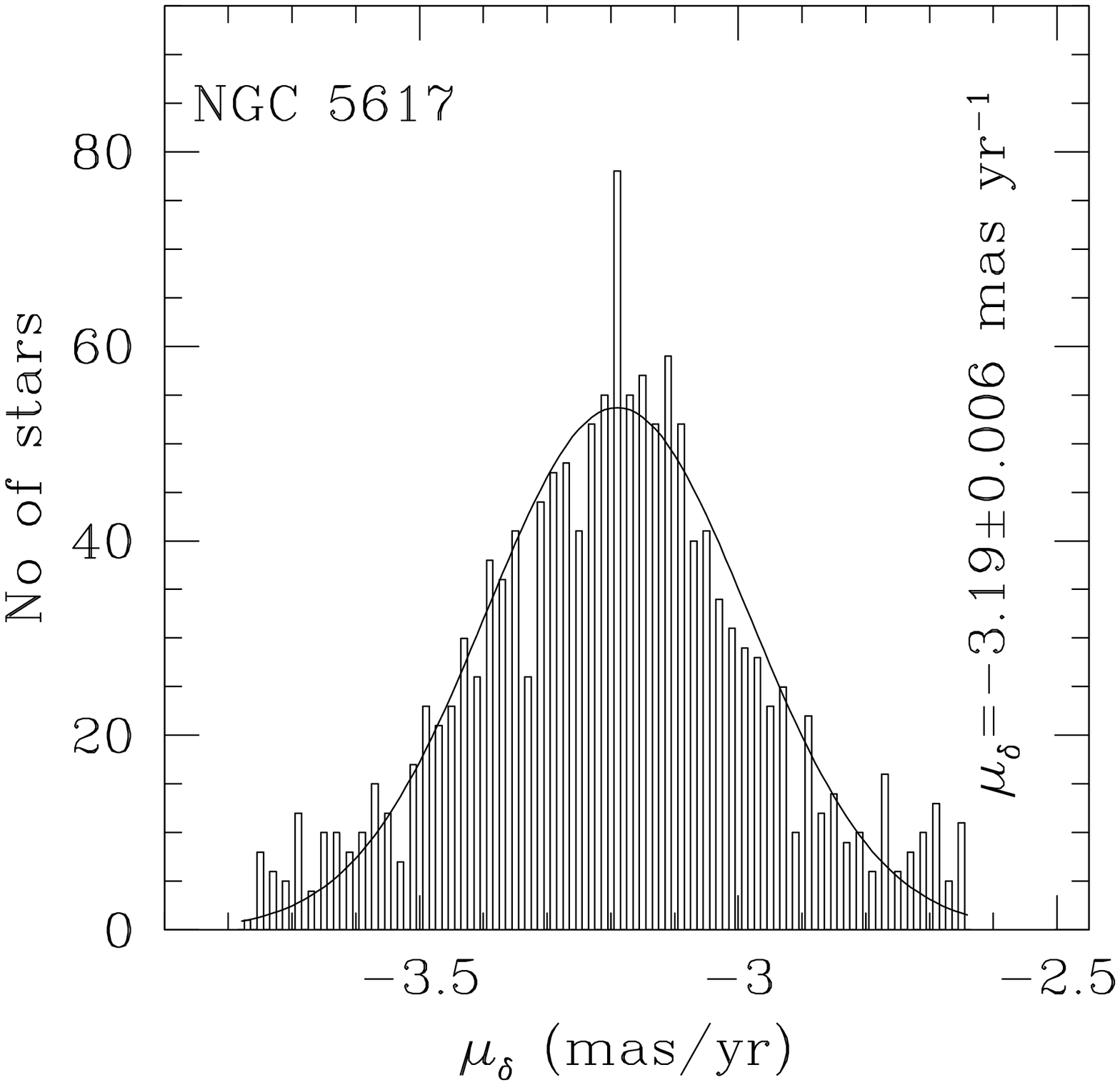}
}
\hbox{
\includegraphics[width=4.2cm, height=4.2cm]{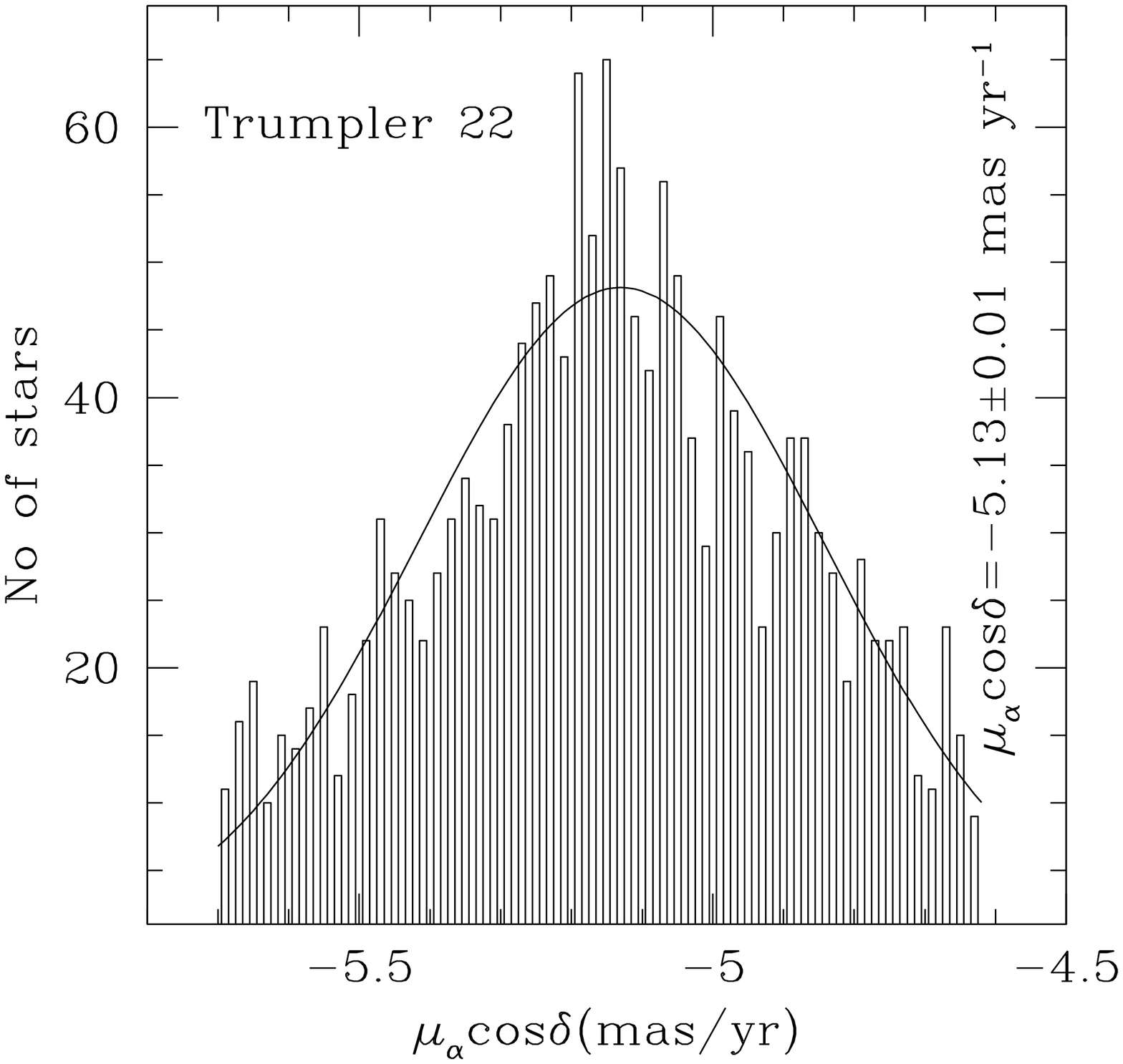}
\includegraphics[width=4.2cm, height=4.2cm]{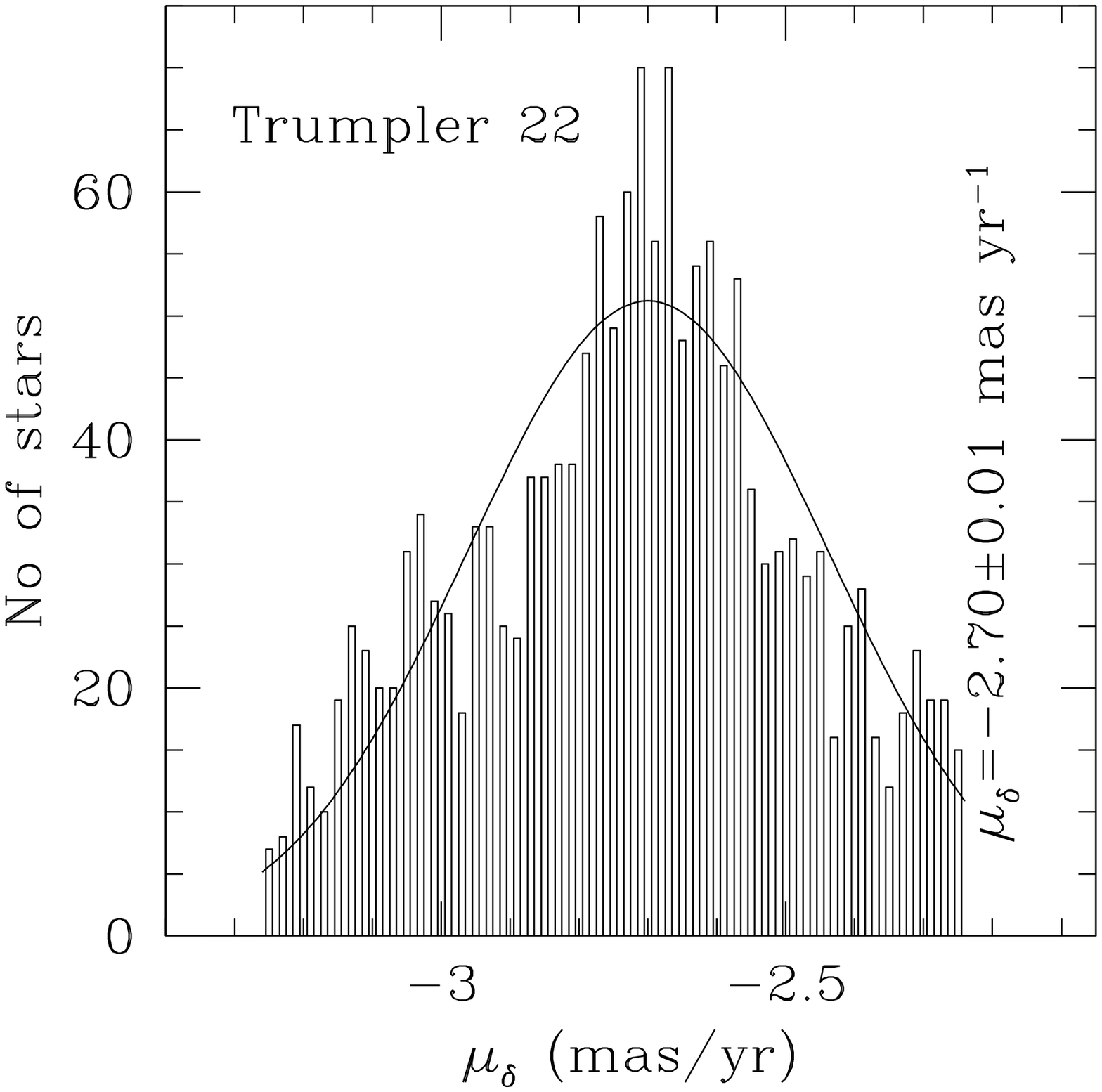}
}
\hbox{
\includegraphics[width=4.2cm, height=4.2cm]{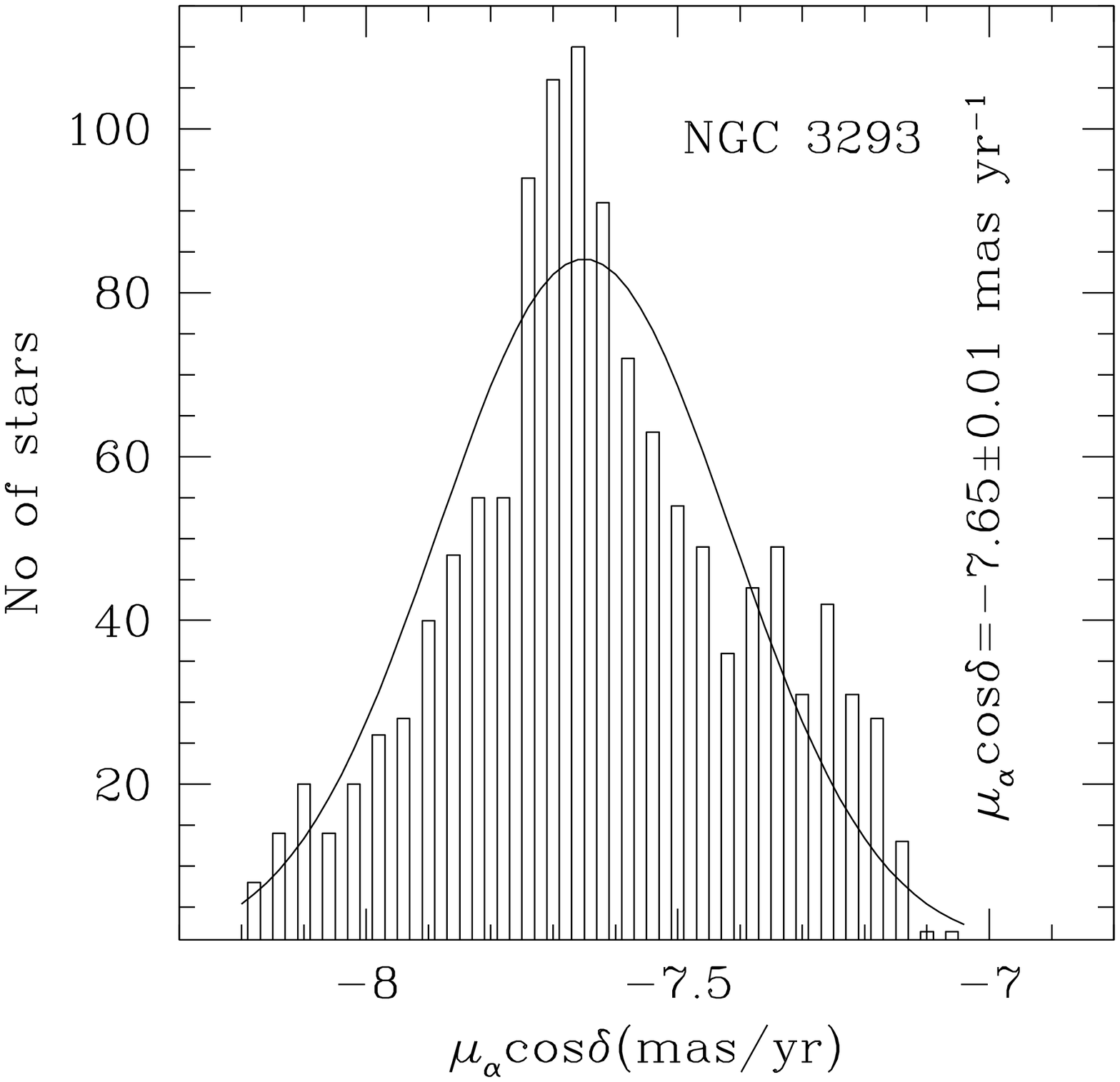}
\includegraphics[width=4.2cm, height=4.2cm]{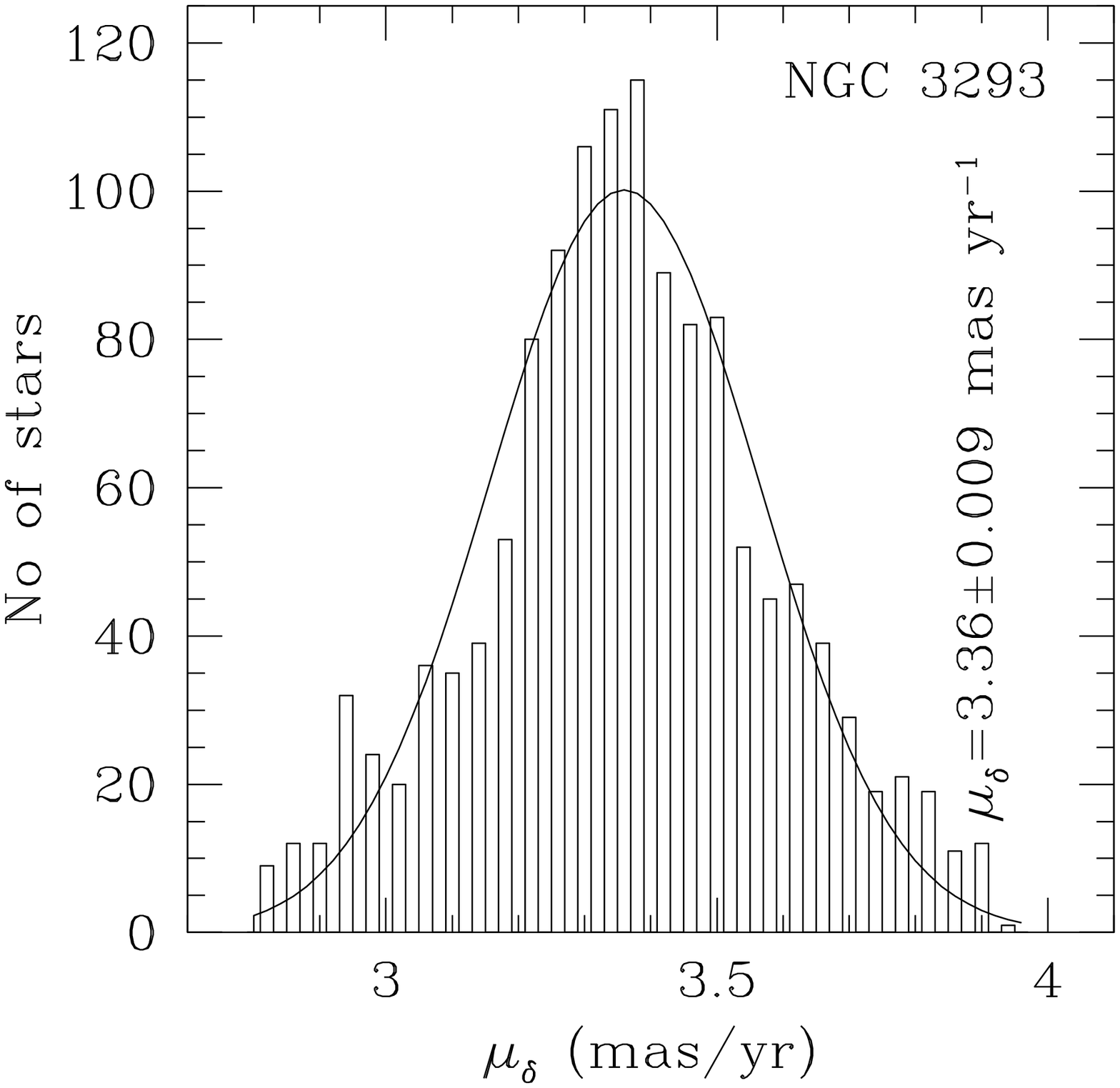}
}
\hbox{
\includegraphics[width=4.2cm, height=4.2cm]{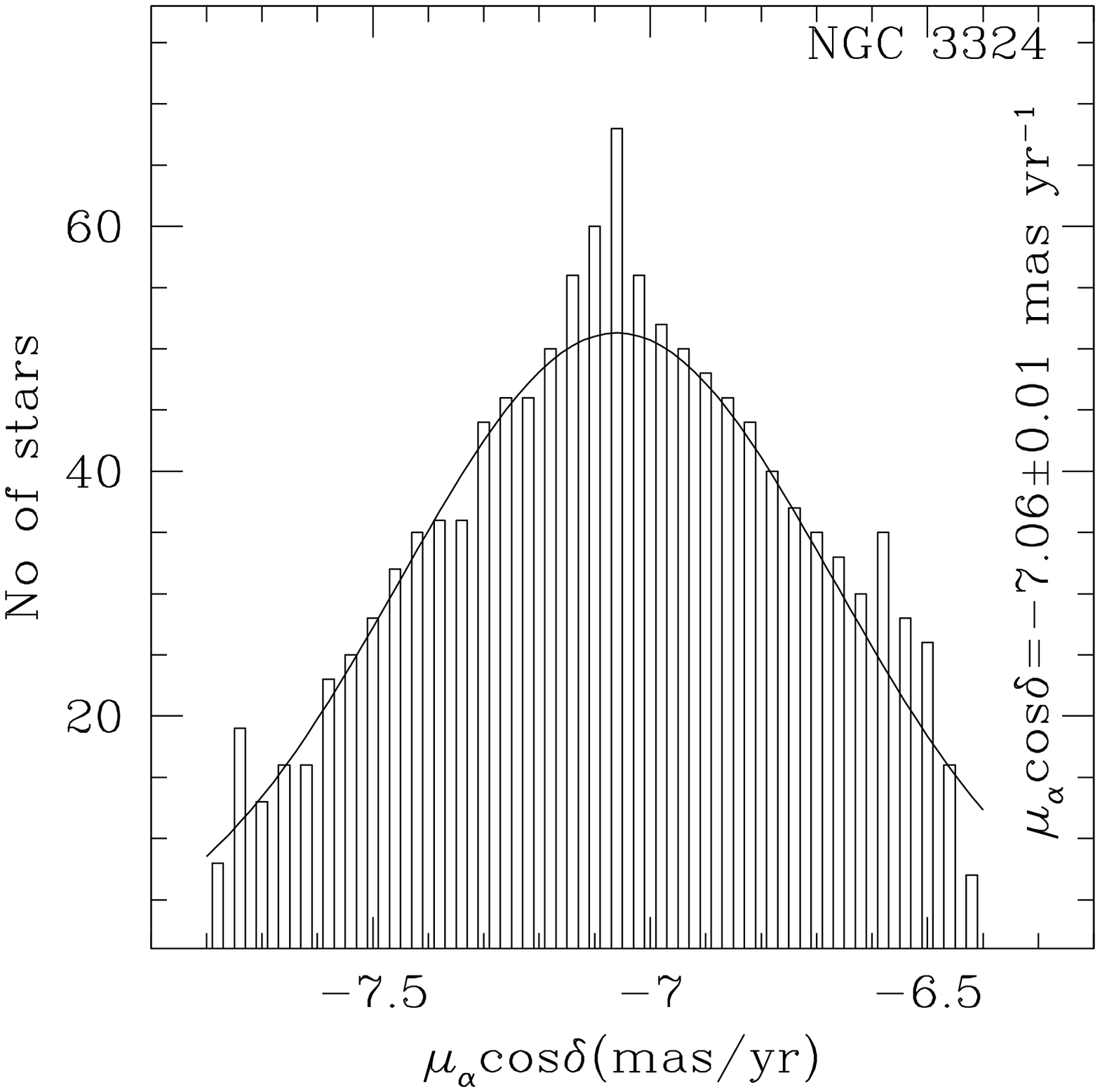}
\includegraphics[width=4.2cm, height=4.2cm]{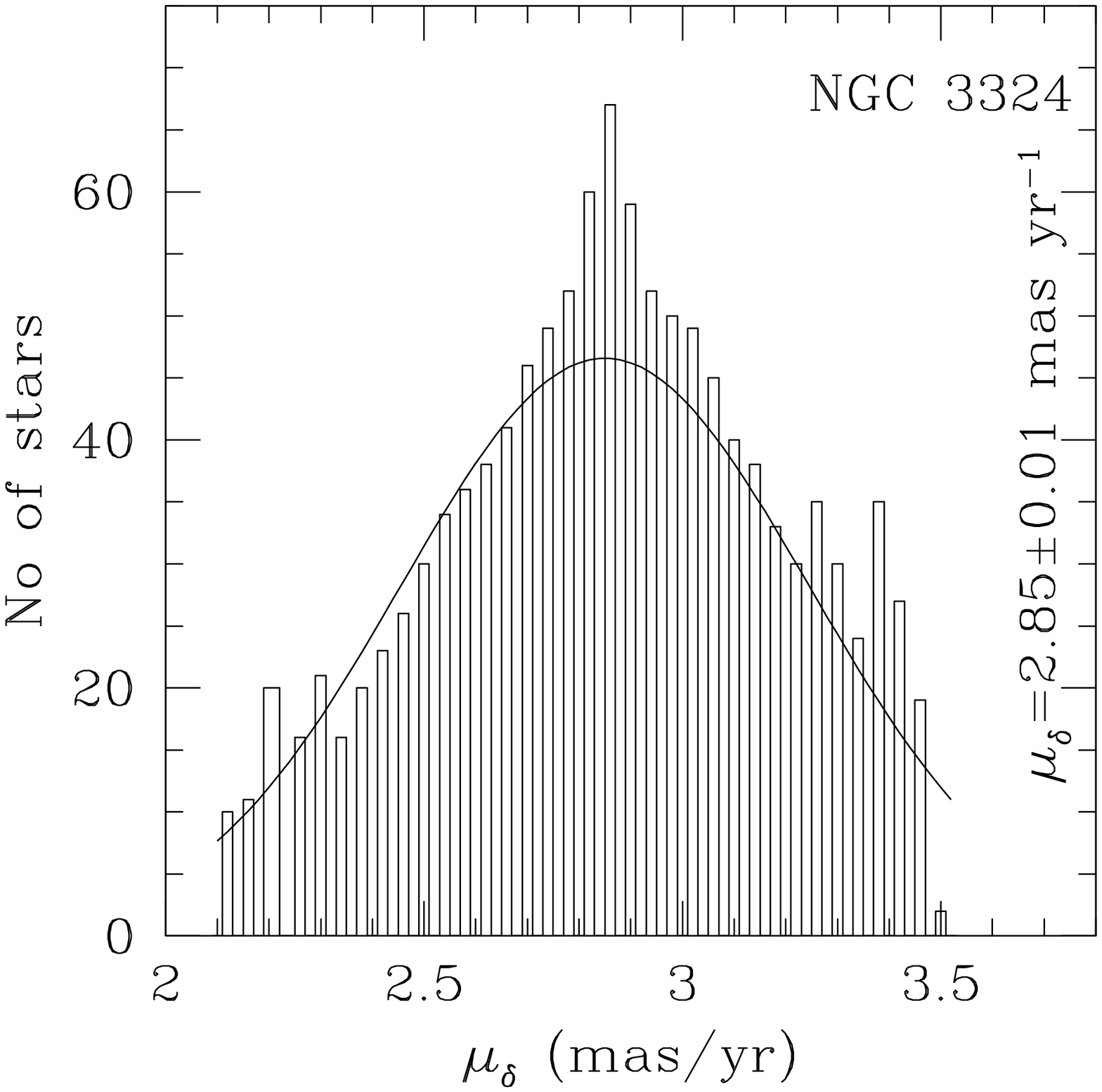}
}
\caption{Proper motion histograms in 0.1 $mas~ yr^{-1}$ bins in $\mu_{\alpha} cos{\delta}$ and $\mu_{\delta}$ of
the clusters. The Gaussian function fit to the central bins provides the mean values in both directions as shown
in each panel.} 
\label{pm_hist}
\end{center}
\end{figure}
The proper motion of stars is very precious to differentiate member stars from field stars. We have used proper motion and
parallax data from the Gaia EDR3 catalog to remove field stars from the clusters NGC 5617, Trumpler 22, NGC 3293 and NGC 3324.
We made a catalog of common stars after matching the Gaia data with the above mentioned photometric data sets in this paper.

PMs, $\mu_{\alpha} cos{\delta}$ and $\mu_{\delta}$ are plotted as vector point diagrams (VPDs) in the top panels of Fig. \ref{vpd} and
Fig. \ref{vpd2} to see the distribution of cluster and field stars. The middle and bottom panels of
Fig. \ref{vpd} show the corresponding $Z$ versus $(Z-Y)$ and $G$ versus $(G_{BP}-G_{RP})$ CMDs for clusters
NGC 5617 and Trumpler 22. In Fig. \ref{vpd2}, we used proper motion distributions of stars in the upper panels while
their corresponding $G, G_{BP}-G_{RP}$ CMDs are plotted in the lower panels for clusters NGC 3293 and NGC 3324. The left
panel in the CMDs show all stars present in the cluster's area, while the middle and right panels show the probable
cluster members and non-member stars respectively. By visual inspection we define the center and radius of the
cluster members in VPD for a preliminary analysis. This selection was performed in a way to minimize the field star contamination and to keep
the maximum possible number of lower mass stars. A circle of 0.6  $mas~ yr^{-1}$ for NGC 5617, NGC 3293 and NGC 3324 while
0.4 $mas~ yr^{-1}$ for Trumpler 22 around the center of the member stars distribution in the VPDs characterize
our membership criteria. The picked radius is an agreement between losing cluster members with poor PMs and the involvement
of non-member stars. We have also used parallax for the reliable estimation of cluster members. A star is considered
as probable cluster member if it lies inside the circle in VPD and has a parallax value within 3$\sigma$ from the mean cluster parallax.
The CMDs of the probable members are shown in the middle and bottom row panels in each cluster CMDs as shown in Fig. \ref{vpd}
and Fig. \ref{vpd2}. The main sequence of the cluster is separated. These stars have a PM error of $\le 0.4 ~ mas~ yr^{-1}$.

For the precise estimation of mean proper motion, we deal with only probable cluster members based on clusters VPDs and
CMDs as shown in Fig. \ref{pm_hist}. By fitting the Gaussian function into the constructed histograms, we determined the
mean proper motion in the directions of RA and DEC, as shown in Fig. \ref{pm_hist}. From the peak of the Gaussian distribution
we found mean-proper motion in RA and DEC directions for all clusters and are listed in Table \ref{para}.
The estimated values of mean proper motions for each cluster are in fair agreement with the values given by Cantat-Gaudin et al. (2018).
Cantat-Gaudin catalog (2018) reports the membership probabilities of few stars towards the region of clusters under study.
We derived membership probabilities of each star in all the studied clusters and the adopted method has been described in
the next section.

\subsection{Membership Probability}
\label{MP}

Open clusters are located within the densely populated Galactic plane and contaminated by a large number of foreground/background
stars. It is necessary to differentiate between cluster members and non-members, in order to derive reliable cluster fundamental
parameters. In this paper, we used the membership estimation criteria for clusters NGC 5617, Trumpler 22, NGC 3293 and
NGC 3324 as given by Balaguer-N\'{u}\~{n}ez et al. (1998). This method has been previously used by many authors
(Bellini et al. 2009, Bisht et al. 2020; Sariya et al. 2021a,2021b; Yadav et al. 2013; Sariya \& Yadav 2015). For the
cluster and field star distributions, two different distribution functions ($\phi_c^{\nu}$) and ($\phi_f^{\nu}$) are constructed
for a particular i$^{th}$ star. The values of frequency distribution functions are given as follows:

\begin{center}
   $\phi_c^{\nu} =\frac{1}{2\pi\sqrt{{(\sigma_c^2 + \epsilon_{xi}^2 )} {(\sigma_c^2 + \epsilon_{yi}^2 )}}}$

$\times$ exp$\{{ -\frac{1}{2}[\frac{(\mu_{xi} - \mu_{xc})^2}{\sigma_c^2 + \epsilon_{xi}^2 } + \frac{(\mu_{yi} - \mu_{yc})^2}{\sigma_c^2 + \epsilon_{yi}^2}] }\}$ \\
\end{center}
\begin{center}
and\\
\end{center}
\begin{center}
$\phi_f^{\nu} =\frac{1}{2\pi\sqrt{(1-\gamma^2)}\sqrt{{ (\sigma_{xf}^2 + \epsilon_{xi}^2 )} {(\sigma_{yf}^2 + \epsilon_{yi}^2 )}}}$

$\times$ exp$\{{ -\frac{1}{2(1-\gamma^2)}[\frac{(\mu_{xi} - \mu_{xf})^2}{\sigma_{xf}^2 + \epsilon_{xi}^2}}
-\frac{2\gamma(\mu_{xi} - \mu_{xf})(\mu_{yi} - \mu_{yf})} {\sqrt{(\sigma_{xf}^2 + \epsilon_{xi}^2 ) (\sigma_{yf}^2 + \epsilon_{yi}^2 )}} + \frac{(\mu_{yi} - \mu_{yf})^2}{\sigma_{yf}^2 + \epsilon_{yi}^2}]\}$\\
\end{center}

\begin{figure}
\begin{center}
\hbox{
\includegraphics[width=4.2cm, height=4.2cm]{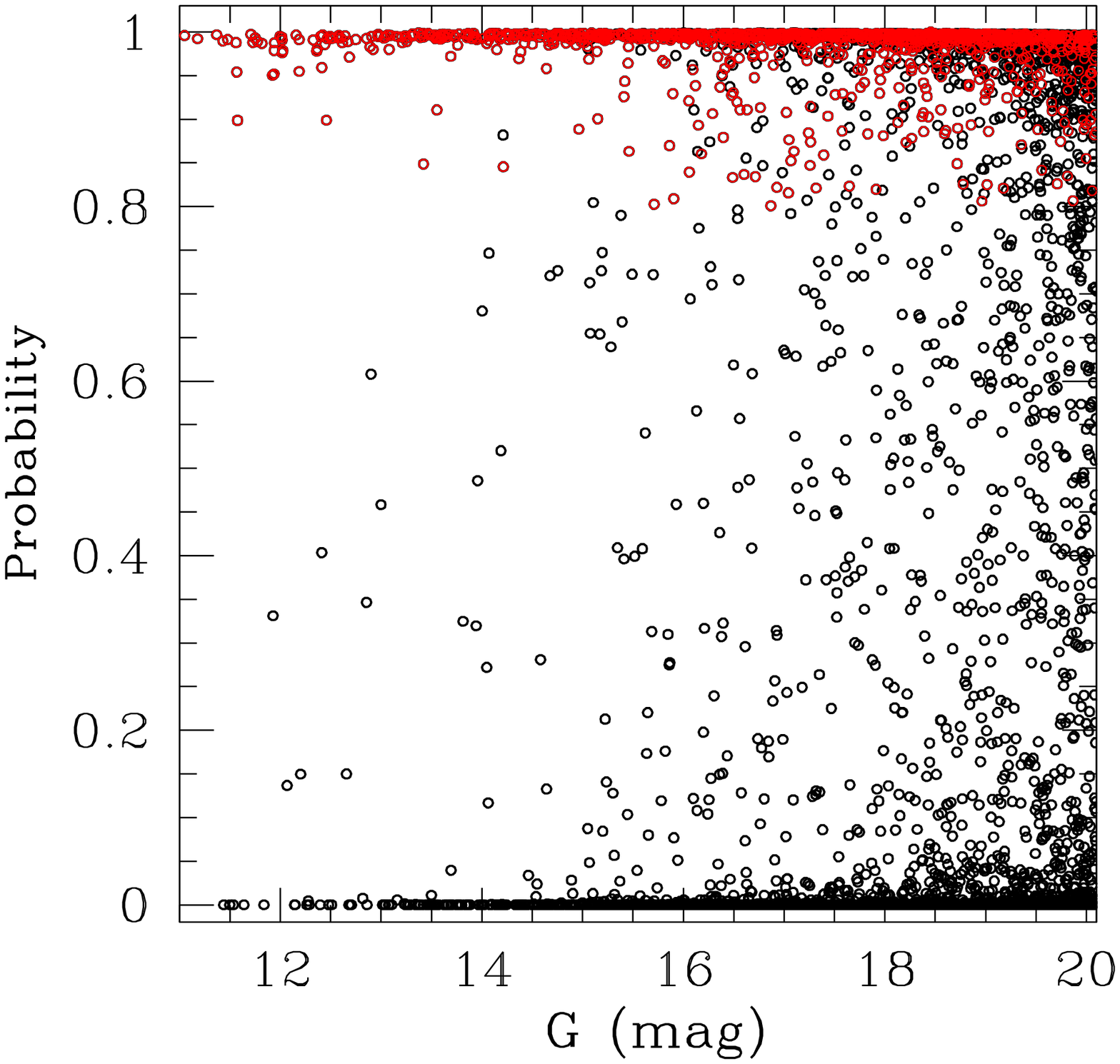}
\includegraphics[width=4.2cm, height=4.2cm]{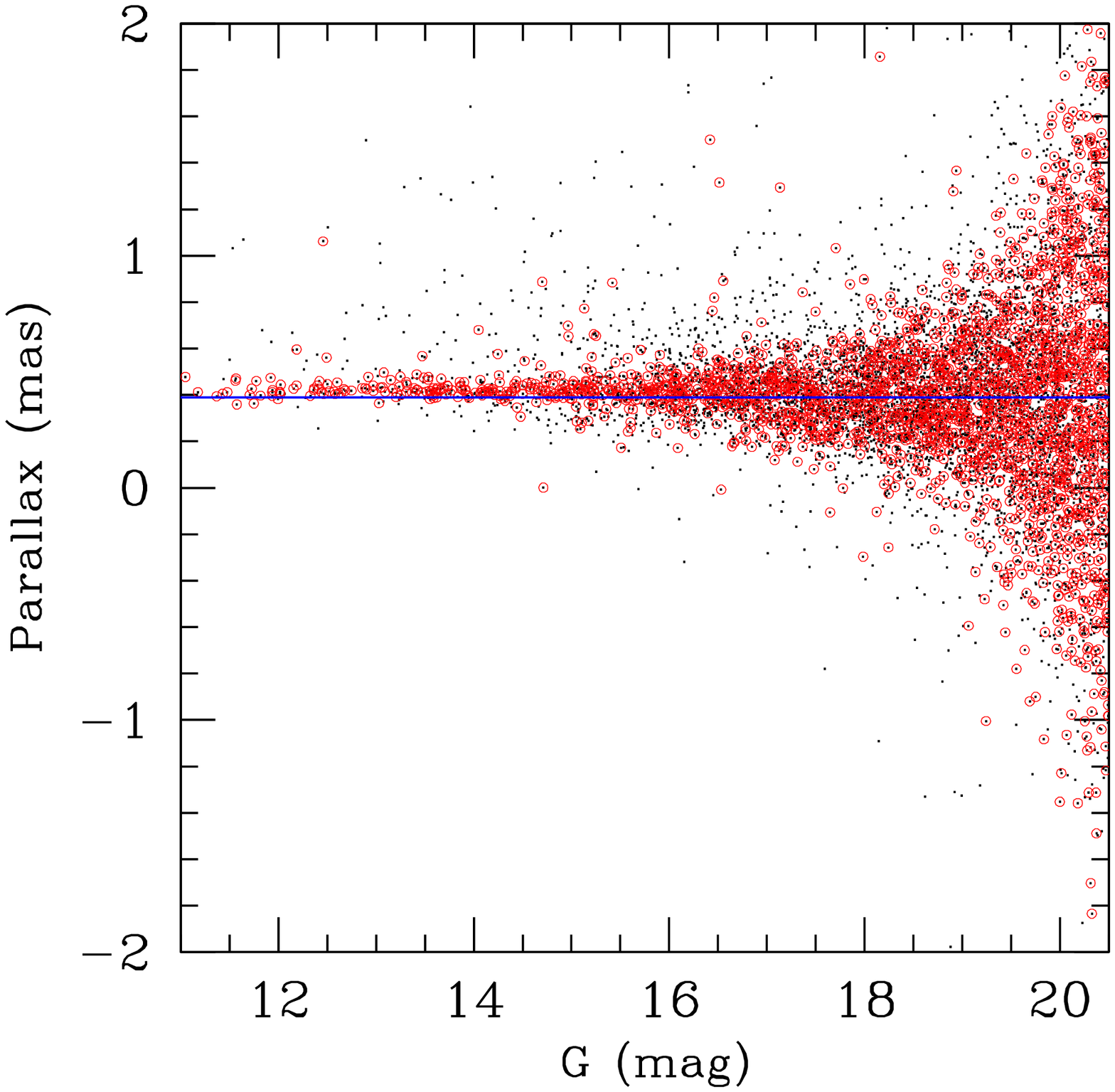}
}
\hbox{
\includegraphics[width=4.2cm, height=4.2cm]{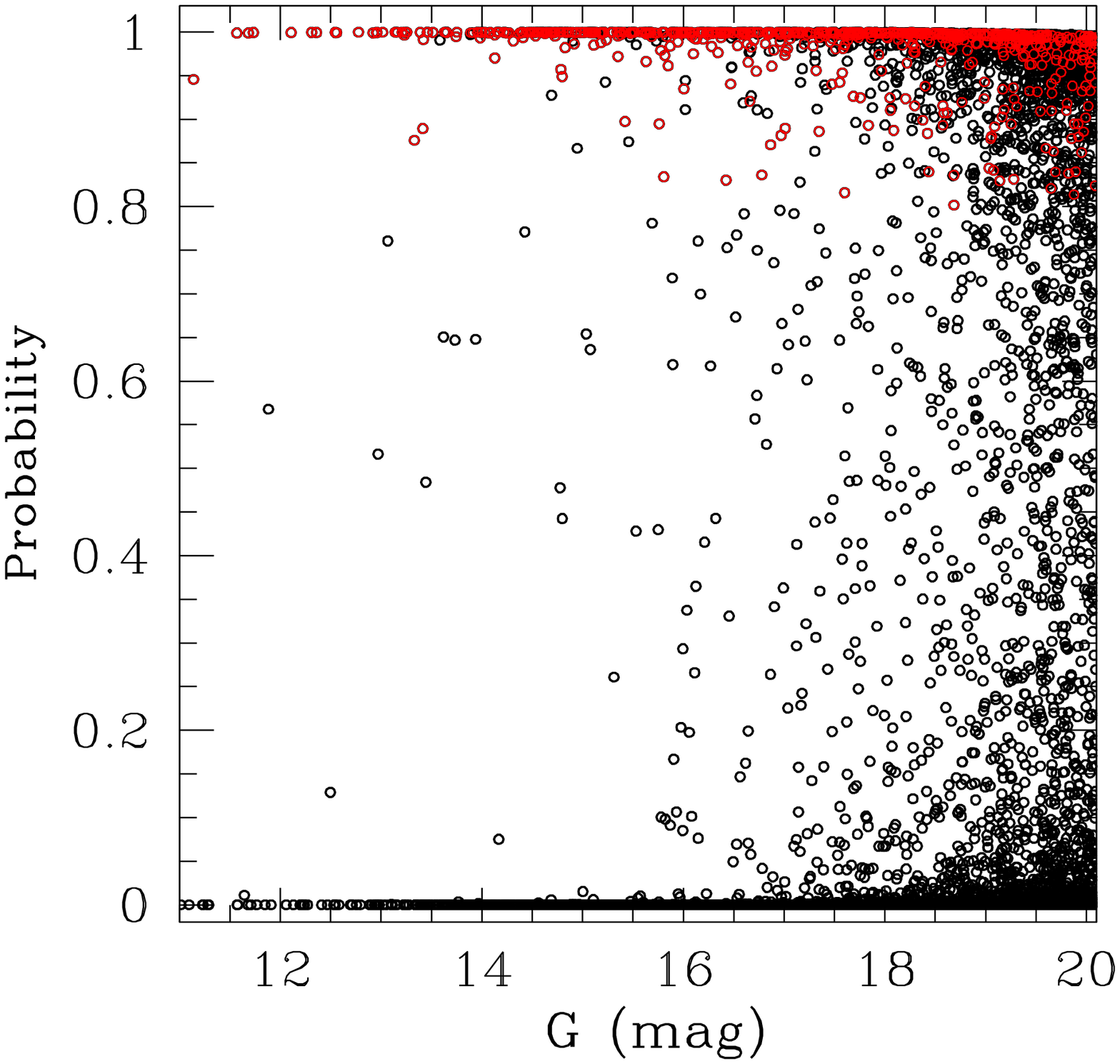}
\includegraphics[width=4.2cm, height=4.2cm]{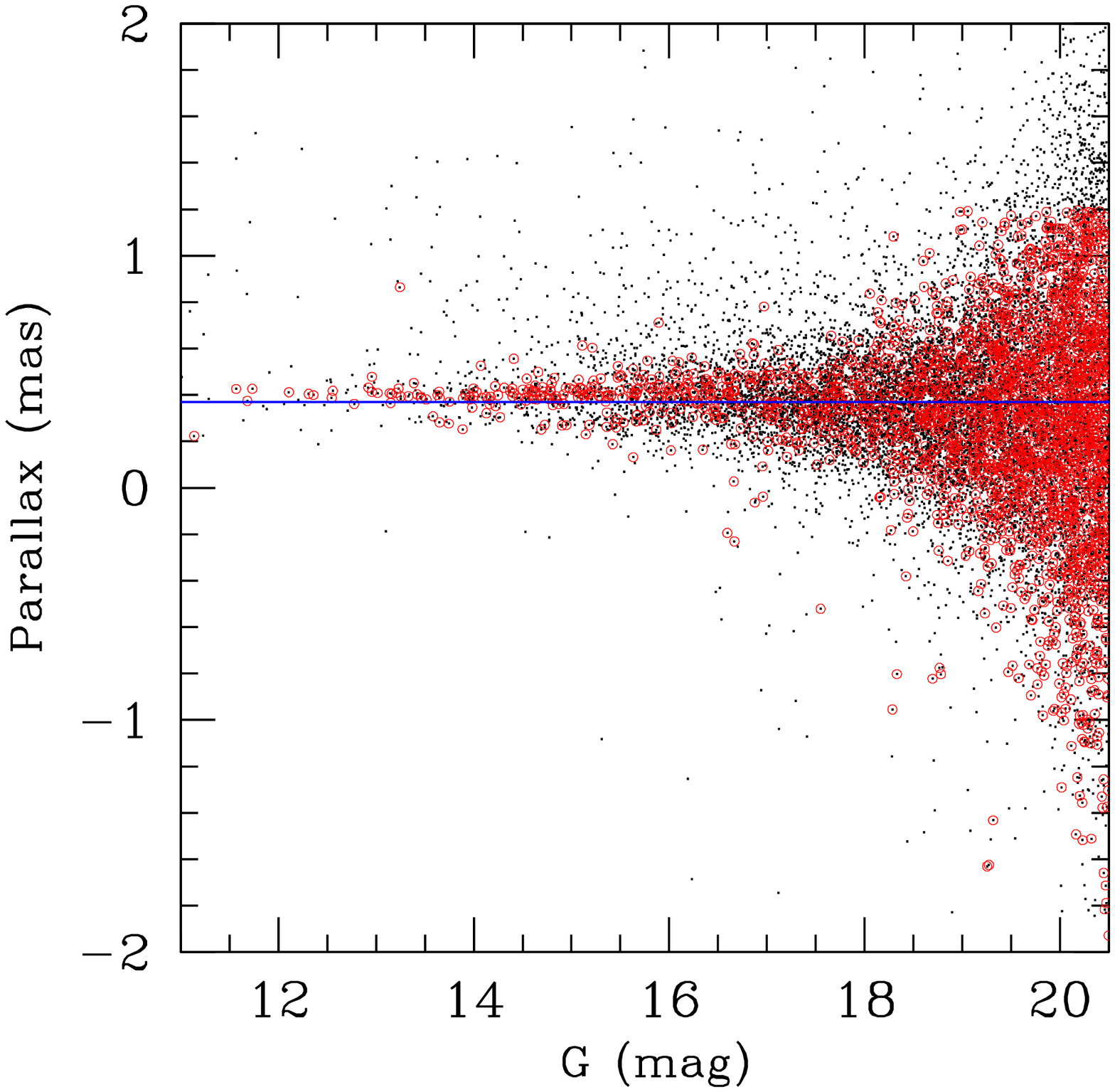}
}
\hbox{
\includegraphics[width=4.2cm, height=4.2cm]{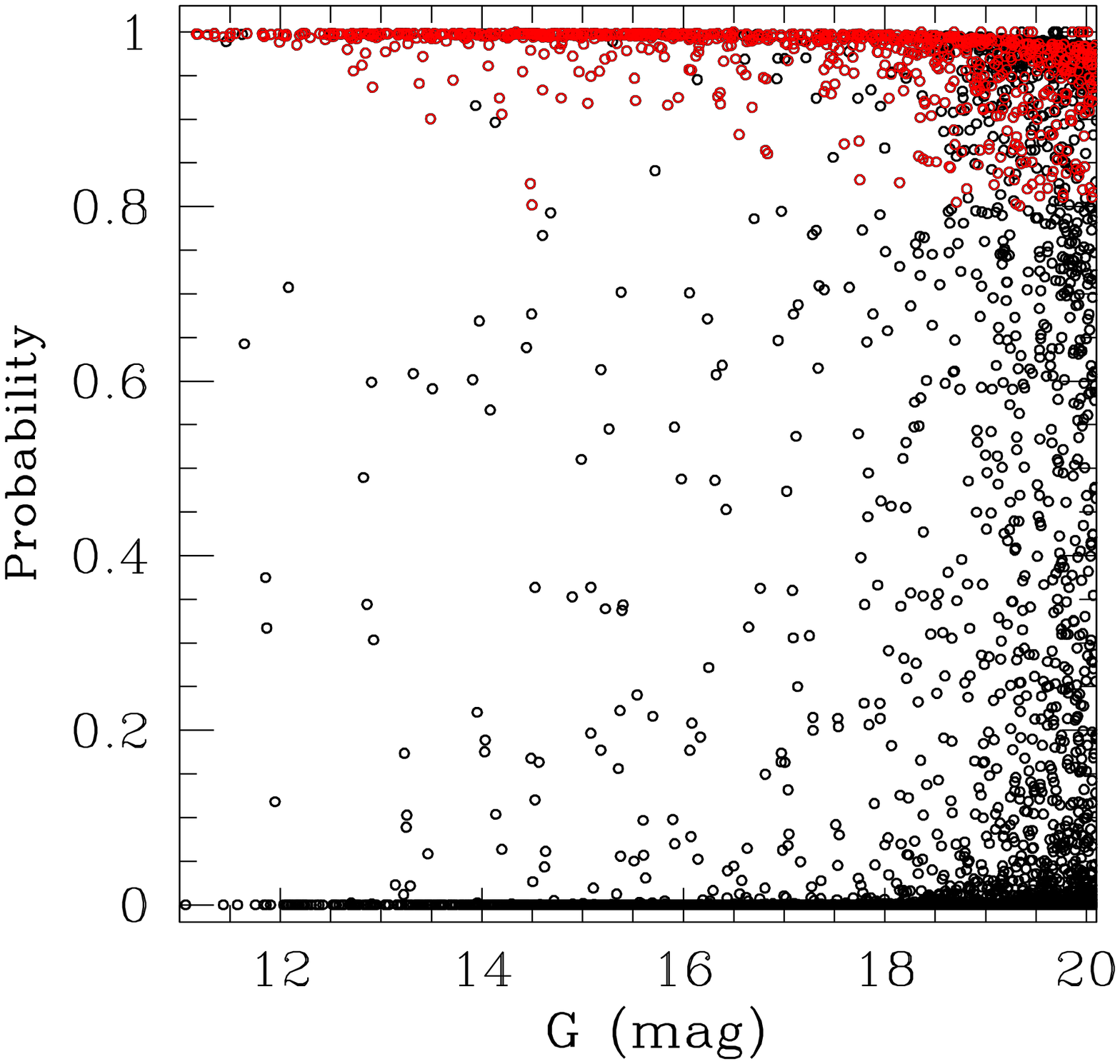}
\includegraphics[width=4.2cm, height=4.2cm]{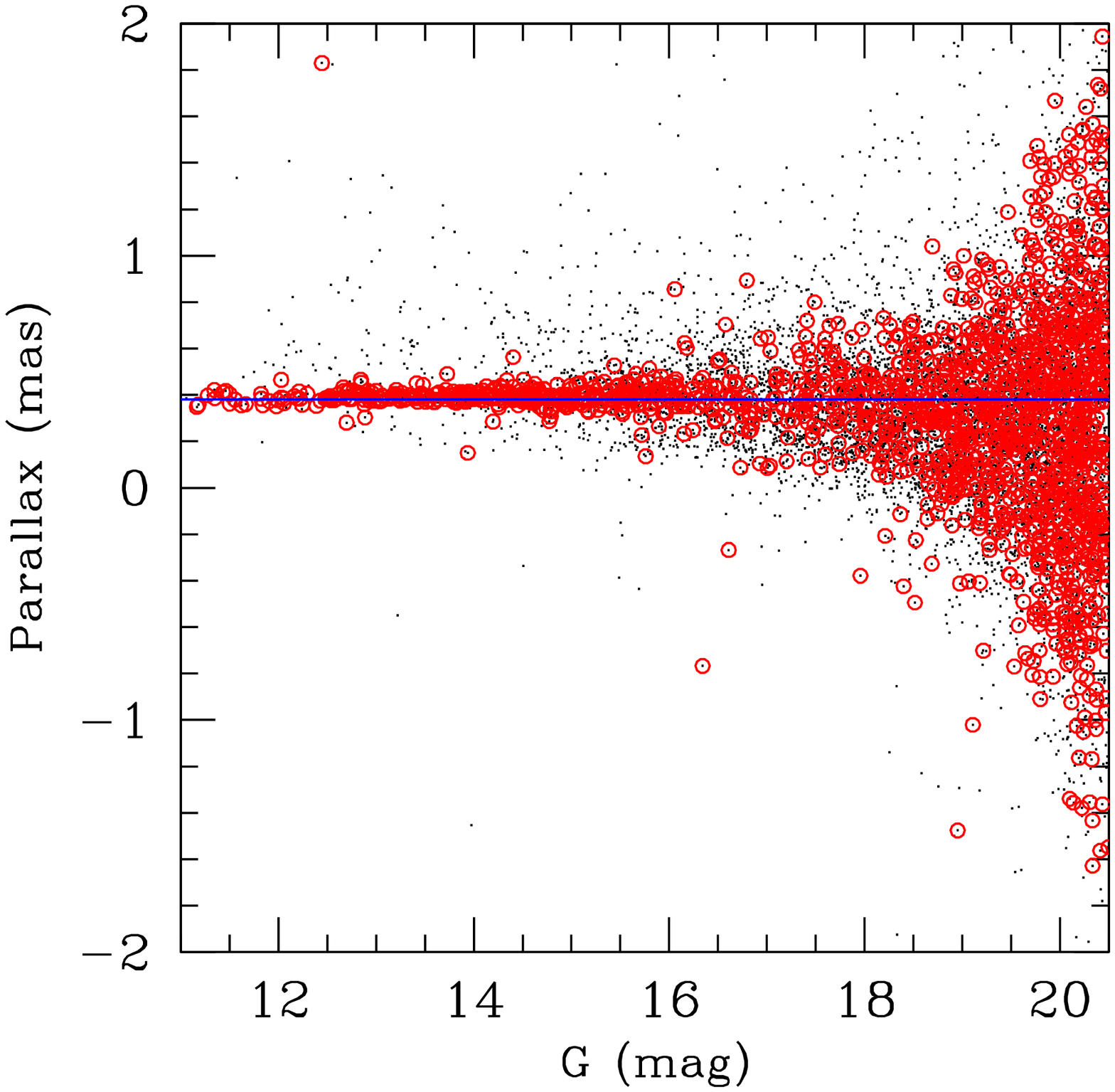}
}
\hbox{
\includegraphics[width=4.2cm, height=4.2cm]{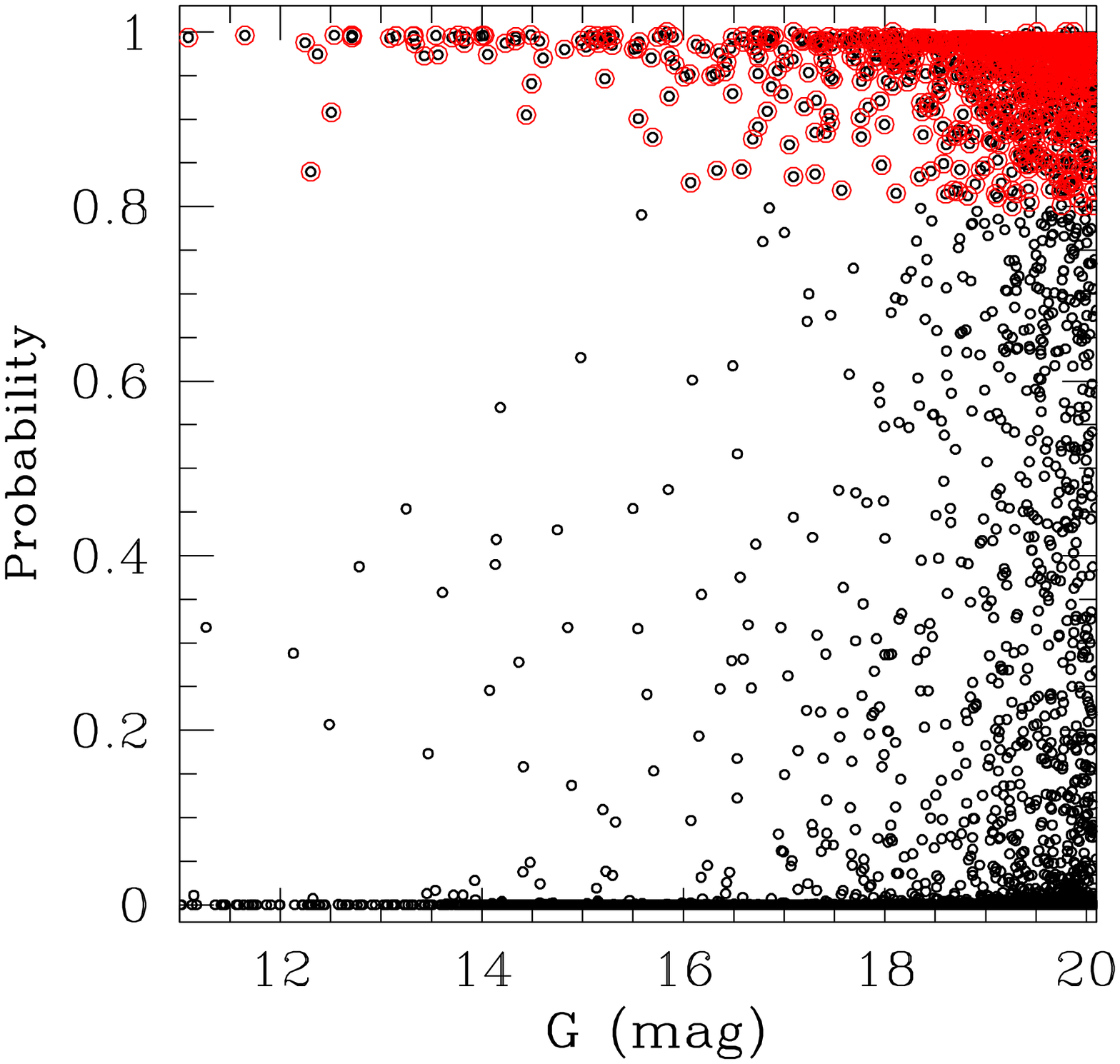}
\includegraphics[width=4.2cm, height=4.2cm]{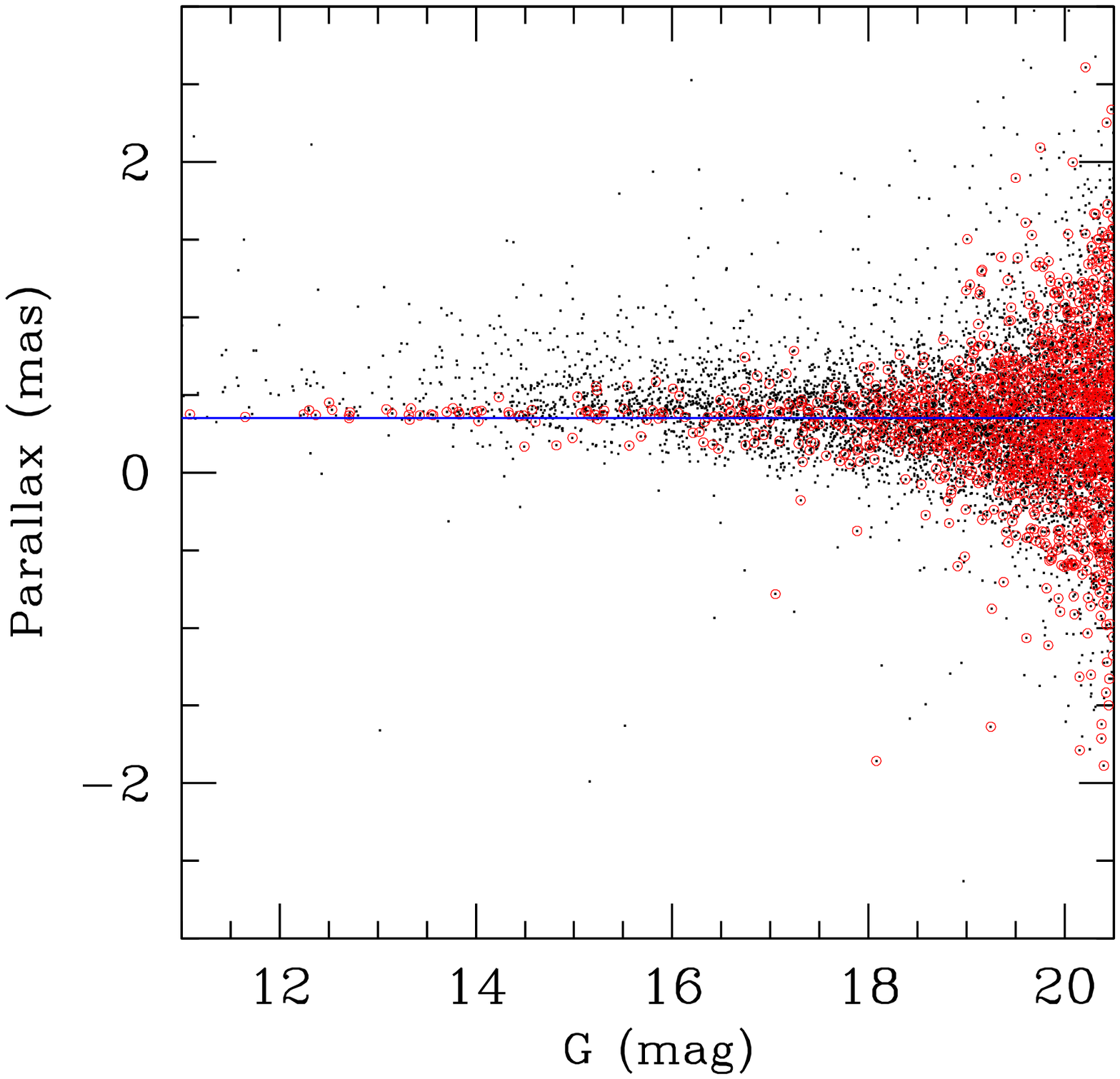}
}
\caption{(Top left panel) Membership probability as a function of G magnitude for NGC 5617. (Top right panel) Parallax as
a function of G magnitude for NGC 5617. Same have been plotted for clusters Trumpler 22, NGC 3293 and NGC 3324 from
top to bottom. Red dots are cluster members with membership probability higher than 80$\%$.} 
\label{mp}
\end{center}
\end{figure}

where ($\mu_{xi}$, $\mu_{yi}$) are the PMs of $i^{th}$ star. The PM errors are represented by ($\epsilon_{xi}$, $\epsilon_{yi}$).
The cluster's PM center is given by ($\mu_{xc}$, $\mu_{yc}$) and ($\mu_{xf}$, $\mu_{yf}$) represent the center of field PM values.
The intrinsic PM dispersion for the cluster stars is denoted by $\sigma_c$, whereas $\sigma_{xf}$ and $\sigma_{yf}$ provide the
intrinsic PM dispersion's for the field populations. The correlation coefficient $\gamma$ is calculated as:\\

\begin{center}
$\gamma = \frac{(\mu_{xi} - \mu_{xf})(\mu_{yi} - \mu_{yf})}{\sigma_{xf}\sigma_{yf}}$.
\end{center}

Stars with PM errors $\le$0.5 mas~yr$^{-1}$ have been used to determine $\phi_c^{\nu}$ and $\phi_f^{\nu}$. A group of stars
is found at $\mu_{xc}$=$-$5.66 mas~yr$^{-1}$, $\mu_{yc}$=$-$3.19 mas~yr$^{-1}$ for NGC 5617,
$\mu_{xc}$=$-$5.13 mas~yr$^{-1}$, $\mu_{yc}$=$-$2.70 mas~yr$^{-1}$ for Trumpler 22, $\mu_{xc}$=$-$7.65 mas~yr$^{-1}$,
$\mu_{yc}$=3.36 mas~yr$^{-1}$ for NGC 3293 and $\mu_{xc}$=$-$7.06 mas~yr$^{-1}$, $\mu_{yc}$=2.85 mas~yr$^{-1}$ for
NGC 3324. Assuming a distance of 2.52, 2.68, 2.65 and 2.85 kpc for clusters NGC 5617, Trumpler 22, NGC 3293 and NGC 3324, respectively and radial
velocity dispersion of 1 km $s^{-1}$ for open star clusters (Girard et al. 1989), the expected dispersion ($\sigma_c$) in PMs
would be $\sim$0.08 $mas~yr^{-1}$ for clusters NGC 5617 and Trumpler 22 while  $\sim$0.10 $mas~yr^{-1}$ for other two
clusters. For field region stars, we have estimated ($\mu_{xf}$, $\mu_{yf}$) = ($-$3.5, $-$5.2) mas yr$^{-1}$ for NGC 5617,
($\mu_{xf}$, $\mu_{yf}$) = ($-$3.2, $-$4.5) mas yr$^{-1}$ for Trumpler 22,  ($\mu_{xf}$, $\mu_{yf}$) = ($-$ 5.5, 2.0) mas yr$^{-1}$
for NGC 3293, ($\mu_{xf}$, $\mu_{yf}$) = ($-$ 5.3, 1.3) mas yr$^{-1}$ for NGC 3324 and ($\sigma_{xf}$, $\sigma_{yf}$) = (4.5, 4.9),
(3.8, 4.1),  (5.5, 4.8), (4.9, 3.7) mas yr$^{-1}$ for NGC 5617, Trumpler 22, NGC 3293 and NGC 3324, respectively.

Considering the normalized numbers of cluster stars and field stars as $n_{c}$ and $n_{f}$ respectively (i.e., $n_c + n_f = 1$), the
total distribution function can be calculated as\\

\begin{center}
$\phi = (n_{c}~\times~\phi_c^{\nu}) + (n_f~\times~\phi_f^{\nu})$,  \\
\end{center}

As a result, the membership probability for the $i^{th}$ star is given by:\\
\begin{center}
$P_{\mu}(i) = \frac{\phi_{c}(i)}{\phi(i)}$. \\
\end{center}

\begin{figure*}
\begin{center}
\centering
\hbox{
\includegraphics[width=6.5cm, height=6.5cm]{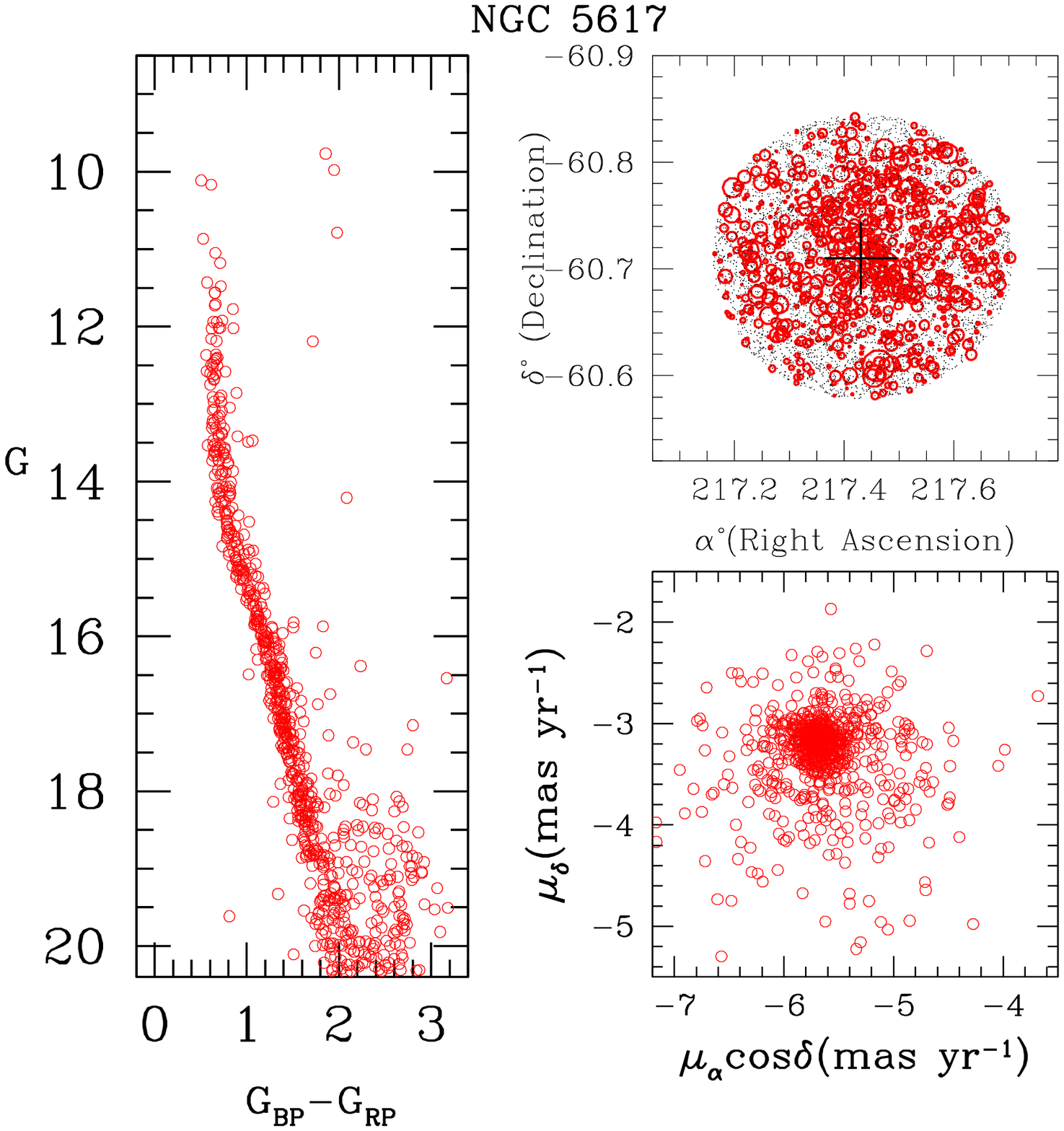}
\includegraphics[width=6.5cm, height=6.5cm]{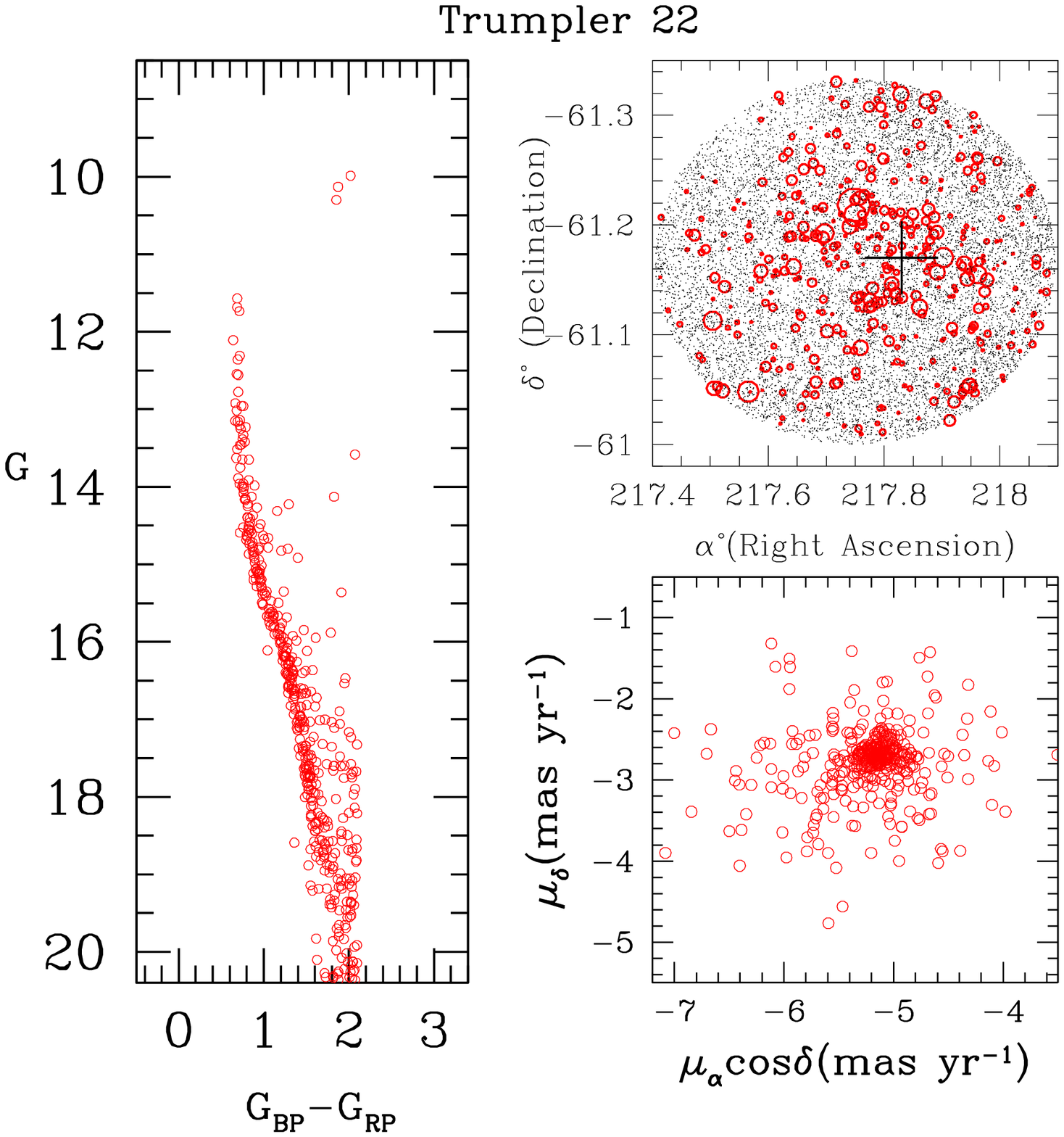}
}
\hbox{
\includegraphics[width=6.5cm, height=6.5cm]{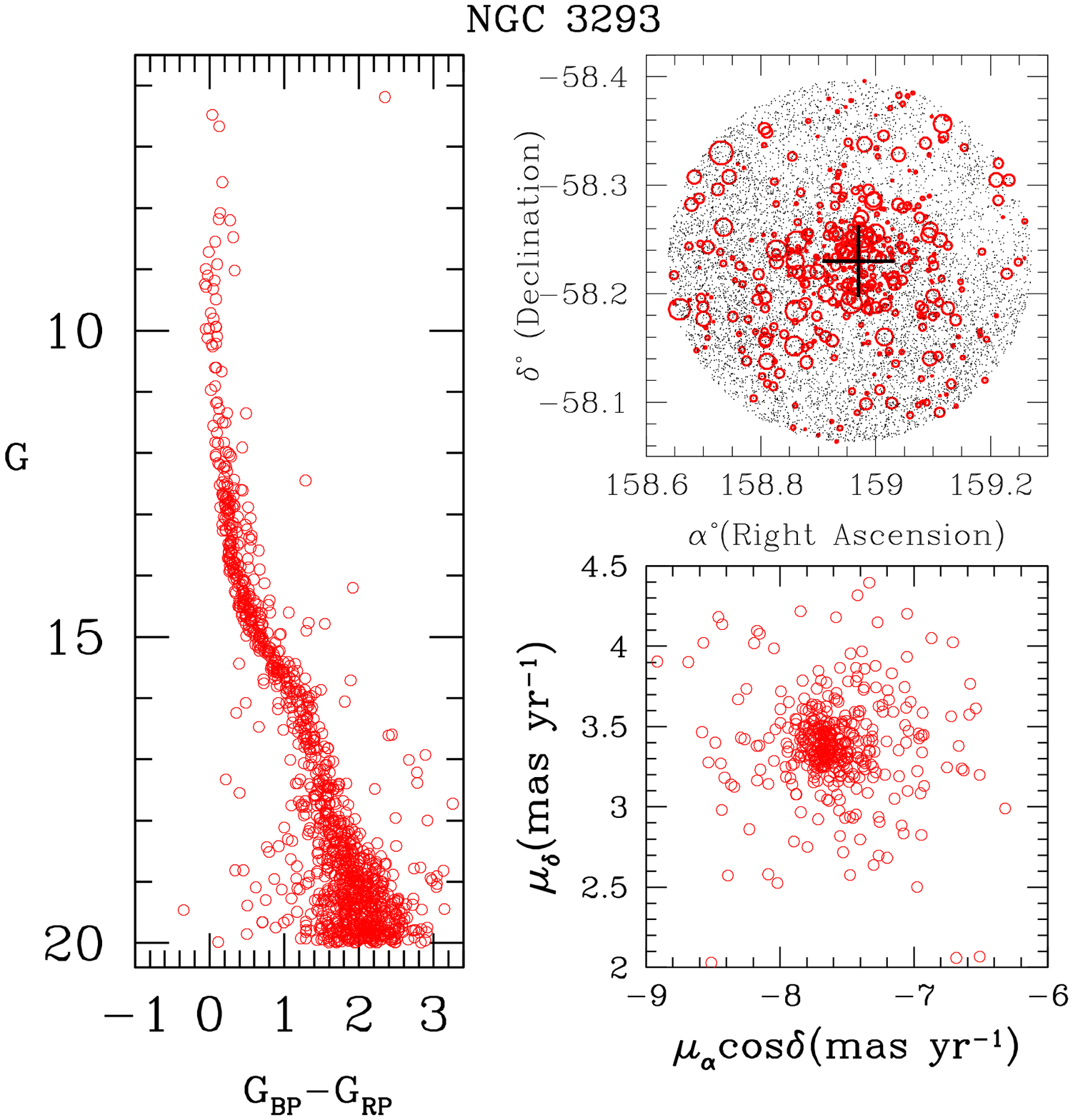}
\includegraphics[width=6.5cm, height=6.5cm]{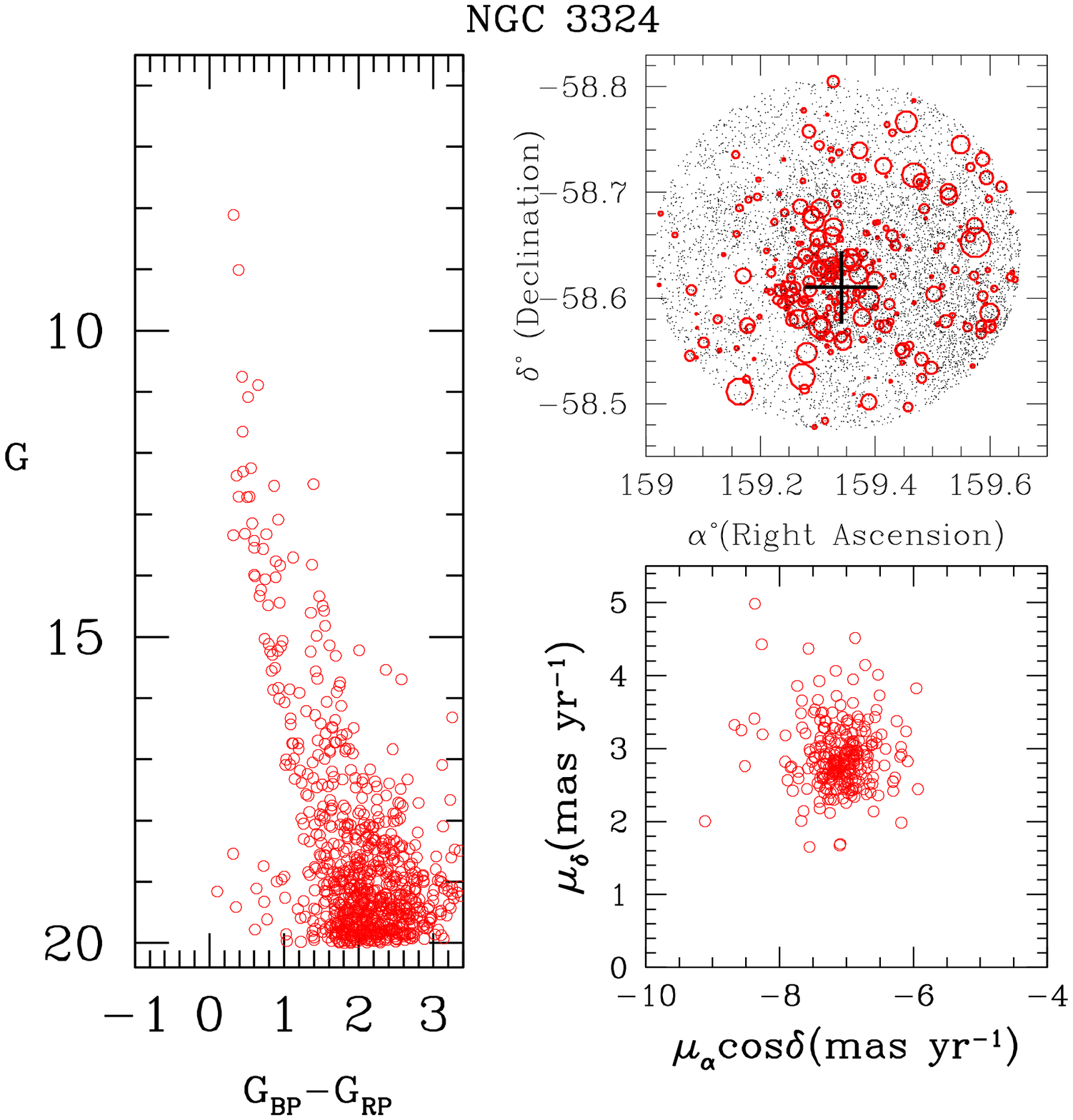}
}
\caption{($G, G_{BP}-G_{RP}$) CMDs, identification charts and proper motion distribution of member stars with membership
probability higher than $80\%$ for clusters under study. The `plus' sign indicates the cluster center.}
\label{mpms}
\end{center}
\end{figure*}

In this way, we identified 584, 429, 692 and 273 stars as cluster members for NGC 5617, Trumpler 22, NGC 3293 and NGC 3324,
respectively with membership probability higher than $80\%$ and $G\le20$ mag. In the top left panel and top right panel of
Fig. \ref{mp}, we plotted membership probability versus $G$ magnitude and parallax versus $G$ magnitude, respectively for
cluster NGC 5617. Same have been plotted from top to bottom panels in Fig. \ref{mp} for clusters Trumpler 22, NGC 3293 and
NGC 3324. For all clusters, we have plotted $G$ versus ($G_{BP}-G_{RP}$) CMD, the identification chart and proper motion
distribution using stars with membership probability higher than $80\%$ in Fig. \ref{mpms}. The Cantat-Gaudin et al. (2018)
catalog reports membership probabilities for all clusters under study. We matched our likely members with this catalog
having membership probability higher than 80$\%$ and those have been denoted by blue dots in CMDs as shown in Fig. \ref{dist_age}
and Fig. \ref{dist_age1}.

\subsection{Determination of the effectiveness of probabilities}

The stellar density of the cluster region is affected by the presence of foreground and background
stars. In this regard, we have calculated the effectiveness of membership determination for the clusters under study
using the expression given below (Shao \& Zhao, 1996):\\

$E=1-\frac{N\times\Sigma[P_{i}(1-P_{i})]}{\Sigma P_{i}\Sigma(1-P_{i})}$\\

where $N$ is the total number of cluster members and $P_{i}$ indicates the probability of $i^{th}$ star of the cluster.
We have found the effectiveness $(E)$ values as 0.52, 0.55, 0.62, 0.59 for clusters NGC 5617, Trumpler 22, NGC 3293
and NGC 3324, respectively. Shao \& Zhao (1996) shows that the effectiveness of membership determination of 43 open clusters
ranges from 0.20 to 0.90 and the peak value is 0.55 (Wu et al. 2002). Our estimated value of effectiveness of membership
determination are on the higher side for all objects.

\section{Orbit analysis of clusters}

\subsection{Galactic potential model} \label{inic}

We adopted the approach given by Allen \& Santillan (1991) for Galactic potentials in clusters NGC 5617, Trumpler 22, NGC 3293 and NGC 3324.
According to their model, the mass of Galaxy is described by three components: spherical central bulge,
massive spherical halo, and disc. Recently Bajkova \& Bobylev (2016) and Bobylev et. al (2017) refined the parameters of Galactic
potential models with the help of new observational data for a distance R $\sim$ 0 to 200 kpc. These potentials are given as    \\

$ \Phi_{b}(r,z) = -\frac{M_{b}}{\sqrt{r^{2} + b_{b}^{2}}} $   \\

$ \Phi_{d}(r,z) = - \frac{M_{d}}{\sqrt{r^{2} + (a_{d} + \sqrt{z^{2} + b_{d}^{2}})^{2}}}  $ \\

$ \Phi_{h}(r,z) = - \frac{M_{h}}{a_{h}} ln(\frac{\sqrt{r^{2} + a_{h}^{2}} + a_{h}}{r}) $    \\

Where  $ \Phi_{b} $ , $ \Phi_{d} $ and $ \Phi_{h} $ are the potentials of central bulge, disc and halo of Galaxy respectively. $r$ and $z$ are
the distances of objects from Galactic center and Galactic disc respectively. The halo potential is taken from Wilkinson \& Evans (1999) and
values of the constants are taken from Bajkova and Bobylev (2016).

\subsection{Orbits Calculation}

\begin{figure*}
\begin{center}
\hbox{
\includegraphics[width=3.5cm, height=3.5cm]{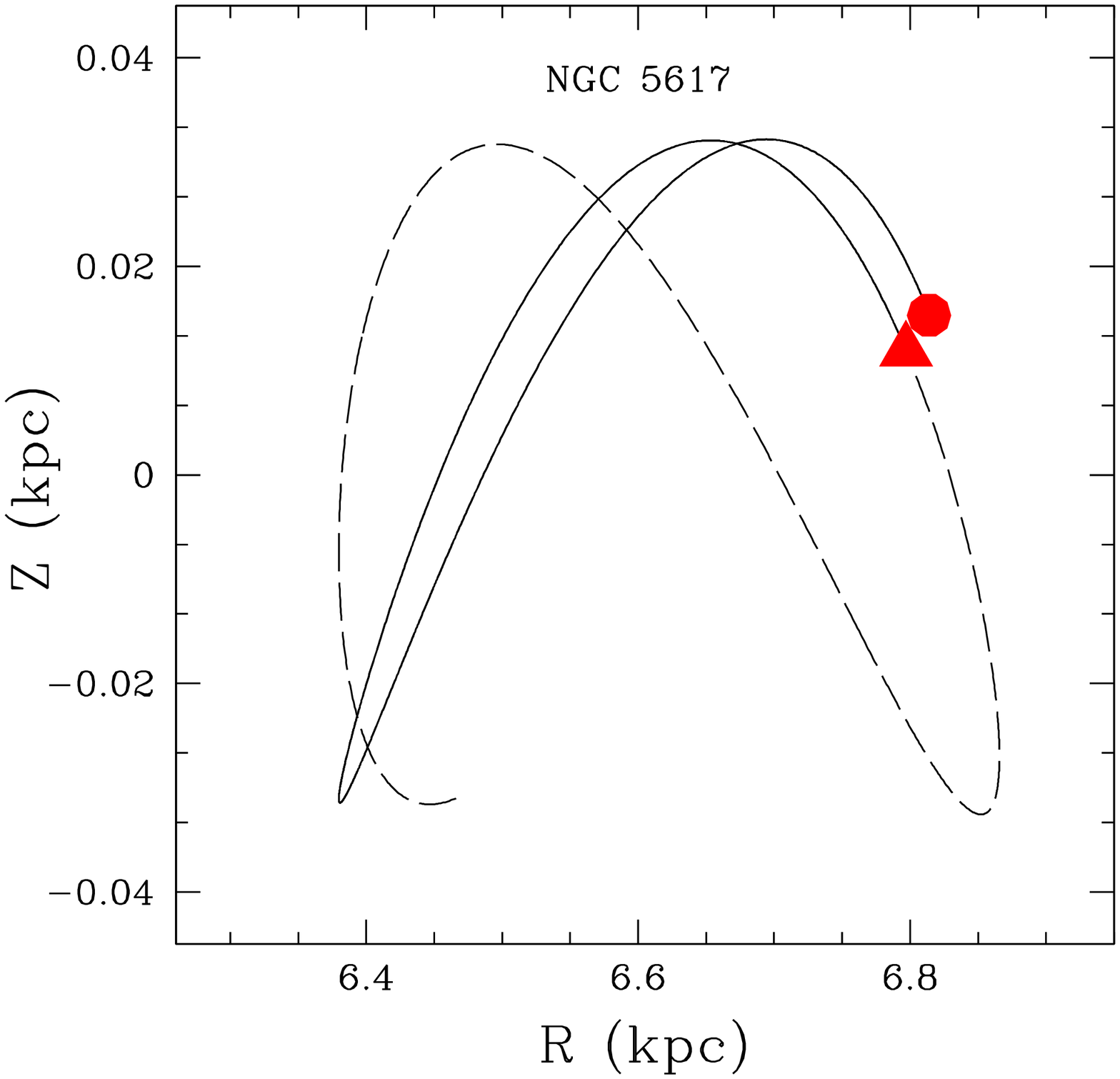}
\hspace{1cm}\includegraphics[width=3.5cm, height=3.5cm]{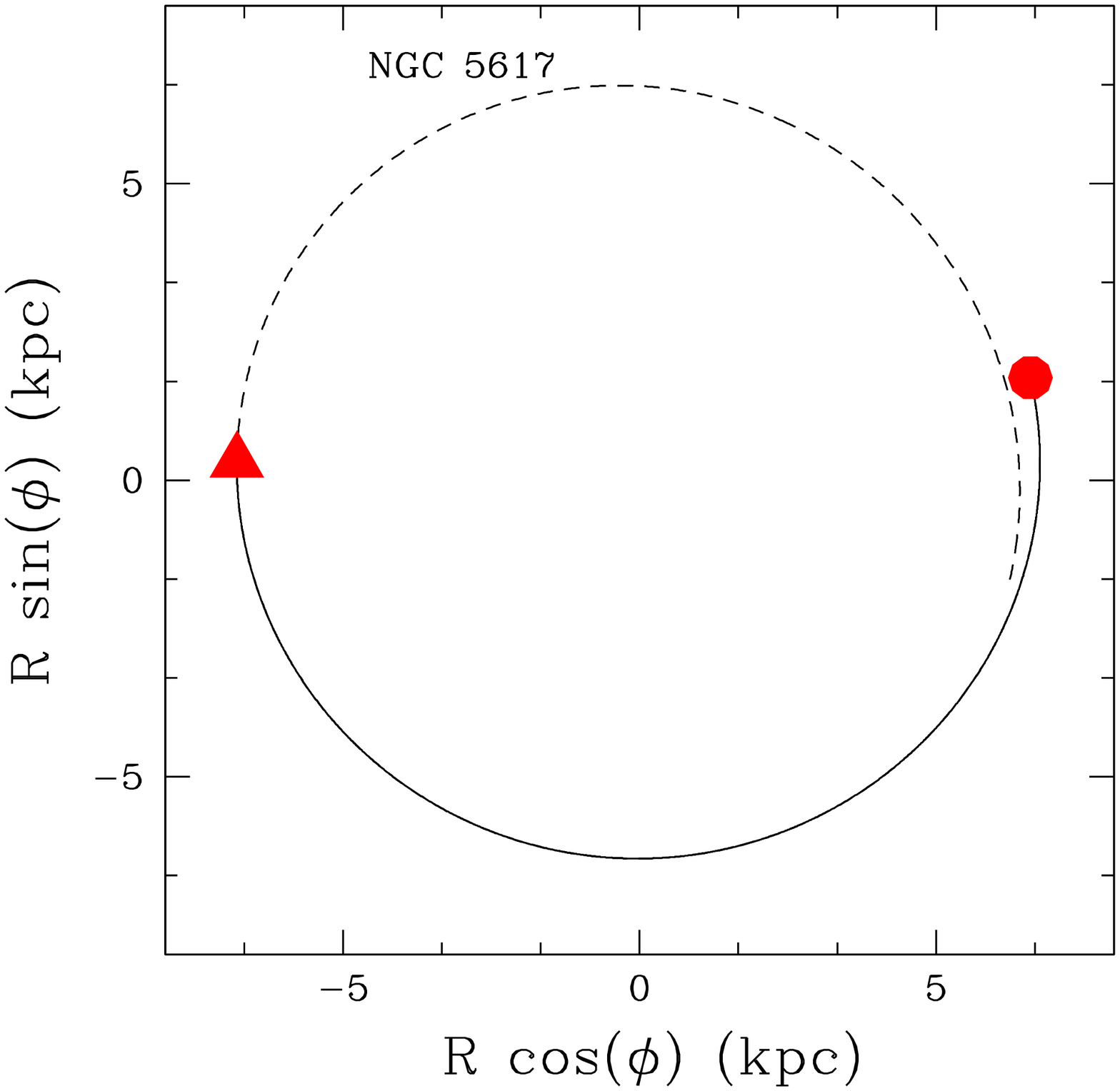}
\hspace{1cm}\includegraphics[width=3.5cm, height=3.5cm]{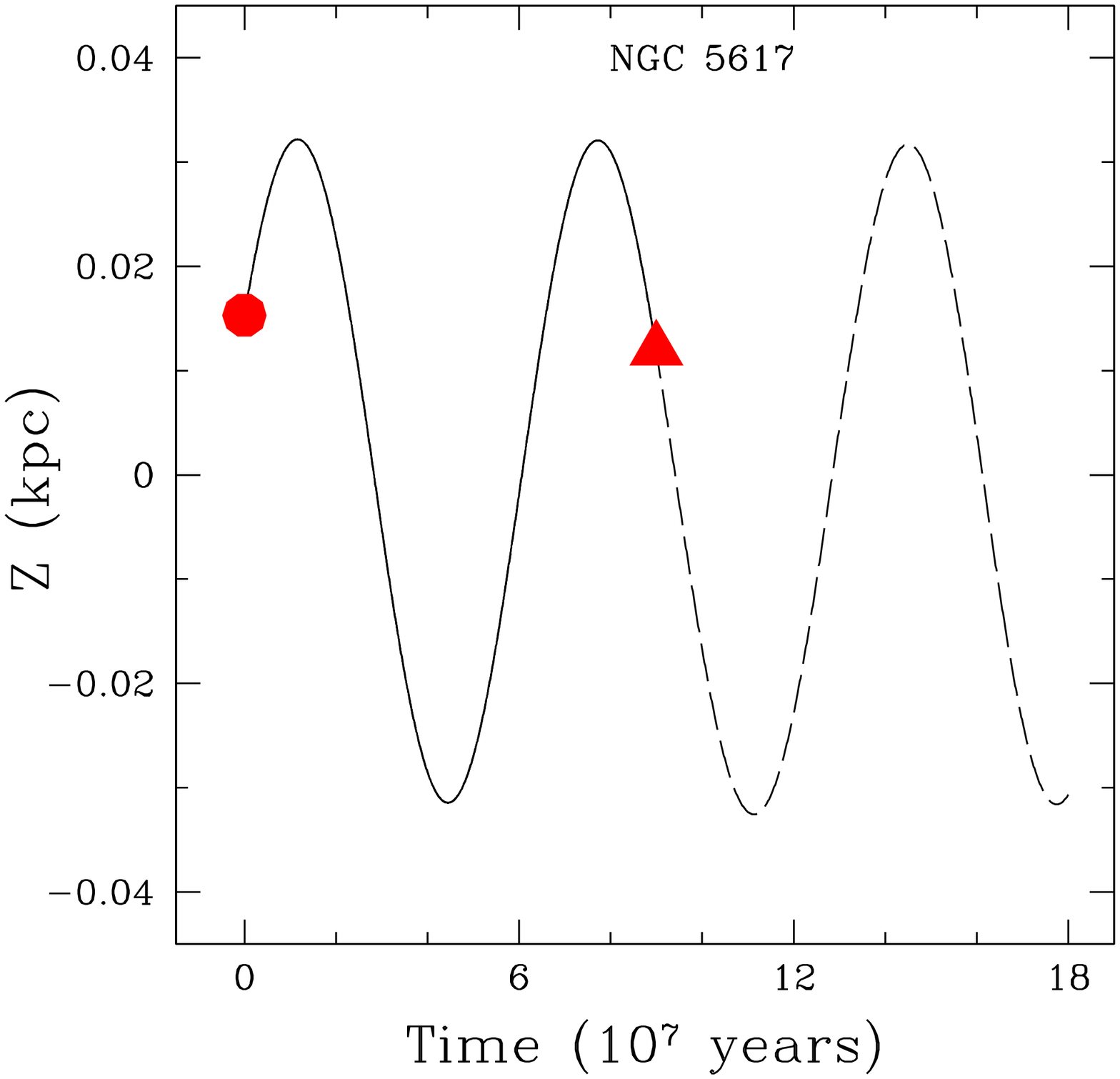}
}
\hbox{
\includegraphics[width=3.5cm, height=3.5cm]{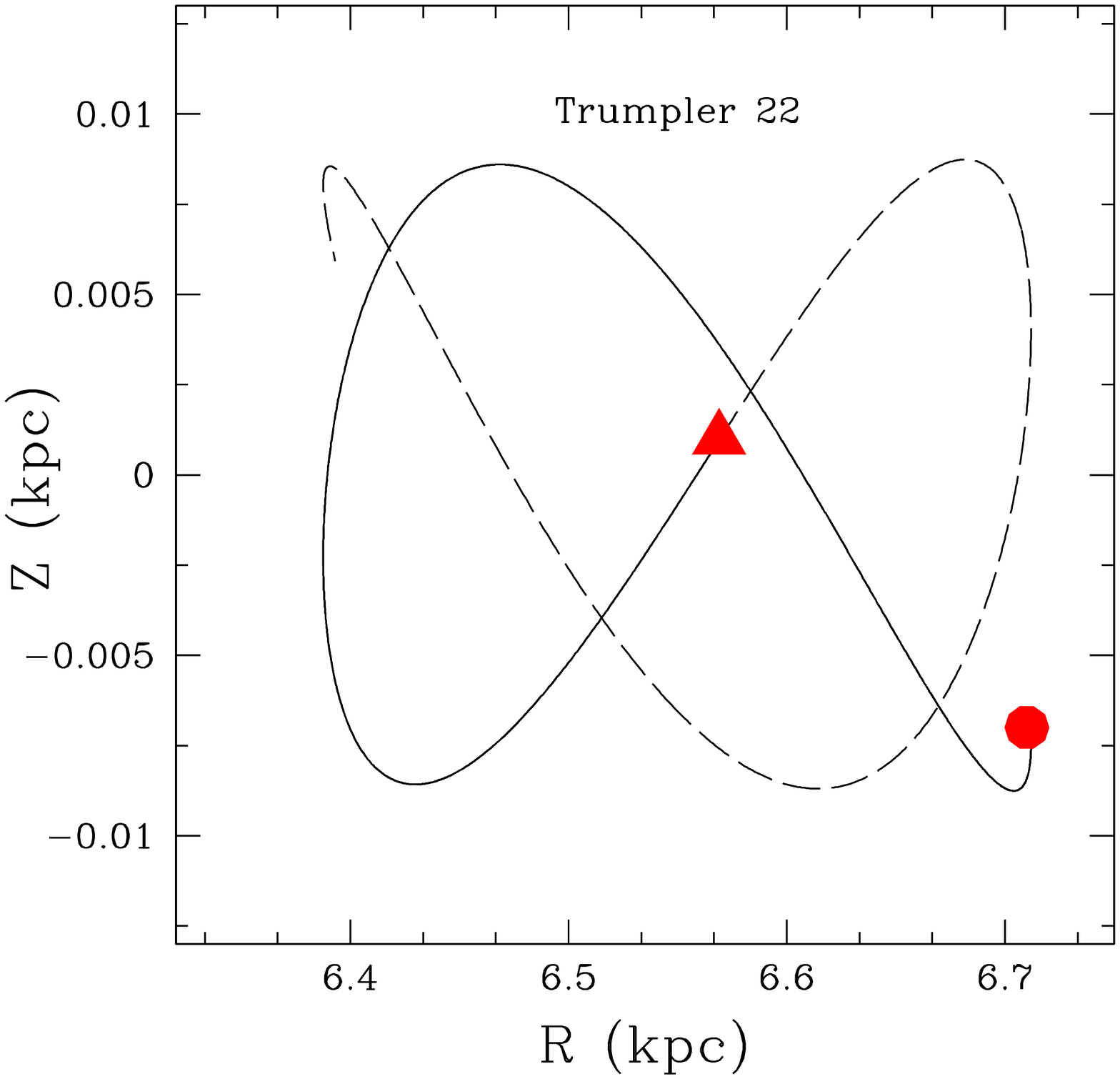}
\hspace{1cm}\includegraphics[width=3.5cm, height=3.5cm]{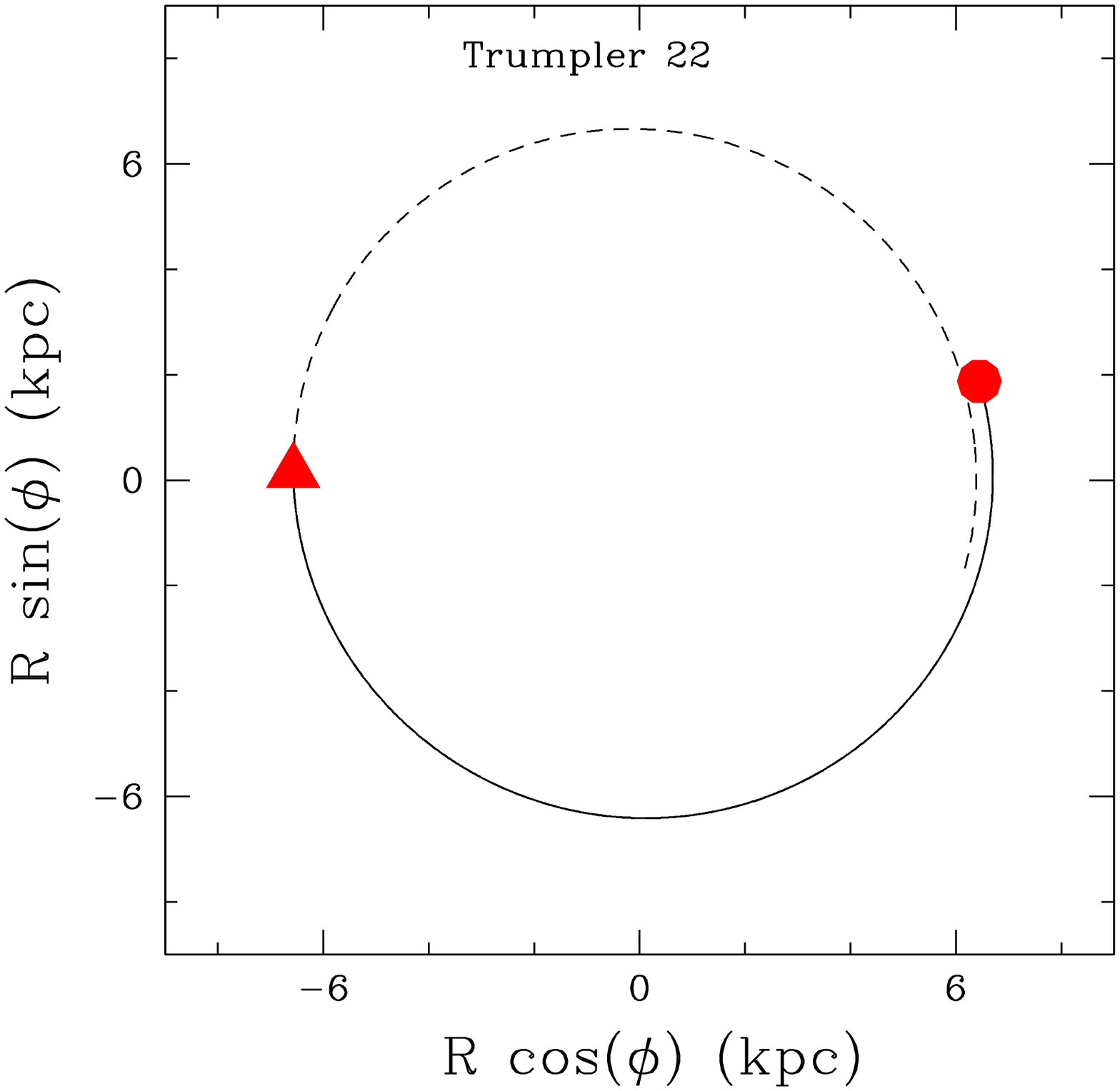}
\hspace{1cm}\includegraphics[width=3.5cm, height=3.5cm]{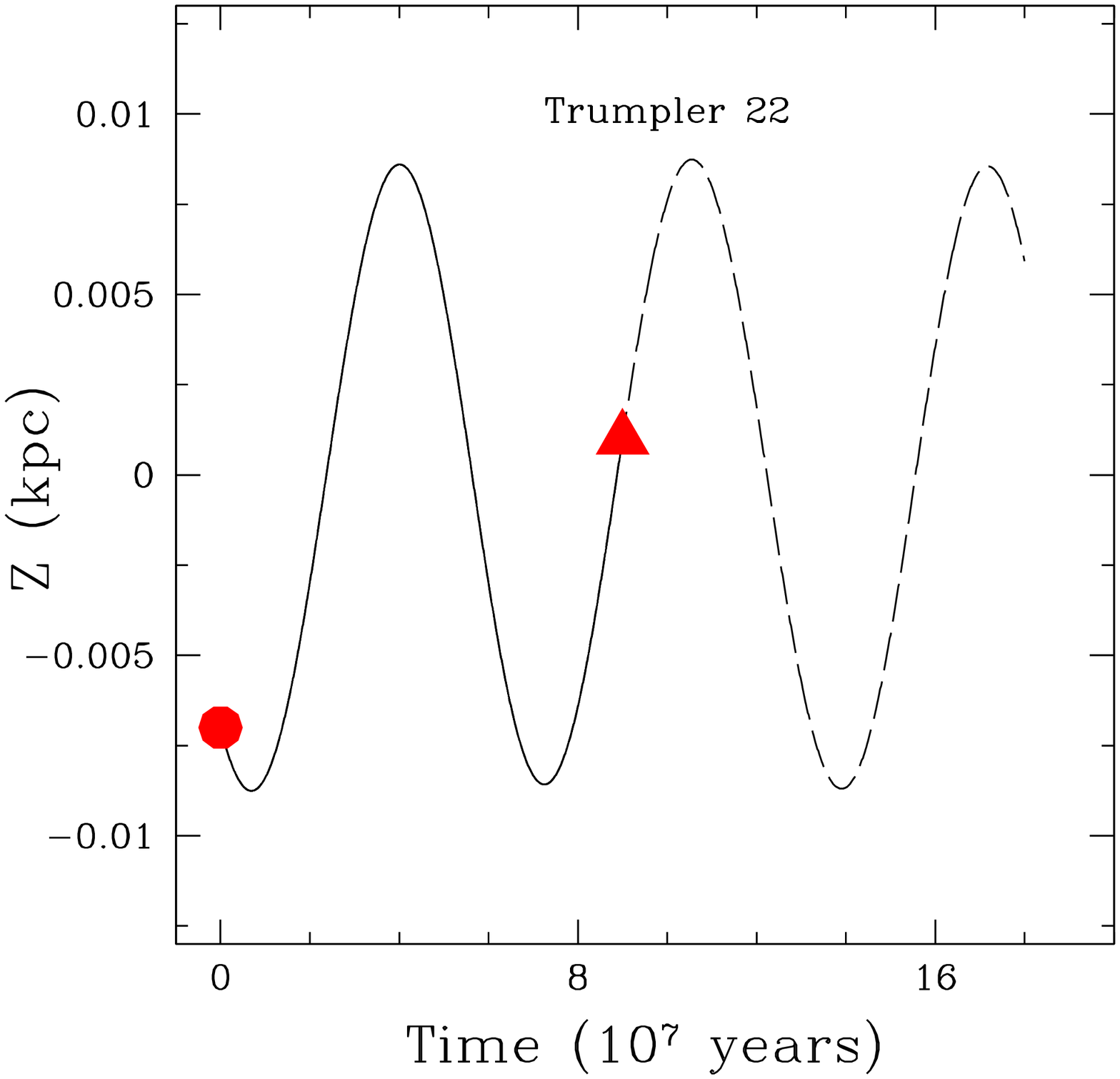}
}
\hbox{
\includegraphics[width=3.5cm, height=3.5cm]{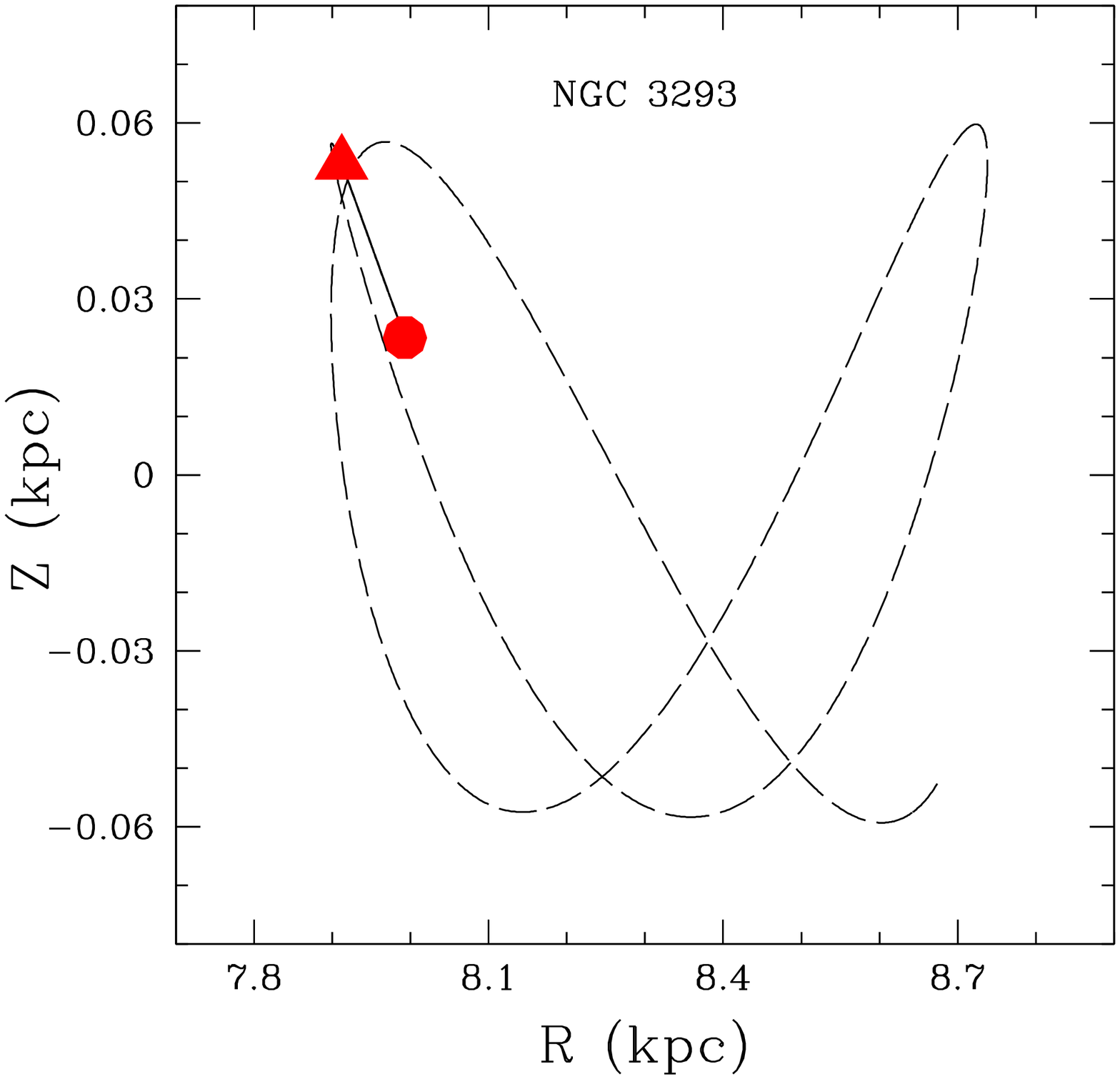}
\hspace{1cm}\includegraphics[width=3.5cm, height=3.5cm]{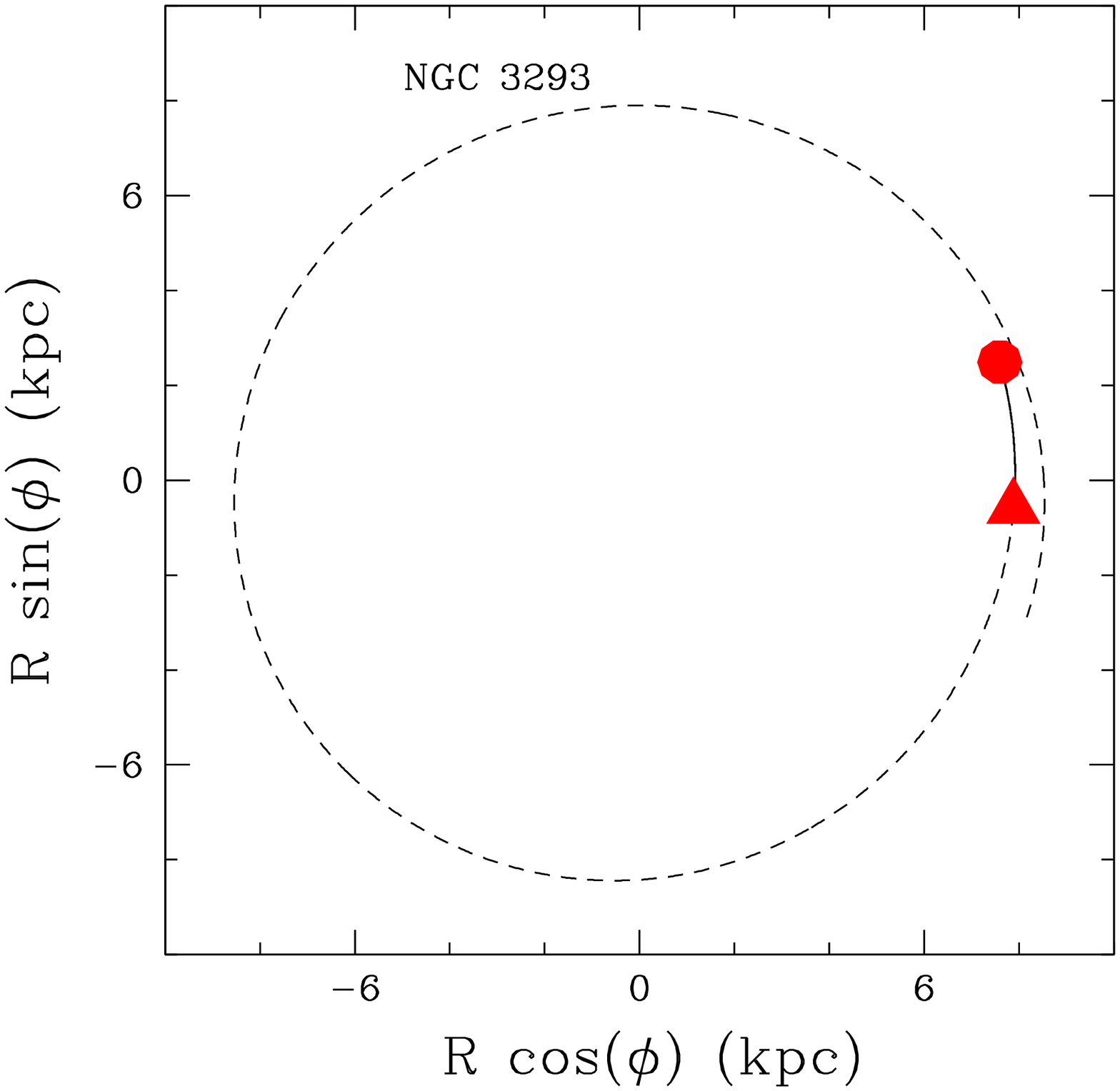}
\hspace{1cm}\includegraphics[width=3.5cm, height=3.5cm]{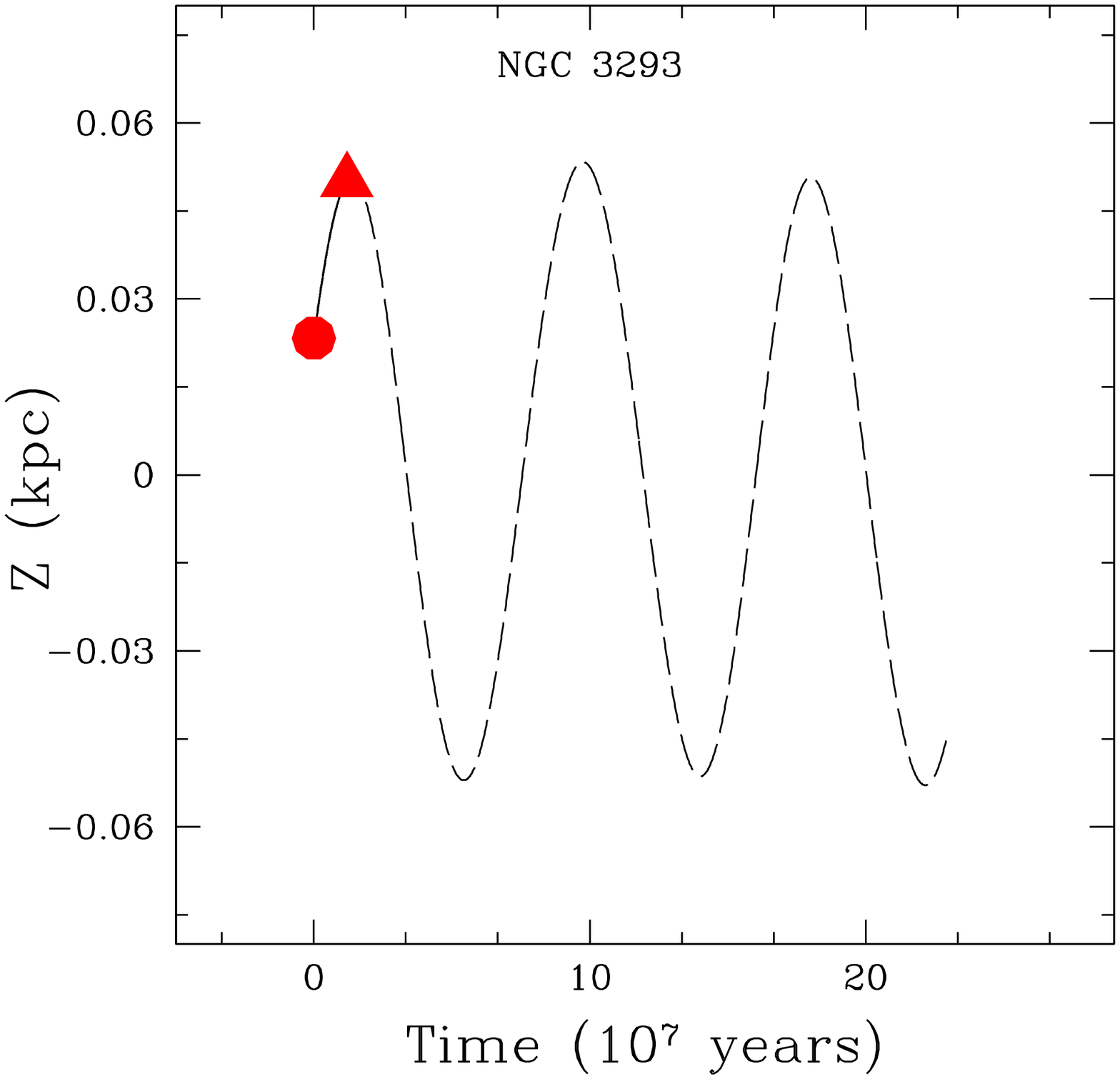}
}
\hbox{
\includegraphics[width=3.5cm, height=3.5cm]{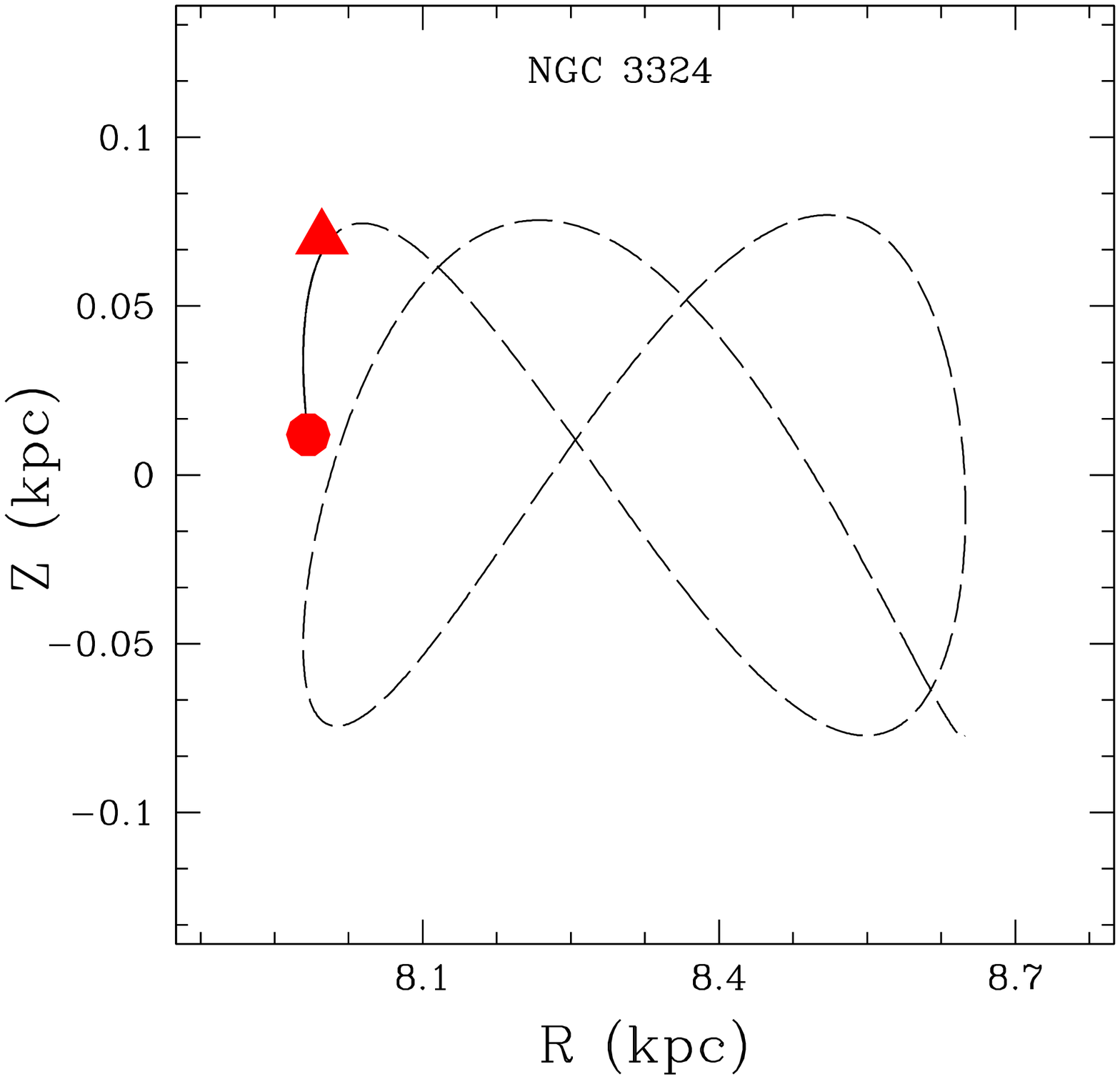}
\hspace{1cm}\includegraphics[width=3.5cm, height=3.5cm]{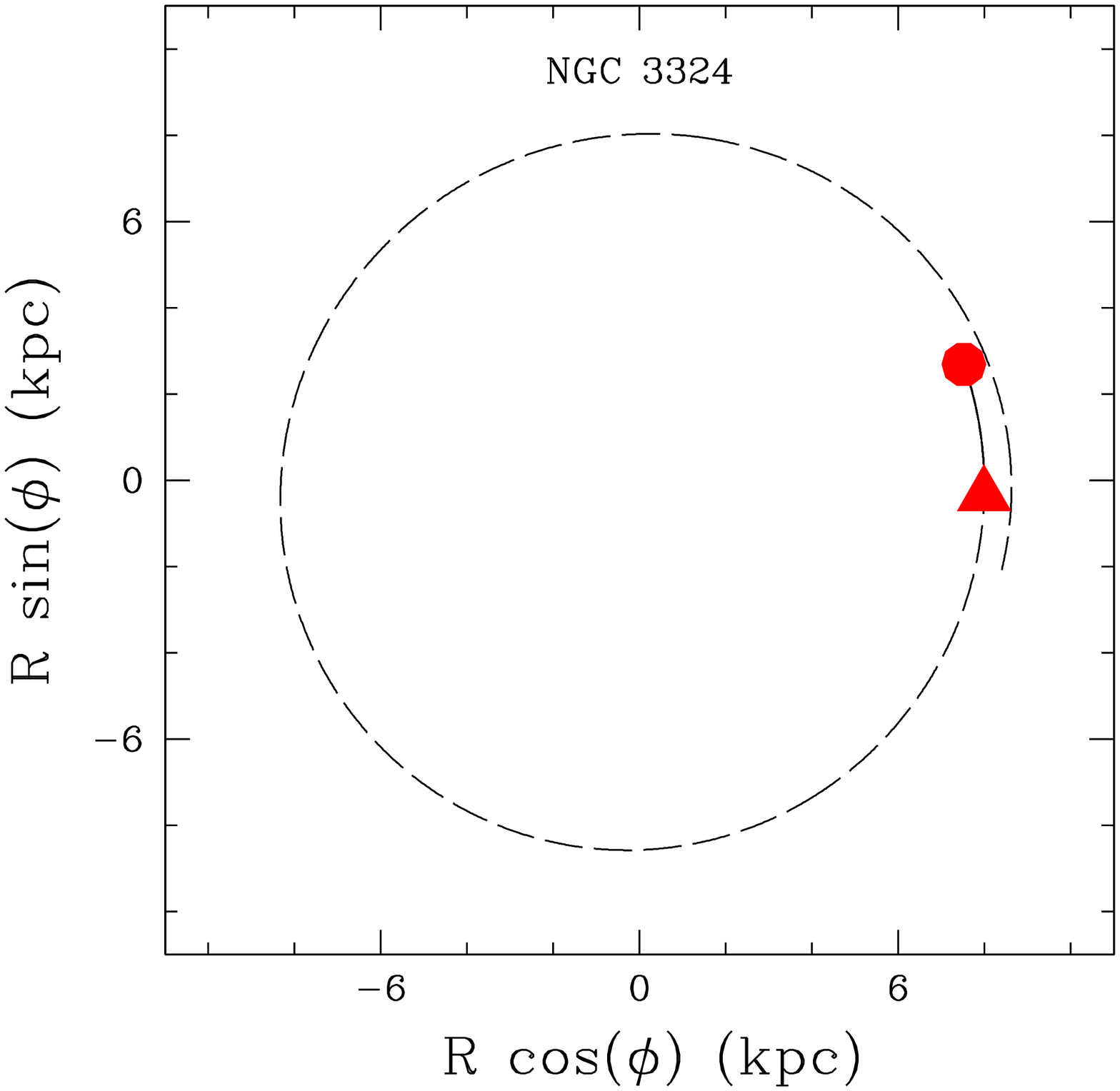}
\hspace{1cm}\includegraphics[width=3.5cm, height=3.5cm]{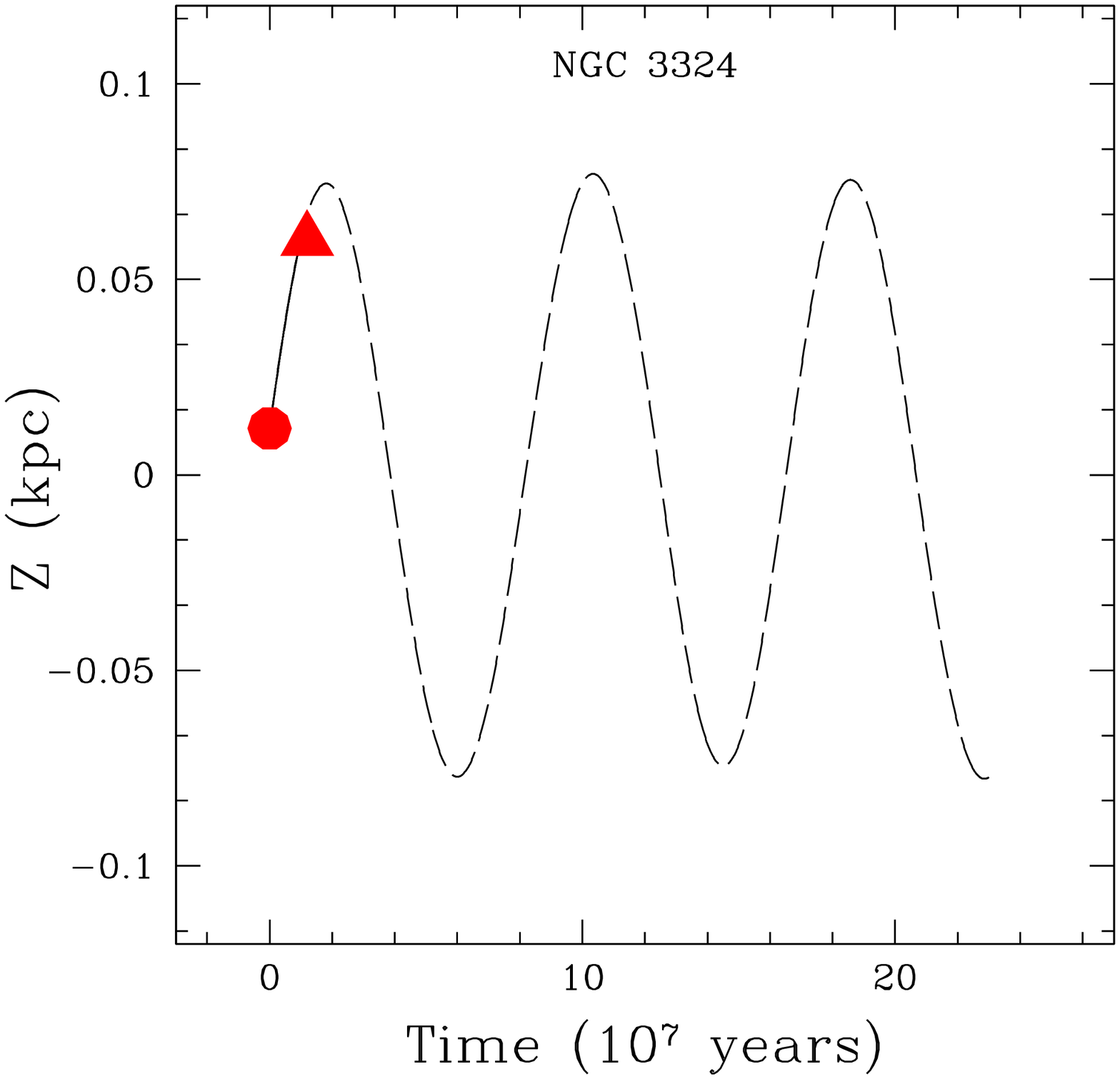}
}
\caption{Galactic orbits of the clusters NGC 5617, Trumpler 22, NGC 3293 and NGC 3324 estimated with the
Galactic potential model described in the text, in the time interval of the age of each cluster. The left panel
shows a side view, the middle panel shows a top view and the right panel shows period of oscillation along Z-axis
of the cluster's orbit. The continuous and dotted lines in this figure are for 90 Myr (age) and 180 Myr for cluster pair
NGC 5617 and Trumpler 22 while 12 Myr (age) and 230 Myr for cluster pair NGC 3293 and NGC 3324, respectively. The 
curves with dotted lines represent the complete cycle. The filled
triangle and filled circle denotes birth and present day position of clusters in the Galaxy.} 
\label{orbit}
\end{center}
\end{figure*}
To estimate the orbits and orbital parameters for clusters under study, we have used the Galactic potential models. The input
parameters, such as central coordinates ($\alpha$ and $\delta$), mean proper motions ($\mu_{\alpha}cos\delta$, $\mu_{\delta}$),
parallax values, clusters age and heliocentric distances ($d_{\odot}$) for the clusters used in this paper have been taken from
Table \ref{para}. The radial velocity of clusters NGC 5617 and Trumpler 22 has been taken as $-38.50\pm2.15$ km/sec calculated
by De Silva et al. (2015). Radial velocity of cluster NGC 3293 has been taken as $-13.16\pm0.55$ km/sec from
Soubiran et al. (2018) while for cluster NGC 3324, we obtained it as $-14.27\pm0.70$ km/sec taking weighted mean of
probable members from Gaia EDR3. 

We transformed position and velocity vectors into the Galacto-centric cylindrical coordinate system using the transformation
matrix given in Johnson \& Soderblom (1987). In this system, $(r,\phi,z)$ indicates the position of an object in Galaxy,
where $r$ is the distance from Galactic center, $\phi$ is the angle relative to Sun's position in the Galactic plane and $z$
is the distance from the Galactic plane.

The right-hand coordinate system is adopted to transform equatorial velocity components into Galactic-space velocity components ($U,V,W$),
where $U$, $V$ and $W$ are radial, tangential, and vertical velocities respectively. In this system, the x-axis is taken positive towards
Galactic-center, the y-axis is along the direction of Galactic rotation and the z-axis is towards Galactic north pole. The Galactic center is
taken at ($17^{h}:45^{m}:32^{s}.224, -28^{\circ}:56^{\prime}:10^{\prime\prime}$) and North-Galactic pole is at ($12^{h}:51^{m}:26^{s}.282,
27^{\circ}:7^{\prime}:42^{\prime\prime}.01$) (Reid \& Brunthaler, 2004). To apply a correction for Standard Solar Motion and Motion of the
Local Standard of Rest (LSR), we used position coordinates of Sun as ($8.3,0,0.02$) kpc and its space-velocity components as
($11.1, 12.24, 7.25$) km/s (Schonrich et al. 2010). Transformed parameters in Galacto-centric coordinate system are listed in
Table \ref{inp}.

In orbit determination, we estimated the radial and vertical components of gravitational force, by differentiating total gravitational
potentials with respect to $r$ and $z$. The second order derivatives of gravitational force describe the motion of the clusters. For
orbit determination, the second order derivatives are integrated backward in time, which is equal to the age of clusters. Since potentials
used are axis-symmetric, energy and $z$ component of angular momentum are conserved throughout the orbits.

\begin{table*}
   \centering
   \caption{Position and velocity components in the Galactocentric coordinate system. Here $R$ is the galactocentric
            distance, $Z$ is the vertical distance from the Galactic disc, $U$ $V$ $W$ are the radial tangential and the vertical 
            components of velocity respectively and $\phi$ is the position angle relative to the sun's direction.
}
   \begin{tabular}{ccccccccc}
   \hline\hline
   Cluster   & $R$ &  $Z$ &  $U$  & $V$  & $W$ & $\phi$   \\
   & (kpc) & (kpc) & (km/s) &  (km/s) & (km/s) & (radian)    \\
  \hline
   NGC 5617 & 6.81 & 0.02 & $-7.80 \pm 1.62$  & $-241.30 \pm 1.63$ &  $2.61 \pm 1.54$ & 0.26    \\
   Trumpler 22 & 6.71 & -0.01 & $0.46 \pm 1.66$ & $-242.41 \pm 1.66$ & $-0.49 \pm 1.25$ & 0.28      \\
   NGC 3293 & 7.99 & 0.02 & $-10.98 \pm 0.37$ & $-257.23 \pm 0.55$ & $3.54 \pm 0.56$ & 0.32      \\
   NGC 3324 & 7.98 & 0.01 & $-2.45 \pm 8.42$ & $-257.82 \pm 4.32$ & $5.76 \pm 2.53$ & 0.34      \\
\hline
  \end{tabular}
  \label{inp}
  \end{table*}

\begin{figure}
\begin{center}
\hbox{
\includegraphics[width=4.2cm, height=4.2cm]{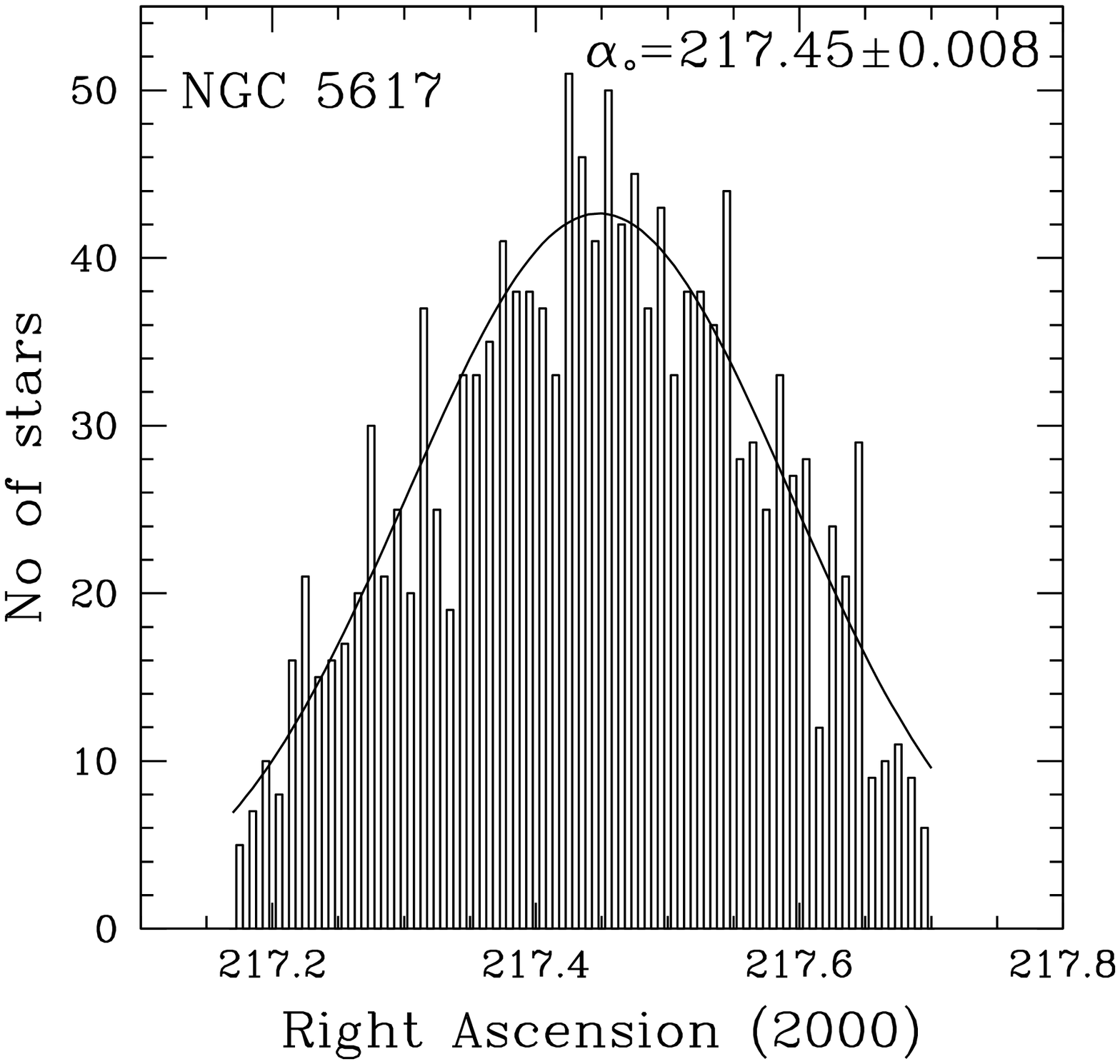}
\includegraphics[width=4.2cm, height=4.2cm]{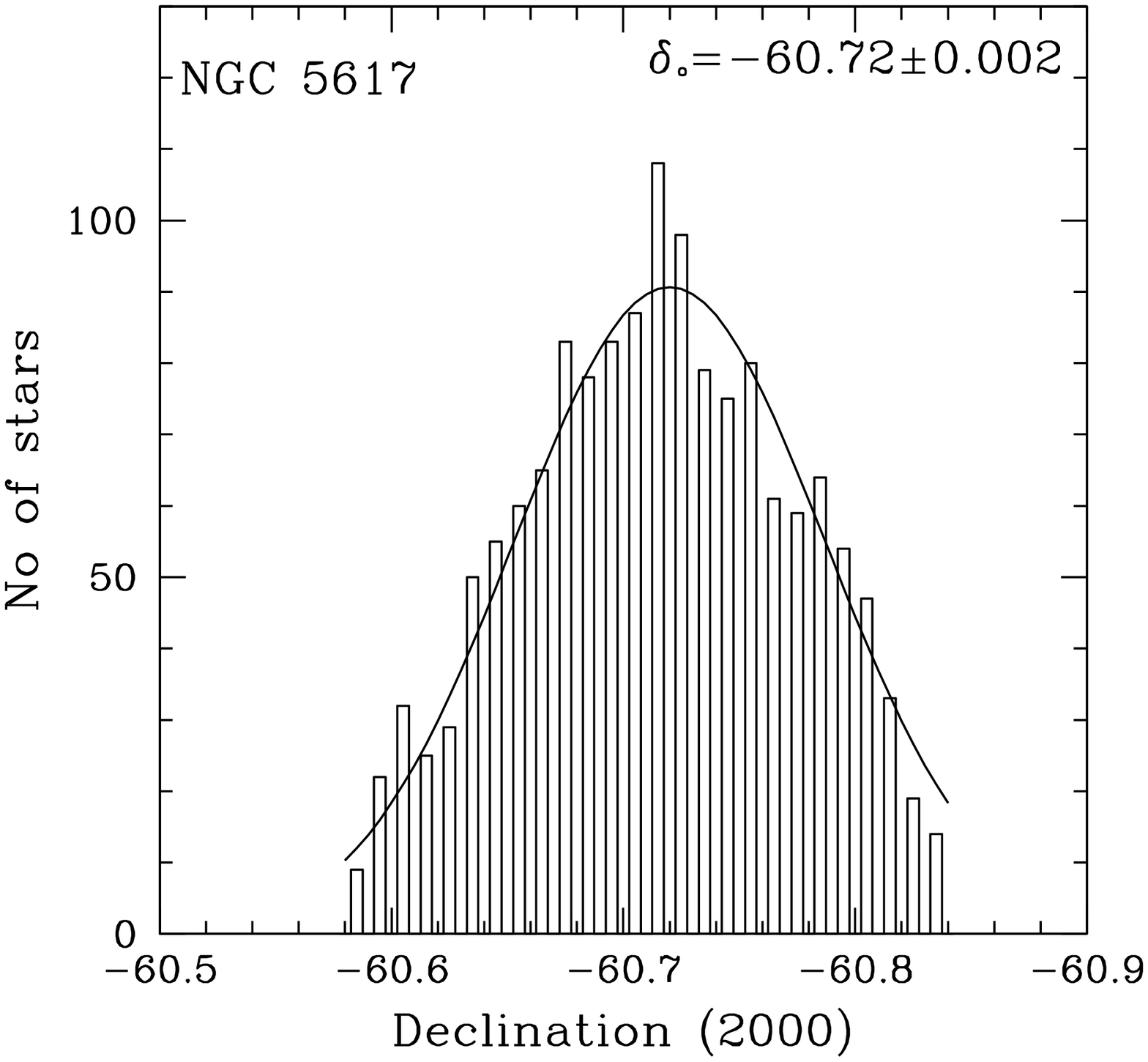}
}
\hbox{
\includegraphics[width=4.2cm, height=4.2cm]{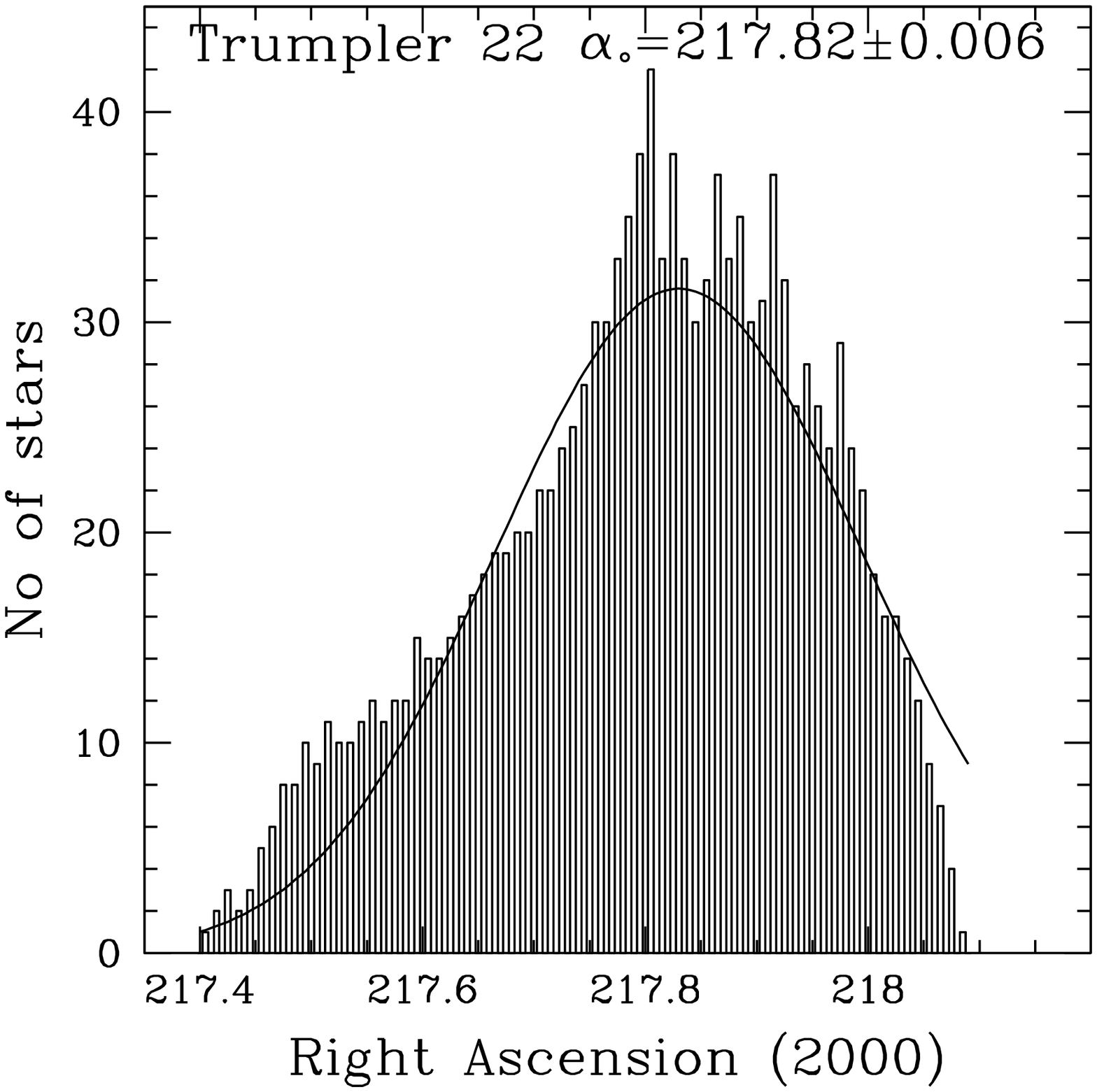}
\includegraphics[width=4.2cm, height=4.2cm]{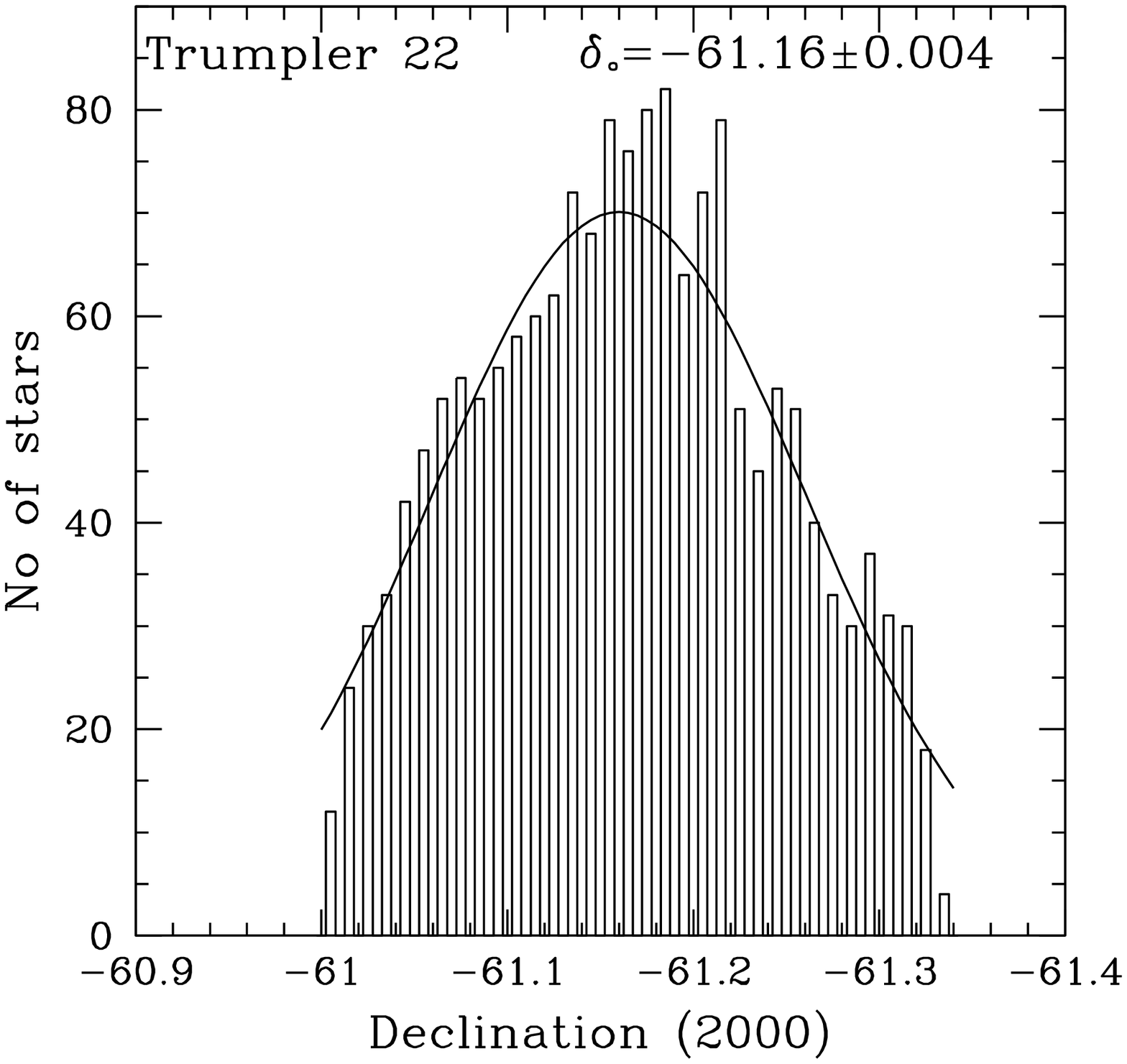}
}
\hbox{
\includegraphics[width=4.2cm, height=4.2cm]{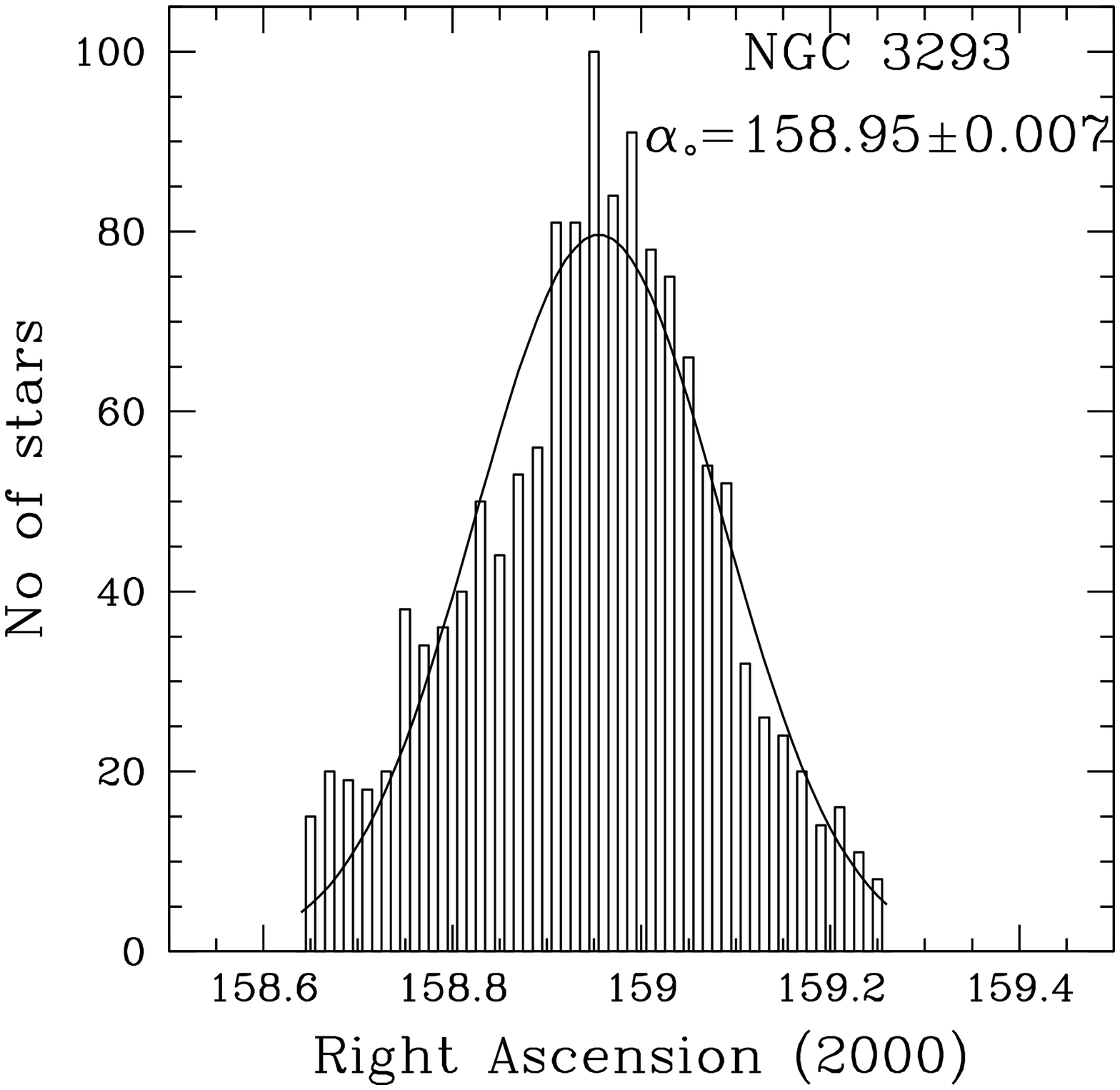}
\includegraphics[width=4.2cm, height=4.2cm]{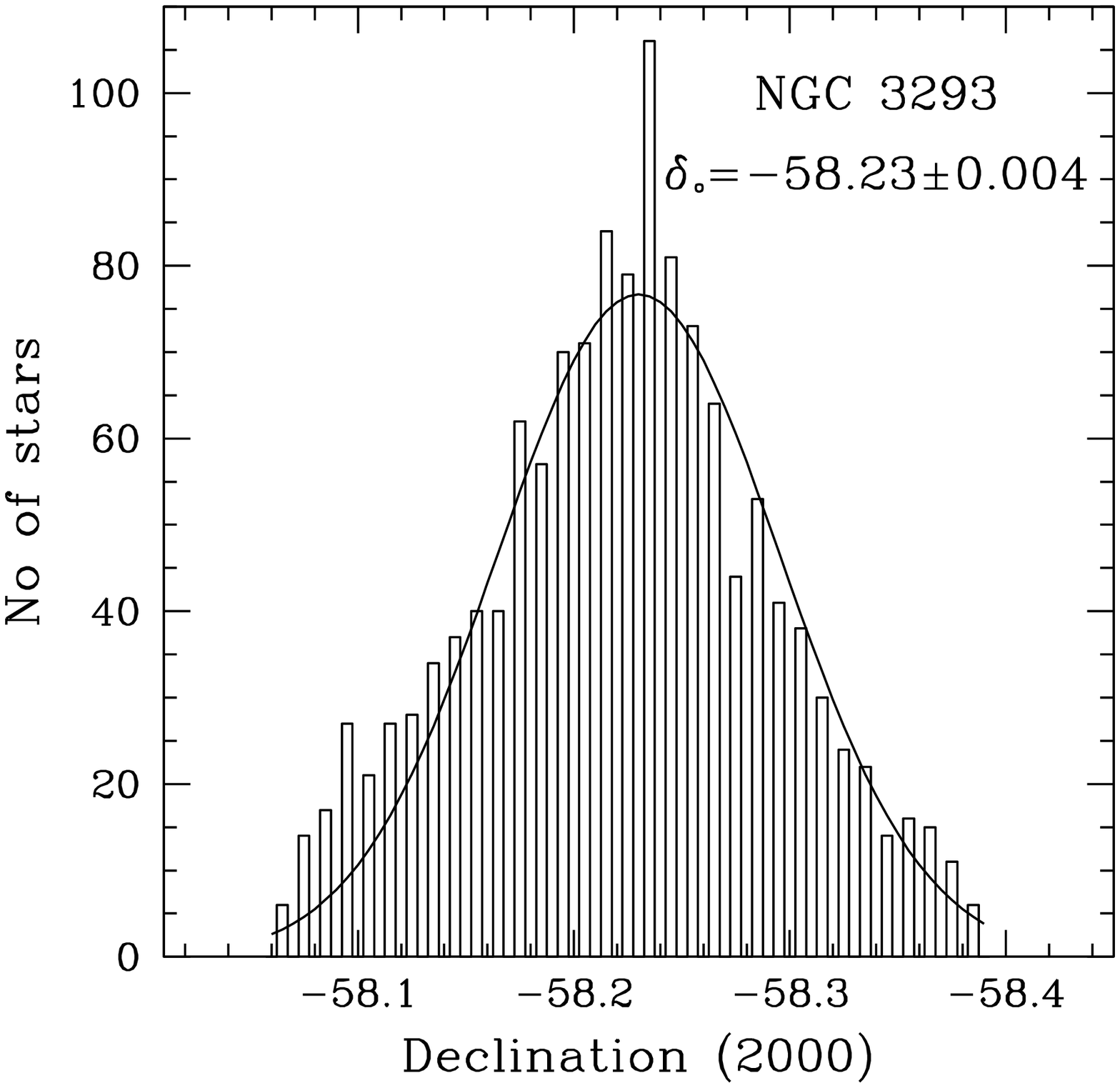}
}
\hbox{
\includegraphics[width=4.2cm, height=4.2cm]{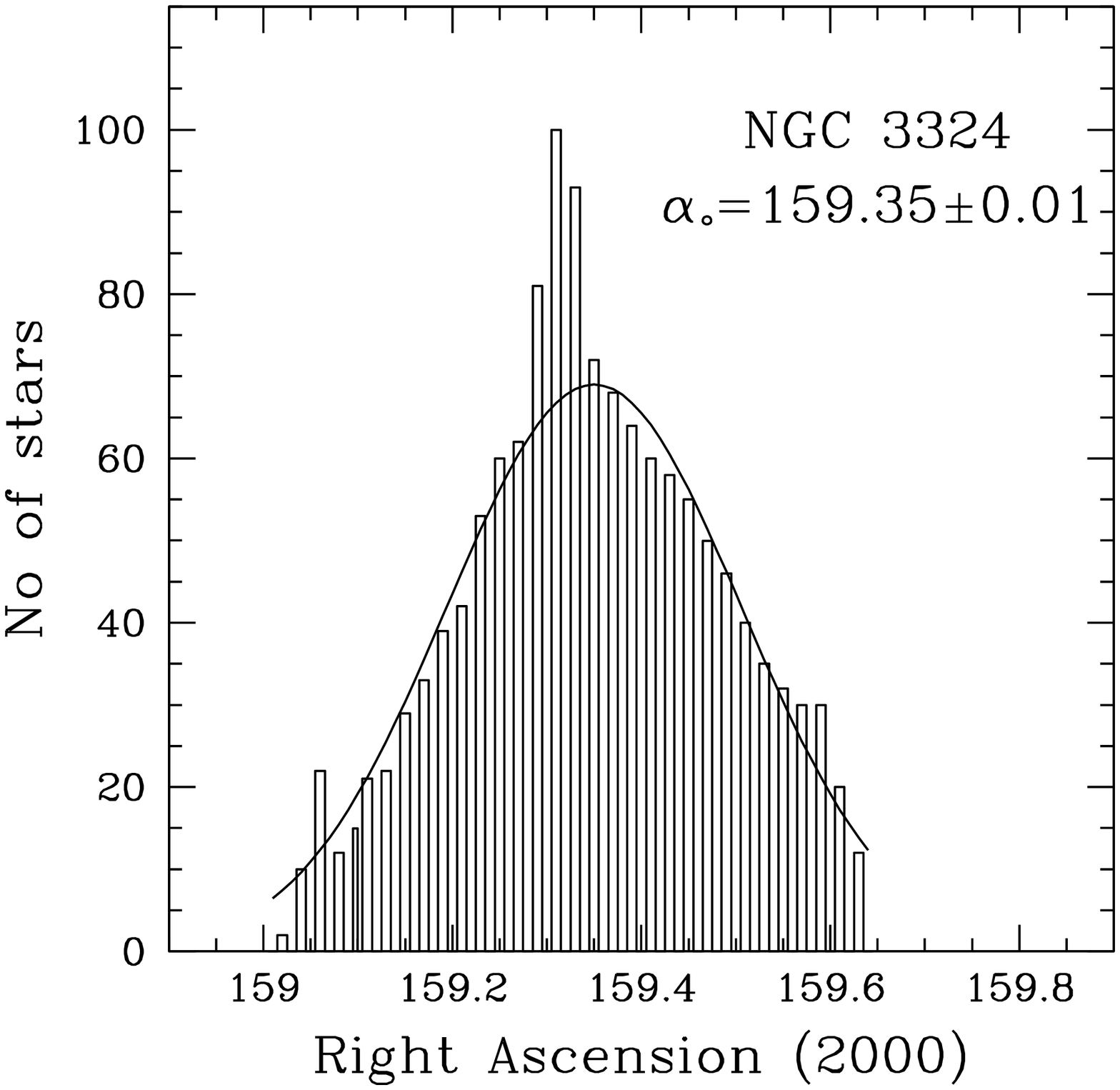}
\includegraphics[width=4.2cm, height=4.2cm]{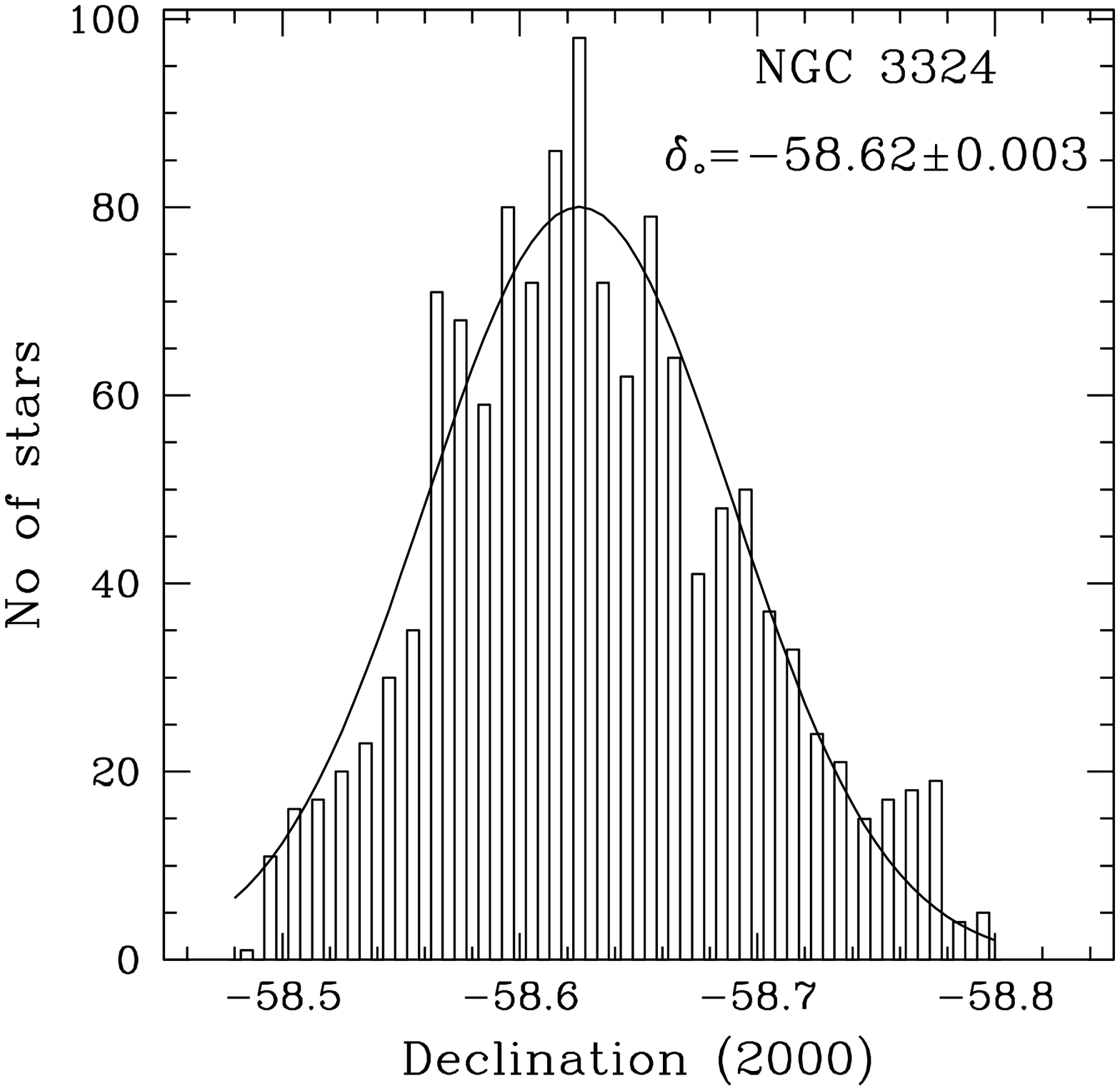}
}
\caption{Profiles of stellar counts across clusters NGC 5617, Trumpler 22, NGC 3293 and NGC 3324 using Gaia~EDR3.
The Gaussian fits have been applied. The center of symmetry about the peaks of Right Ascension and Declination is taken to be
the position of the cluster's center.} 
\label{center}
\end{center}
\end{figure}

Fig. \ref{orbit} show orbits of the clusters NGC 5617, Trumpler 22, NGC 3293 and NGC 3324. In left panels, the motion of
clusters is shown in terms of distance from the Galactic center and Galactic plane, which manifests two dimensional side view
of the orbits. In middle panels, cluster motion is described in terms of $x$ and $y$ components of Galactocentric distance,
which shows a top view of orbits. The clusters are also oscillating along 
Z-axis as shown in right panels of these figures.
(NGC 5617 and Trumpler 22) and (NGC 3293 and NGC 3324) are oscillating
along Z-axis within the limit of -0.04 to 0.04 and -0.1 to 0.1 kpc, with a time
period of 67 and 79 Myr respectively.
The time period of revolution around the Galactic center is 176, 172, 194
and 193 Myr for NGC 5617, Trumpler 22, NGC 3293 and NGC 3324 respectively. Time period of these oscillations for both the 
binary clusters are very similar. We also calculated the orbital parameters for the clusters and are listed in
Table \ref{orpara}. Here $e$ is eccentricity, $R_{a}$ is the apogalactic distance, $R_{p}$ is the perigalactic distance,
$Z_{max}$ is the maximum distance traveled by cluster from Galactic disc, $E$ is the average energy of orbits, $J_{z}$
is $z$ component of angular momentum and $T$ is the time period of the clusters in the orbits.

\begin{table*}
  \centering
   \caption{Orbital parameters for the clusters obtained using the Galactic potential model.
   }
   \begin{tabular}{ccccccccc}
   \hline\hline
   Cluster  & $e$  & $R_{a}$  & $R_{p}$ & $Z_{max}$ &  $E$ & $J_{z}$ & $T$   \\ 
           &    & (kpc) & (kpc) & (kpc) & $(100 km/s)^{2}$ & (100 kpc km/s) & (Myr) \\ 
   \hline\hline  
   NGC 5617 &  0.01  & 6.75  & 6.66  & 0.03 & -12.33 & -16.44  & 176 \\
   Trumpler 22  & 0.003 & 6.70  & 6.66  & 0.01 & -12.40 & -16.27  & 172  \\
   NGC 3293  & 0.04 & 8.54  & 7.90  & 0.05 & -10.95 & -20.56  & 194  \\
   NGC 3324  & 0.04 & 8.62  & 8.03  & 0.08 & -10.94 & -20.58  & 193  \\
 \hline 
  \end{tabular}
  \label{orpara}
  \end{table*}

The orbits of the clusters under study follow a boxy pattern and eccentricities for all the clusters are nearly zero. Hence they
trace a circular path around the Galactic center. We have shown the birth and present day position of clusters in the Galaxy
which are represented by the filled circle and filled triangle respectively in Fig. \ref{orbit}. NGC 5617 and Trumpler 22
are intermediate age while NGC 3293 and NGC 3324 are young age OCs. The orbits are confined
in a box of $\sim$$ 6.6 < R_{gc} \leq 6.7 $ kpc for (NGC 5617 and Trumpler 22) and $\sim 7.9 < R_{gc} \leq 8.6$ kpc for
clusters (NGC 3293 and NGC 3324). Our analysis indicates that all the clusters under study are inside the solar circle in 
the thin disk and may interact with the molecular clouds present in the Galaxy. Carraro \& Chiosi (1994) found that clusters 
orbiting beyond the solar circle survive more as compared to the clusters which are in the inner solar circle. Webb et al. (2014) 
found that clusters having circular orbits evolve slower as compared to the eccentric ones. Orbital parameters determined in 
the present analysis are very much similar to the parameters found by Wu et al. (2009), except their orbits, are more eccentric 
than what we found in the present analysis.

\section{Structural parameters of the clusters}

\subsection{Cluster center}

The central coordinates of OCs play an important role to describe cluster properties more accurately. In the
previous studies, the center has been determined just by the visual inspection (Becker \& Fenkart 1971; Romanishim \& Angel 1980).
We applied the star-count method using probable cluster members based on proper motion and parallax database. The histograms are
built for clusters NGC 5617, Trumpler 22, NGC 3293 and NGC 3324 in both the RA and DEC directions as shown in Fig. \ref{center}.
The Gaussian curve-fitting is performed to the star counts profiles in RA and DEC directions. All estimated center coordinates are
listed in Table \ref{para}.
Our estimated values are in good agreement with the values given by Dias et al. (2002). Our derived center coordinates
for all objects are also coordinated with Cantat-Gaudin et al. (2018) catalog within uncertainty.

\subsection{Radial density profile}

\begin{figure}
\begin{center}
\hbox{
\includegraphics[width=4.5cm, height=4.5cm]{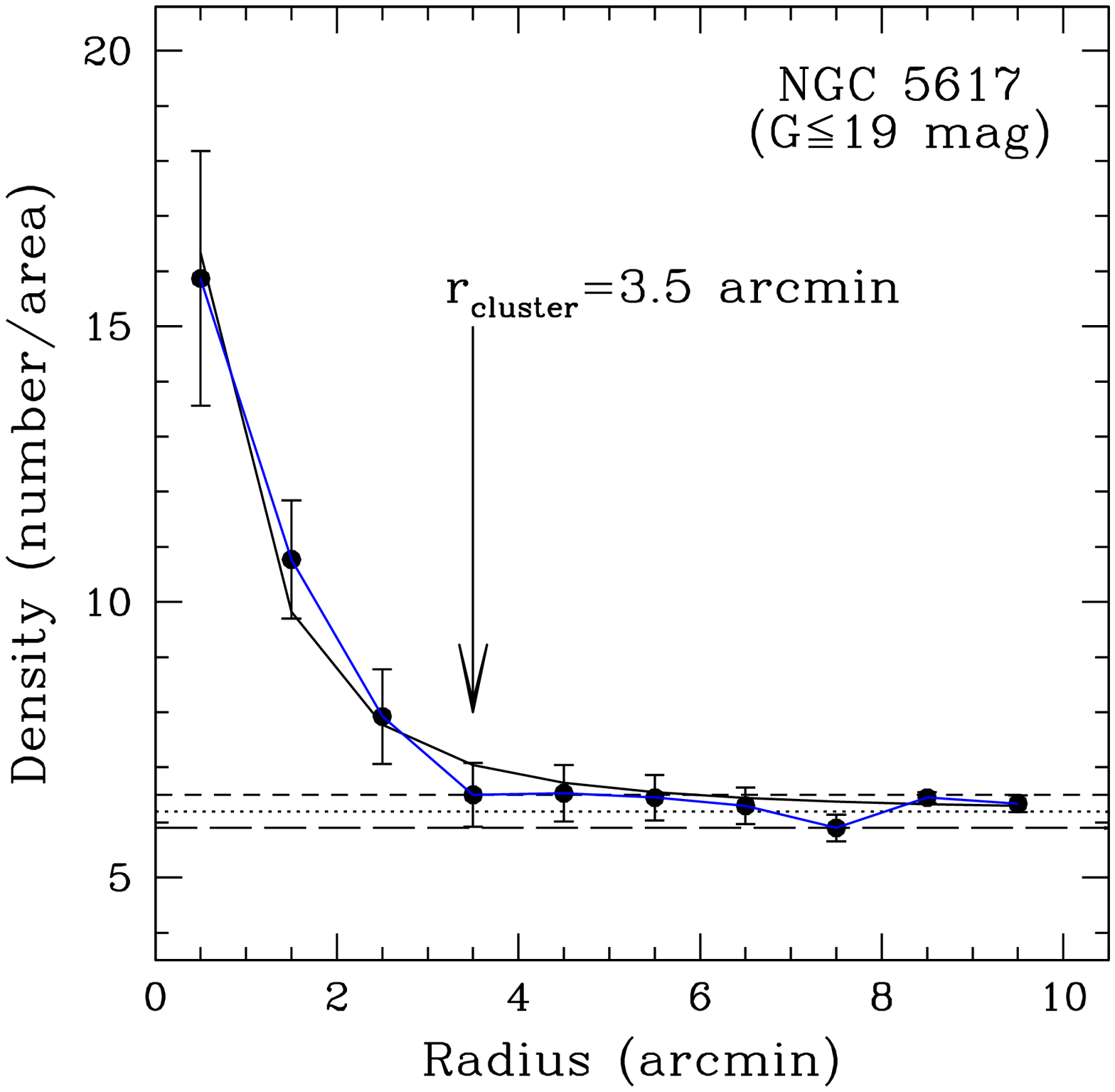}
\includegraphics[width=4.5cm, height=4.5cm]{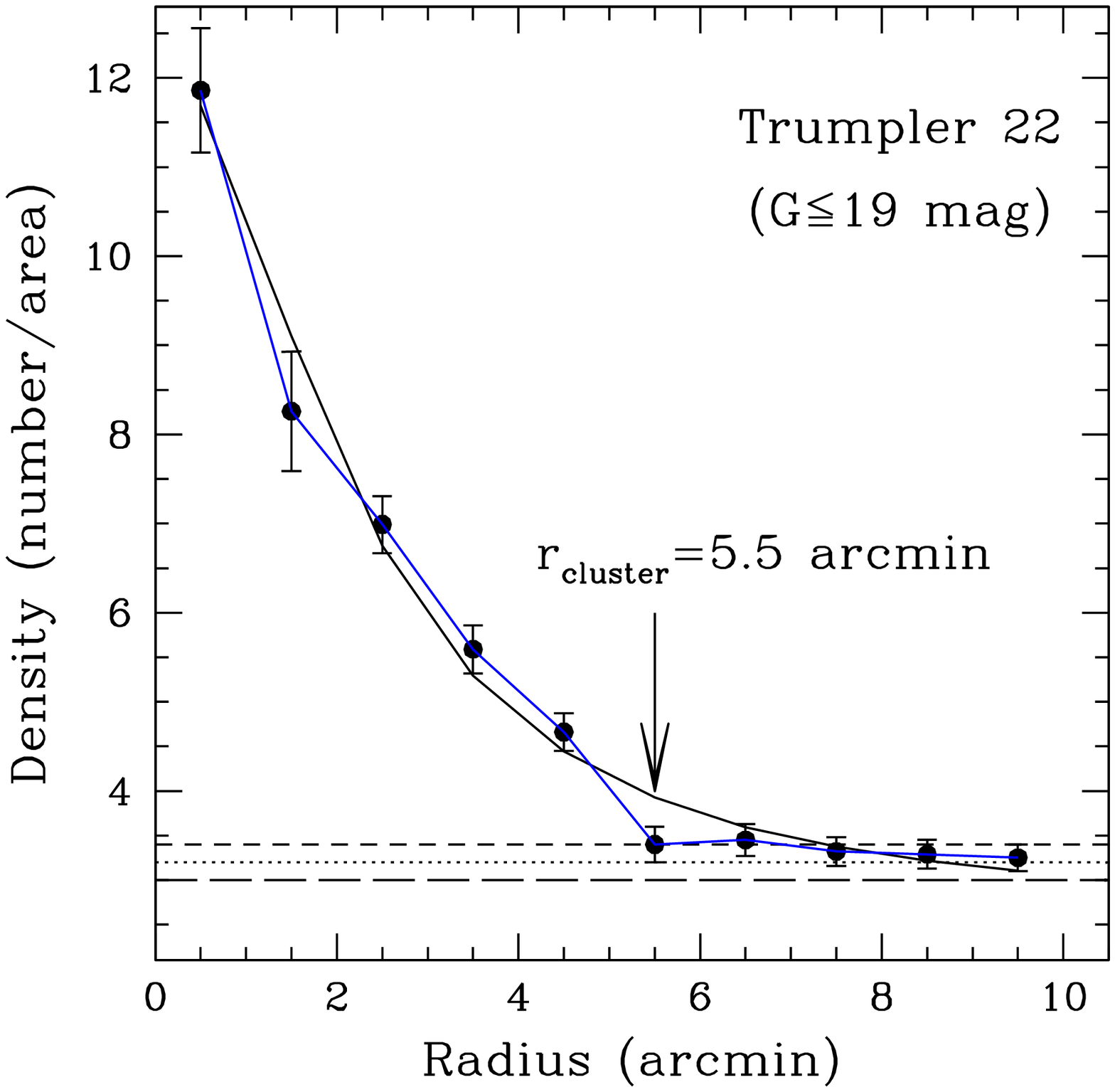}
}
\hbox{
\includegraphics[width=4.5cm, height=4.5cm]{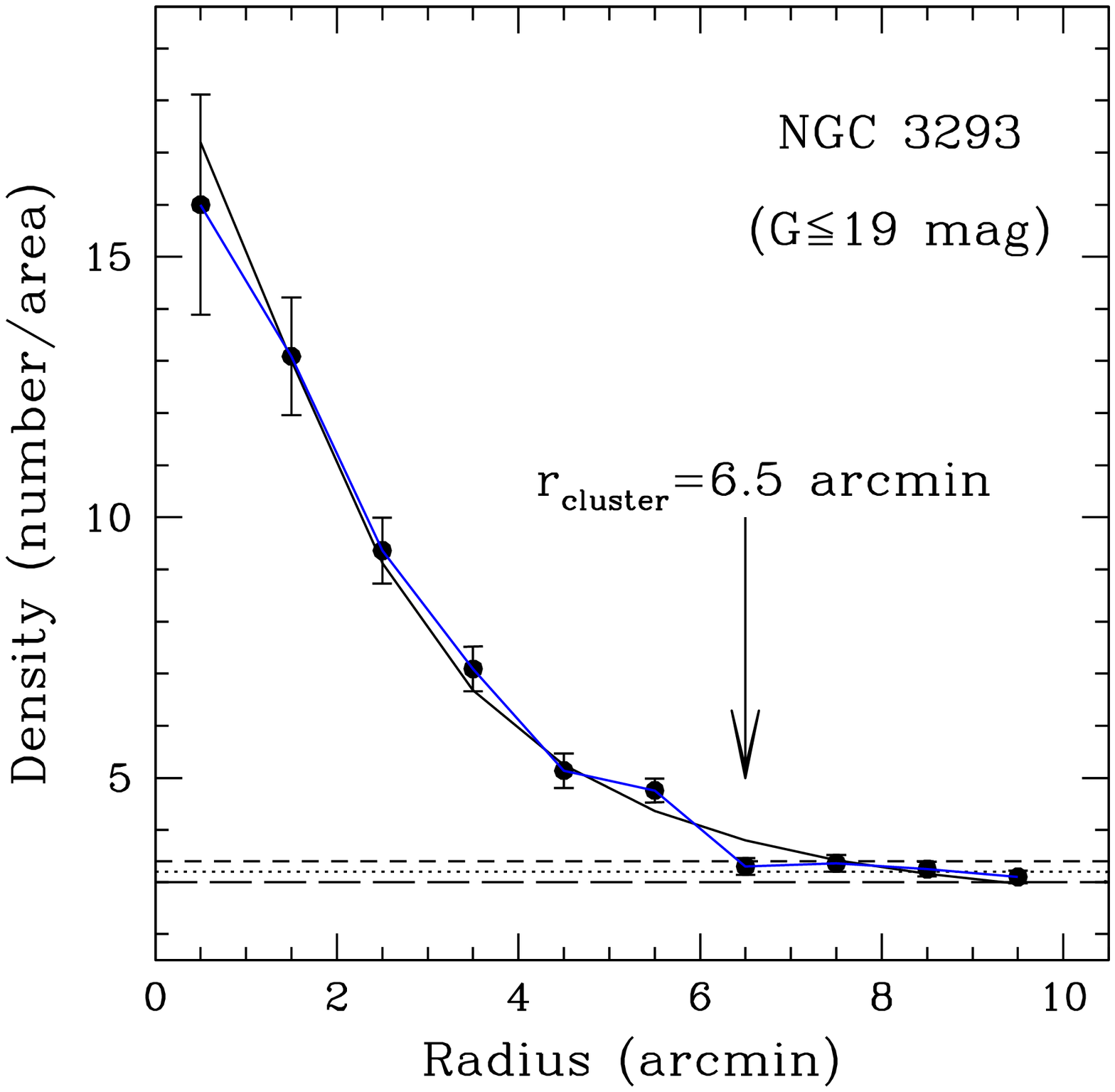}
\includegraphics[width=4.5cm, height=4.5cm]{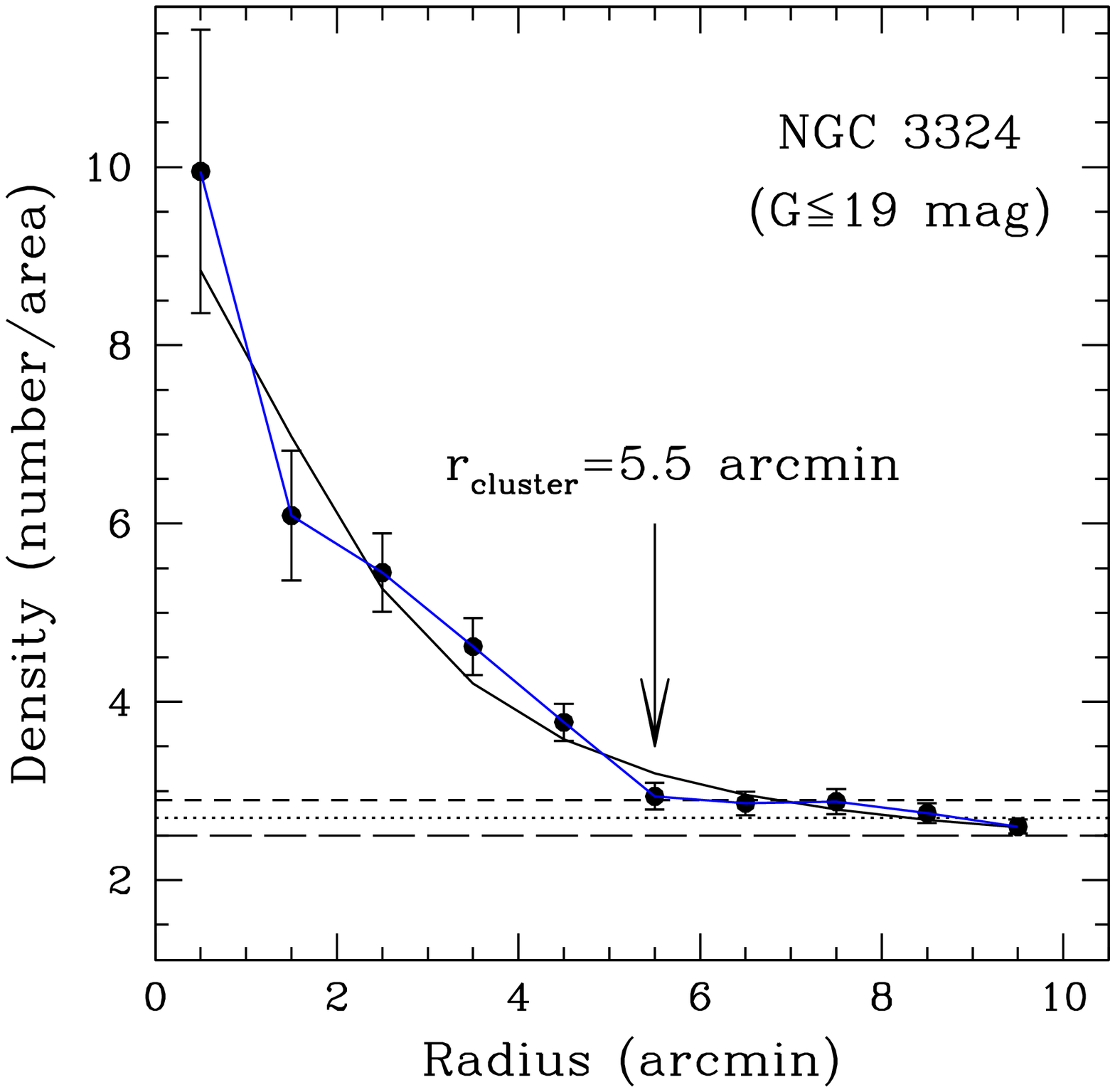}
}
\caption{Surface density distribution of the clusters under study. Errors are determined from
sampling statistics (=$\frac{1}{\sqrt{N}}$ where $N$ is the number of stars used in the density estimation at
that point). The smooth line represent the fitted profile whereas dotted line shows the background 
density level. Long and short dash lines represent the errors in background density.} 
\label{dens}
\end{center}
\end{figure}

To estimate the structural parameters of the cluster, we have plotted the radial density profile (RDP) for OCs
NGC 5617, Trumpler 22, NGC 3293 and NGC 3324. We have organized clusters area in many concentric circles
around the cluster's core having equal incremental radii. The number density, $\rho_{i}$, in the $i^{th}$ zone is
calculated by adopting the formula, $\rho_{i}$ = $\frac{N_{i}}{A_{i}}$, where $N_{i}$ is the number of cluster members
and $A_{i}$ is the area of the $i^{th}$ zone. Based on the visual inspection in clusters RDPs, the distance at which each
distribution flattens is considered as cluster radius. The error in the background density level is shown with dotted lines in
Fig. \ref{dens}. RDP becomes flat at $r\sim$ 3.5$^{\prime}$, 5.5$^{\prime}$, 6.5$^{\prime}$ and 5.5$^{\prime}$ for the
clusters NGC 5617, Trumpler 22, NGC 3293 and NGC 3324, respectively. After this distance from the cluster center,
cluster stars merged with field stars as clearly shown in Fig. \ref{dens}. Accordingly, we considered 3.5$^{\prime}$,
5.5$^{\prime}$, 6.5$^{\prime}$ and 5.5$^{\prime}$  as the cluster radius for clusters under study.
The observed radial density profile was fitted using King (1962) profile:\\

~~~~~~~~~~~~~~~~~~~~~~ $f(r) = f_{b}+\frac{f_{0}}{1+(r/r_{c})^2}$\\

\begin{table*}
\centering
\caption{Structural parameters of the clusters under study. Background and central
density are in the unit of stars per arcmin$^{2}$. Core radius ($r_c$) and tidal radius ($R_t$) are
in arcmin and pc.
}
\vspace{0.5cm}
\begin{center}
\small
\begin{tabular}{cccccccc}
\hline\hline
Name & $f_{0}$ &$f_{b}$& $r_{c}$&$r_{c}$&$\delta_{c}$&$r_{lim}$&$c$ \\
&&& arcmin & parsec  & & arcmin
\\
NGC 5617  & $19.28$&$6.20$&$0.93$&$0.66$&$4.1$&$3.6$&$0.58$ \\
Trumpler 22 & $12.17$&$3.20$&$2.2$&$1.7$&$4.8$&$6.6$&$0.47$ \\
NGC 3293 & $17.96$&$3.20$&$2.5$&$1.9$&$6.5$&$9.3$&$0.57$ \\
NGC 3324 & $9.18$&$2.70$&$1.8$&$1.5$&$4.4$&$4.6$&$0.41$ \\
\hline
\end{tabular}
\label{stru_para}
\end{center}
\end{table*}

where $r_{c}$ , $f_{0}$ , and $f_{bg}$ are the core radius, central density, and the background density level, respectively.
By fitting the King model to RDPs, we have derived the structural parameters for clusters NGC 5617, Trumpler 22,
NGC 3293 and NGC 3324. 


Limiting radius ($r_{lim}$) of each cluster is calculated by comparing $f(r)$ to a background density level, $f_{b}$, defined
as\\

~~~~~~~~~~~~~~~~~~~~$f_{b}=f_{bg}+3\sigma_{bg}$\\

where $\sigma_{bg}$ is uncertainty of $f_{bg}$. Therefore, $r_{lim}$ is calculated according to the following formula
(Bukowiecki et al. 2011)\\

~~~~~~~~~~~~~~~~~~~~$r_{lim}=r_{c}\sqrt(\frac{f_{0}}{3\sigma_{bg}}-1)$\\


Maciejewski \& Niedzielski (2007) suggested that $r_{lim}$ may vary for particular clusters from 2$r_{c}$ to 7$r_{c}$. In
present study, all clusters show a good agreement with Maciejewski \& Niedzielski (2007).

The density contrast parameter ($\delta_{c} = 1 +\frac{f_{0}}{f_{b}}$) is calculated for all the clusters under study using
member stars selected from proper motion data. Current evaluation of $\delta_{c}$ (4.1, 4.8, 6.5 and 4.4 for NGC 5617,
Trumpler 22, NGC 3293 and NGC 3324, respectively) are lower than the values ($7\le \delta_{c}\le 23$) given by Bonatto \& Bica (2009).
This estimation of $\delta_{c}$ indicates that all clusters are sparse.

\section{The fundamental parameters of NGC 5617 and Trumpler 22}

\subsection{Two colour diagrams}

The two-colour diagrams (TCDs) are very useful to determine the relation of various colour-excesses and their
variations towards the cluster region.

\subsubsection{Optical to mid-infrared extinction law}

\begin{figure*}
\begin{center}
\hbox{
\includegraphics[width=7.5cm, height=9.5cm]{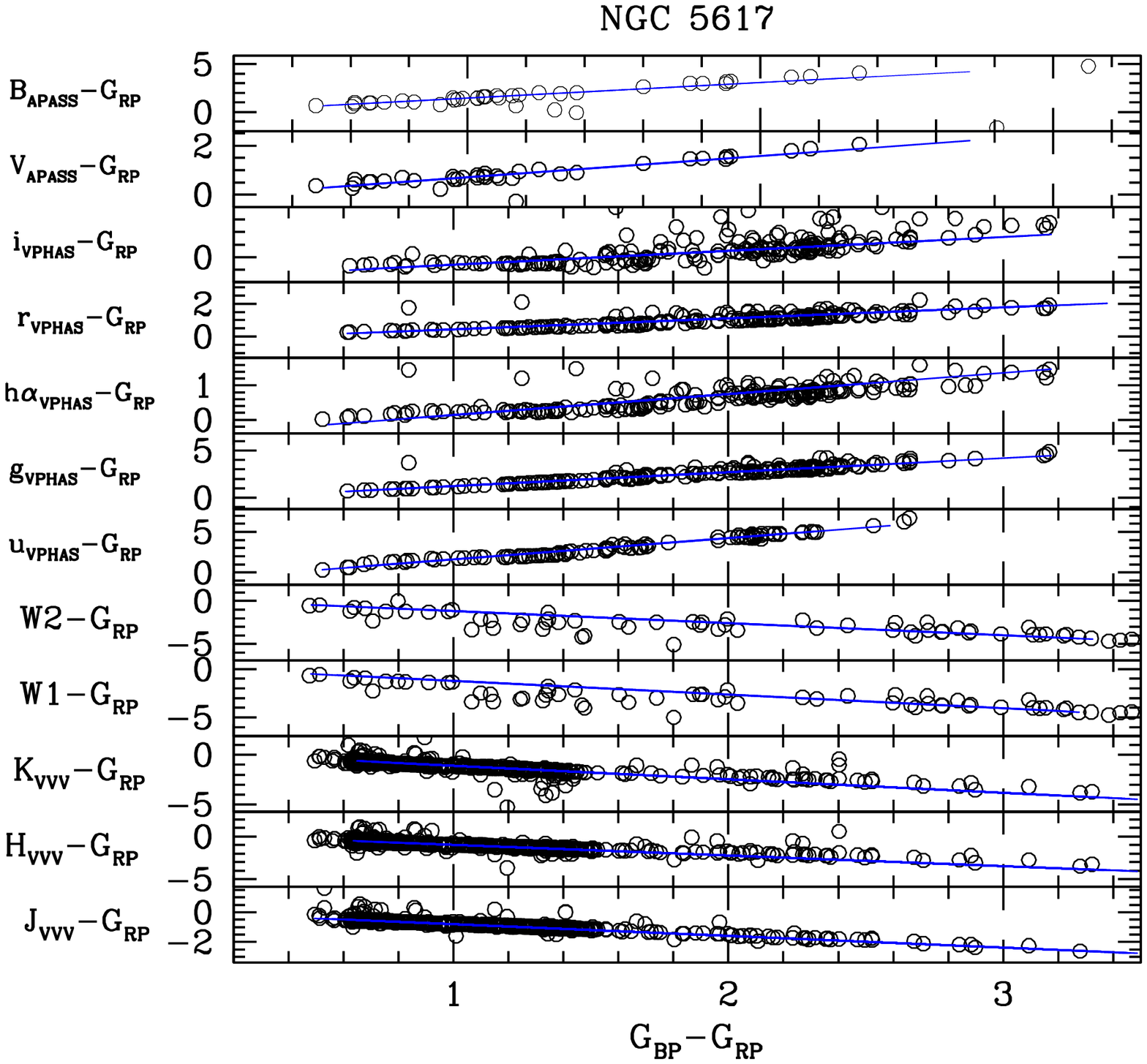}
\includegraphics[width=7.5cm, height=9.5cm]{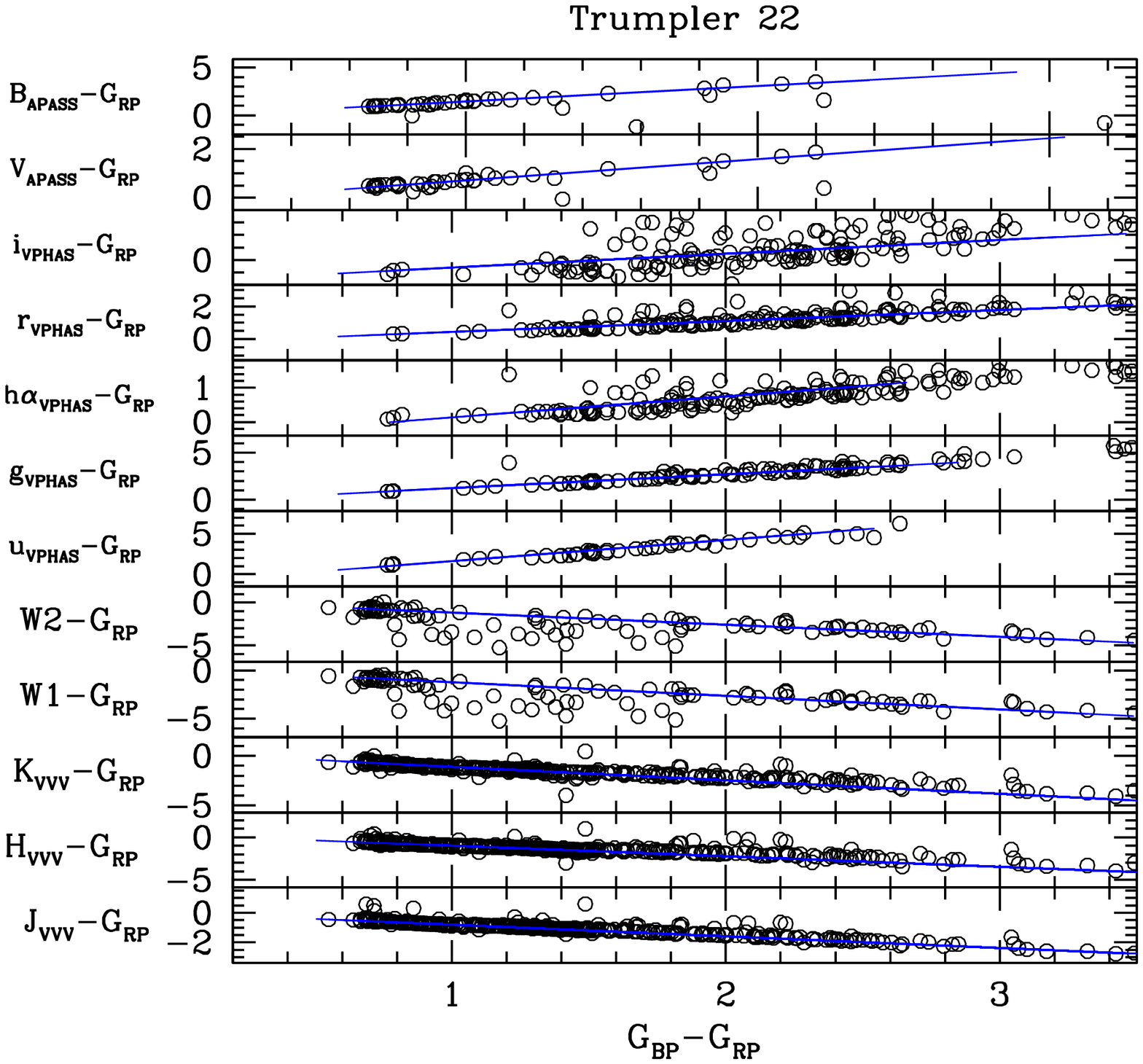}
}
\hbox{
\includegraphics[width=7.5cm, height=9.5cm]{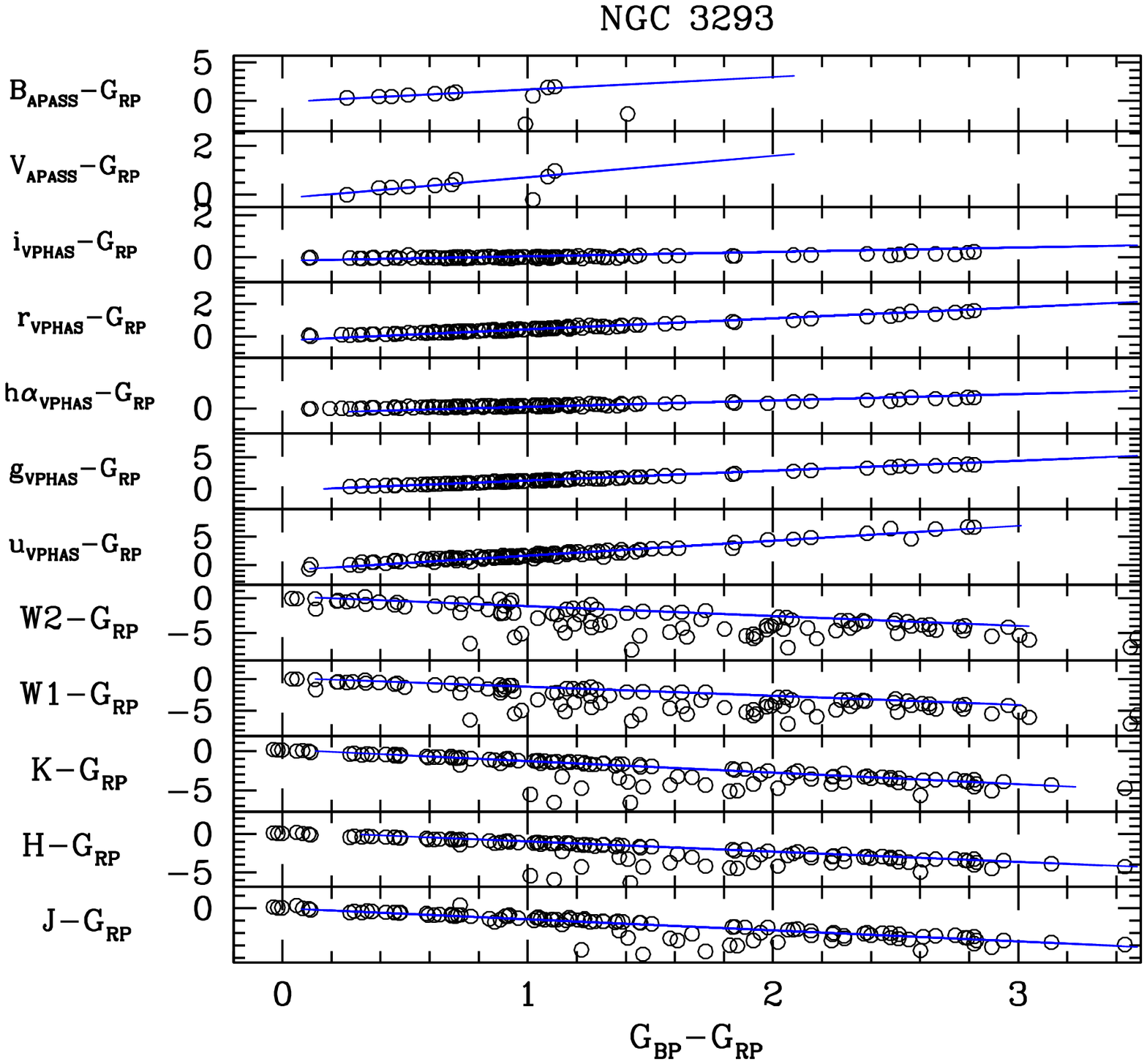}
\includegraphics[width=7.5cm, height=9.5cm]{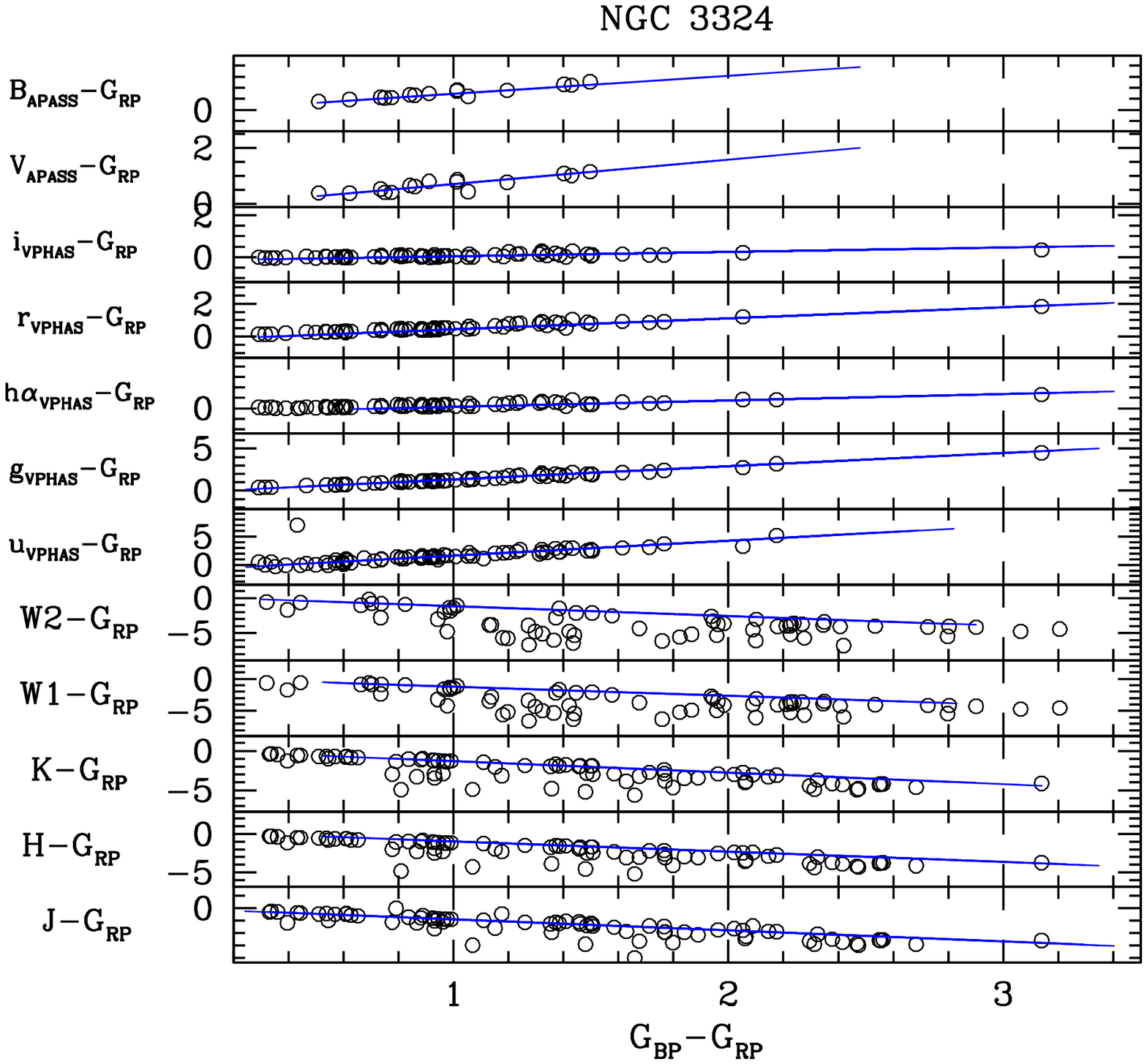}
}
\caption{The $(\lambda-G_{RP})/(G_{BP}-G_{RP})$ TCDs using the stars selected from VPDs of clusters NGC 5617, Trumpler 22,
NGC 3293 and NGC 3324. The continuous lines represent the slope determined through least-squares linear fit.} 
\label{cc_gaia}
\end{center}
\end{figure*}

In this section, we combined multi-wavelength photometric data with Gaia astrometry for clusters under study to check the
extinction law from optical to mid-infrared region. The resultant $(\lambda-G_{RP})/(G_{BP}-G_{RP})$
two colour diagrams (TCDs) are shown in Fig. \ref{cc_gaia} for all the clusters. Here, $\lambda$ denotes the filters other than $G_{RP}$.
All stars showing in Fig. \ref{cc_gaia} are probable cluster members. A linear fit to the data points is performed and slopes
are listed in Table \ref{gaia_slope}. The estimated values of slopes are in good agreement with the value given by Wang
\& Chen (2019) only for binary clusters NGC 5617 and Trumpler 22. We estimated $R=\frac{A_{V}}{E(B-V)}$ as $\sim$ 3.1 for
clusters NGC 5617 and Trumpler 22. Our obtained values of $R$ are $\sim$ 3.8 and 1.9 for clusters
NGC 3293 and NGC 3324. Our analysis indicates that reddening law is normal towards the cluster region of NGC 5617 and
Trumpler 22 while it is abnormal for binary clusters NGC 3293 and NGC 3324.

\begin{table*}
\centering
\caption{Multi-band colour excess ratios in the direction of clusters NGC 5617, Trumpler 22, NGC 3293 and NGC 3324.
}
\vspace{0.5cm}
\begin{center}
\small
\begin{tabular}{cccccc}
\hline\hline
Band $(\lambda)$ & Effective wavelength  &  & $\frac{\lambda-G_{RP}}{G_{BP}-G_{RP}}$ \\
&              (nm) &  NGC 5617 &  Trumpler 22 & NGC 3293 & NGC 3324
\\
\hline\hline
Johnson~ B              &445               &$1.60\pm0.03$ & $1.61\pm0.02$ &$1.27\pm0.03$  & $1.33\pm0.01$   \\
Johnson~ V              &551               &$0.88\pm0.02$ & $0.94\pm0.01$ &$0.65\pm0.02$  & $0.50\pm0.02$   \\
VPHAS~ i                &725               &$0.13\pm0.02$ & $0.12\pm0.04$ &$0.04\pm0.03$  & $0.04\pm0.04$   \\
VPHAS~ r                &620               &$0.68\pm0.04$ & $0.66\pm0.05$ &$0.44\pm0.05$  & $0.44\pm0.05$   \\
VPHAS~ $h_{\alpha}$     &659               &$0.63\pm0.06$ & $0.68\pm0.07$ &$0.24\pm0.07$  & $0.30\pm0.04$   \\
VPHAS~ g                &485               &$1.49\pm0.09$ & $1.52\pm0.08$ &$1.15\pm0.08$  & $1.33\pm0.12$   \\
VPHAS~ u                &380               &$2.60\pm0.10$ & $2.59\pm0.11$ &$2.10\pm0.11$  & $2.17\pm0.10$   \\
 J                      &1234.5            &$-0.77\pm0.03$& $-0.80\pm0.05$ &$-0.90\pm0.04$ & $-1.07\pm0.07$ \\
 H                      &1639.3            &$-1.20\pm0.05$& $-1.24\pm0.05$ &$-1.07\pm0.05$ & $-1.09\pm0.06$ \\
 K                      &2175.7            &$-1.33\pm0.06$& $-1.39\pm0.07$ &$-1.20\pm0.09$ & $-1.31\pm0.09$ \\
WISE ~W1                &3317.2            &$-1.37\pm0.09$& $-1.40\pm0.08$ &$-1.22\pm0.10$ & $-1.21\pm0.08$ \\
WISE~ W2                &4550.1            &$-1.43\pm0.10$& $-1.42\pm0.09$ &$-1.17\pm0.12$ & $-1.15\pm0.11$ \\
\hline
\end{tabular}
\label{gaia_slope}
\end{center}
\end{table*}

\subsubsection{Interstellar reddening from $JHK$ colours}

\begin{figure}
\begin{center}
\includegraphics[width=8.5cm, height=8.5cm]{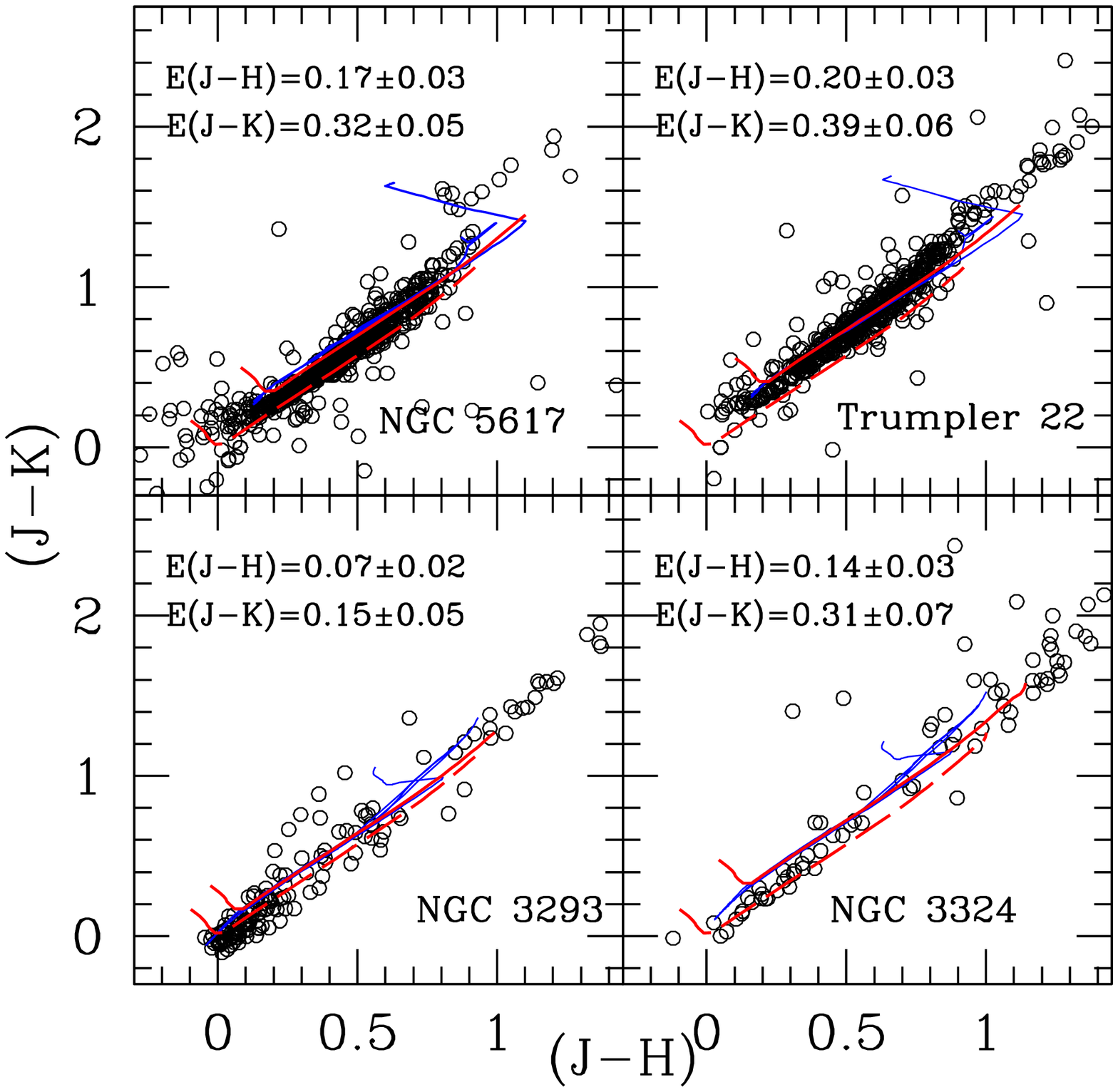}
\caption{The colour-colour diagrams (CCDs) for clusters under study using probable cluster members.
In CCDs, the red solid line is the ZAMS taken from Caldwell et al. (1993) while the red dotted lines are the same
ZAMS shifted by the values as described in the text. The blue line is the theoretical isochrones of log(age)=7.95
and 7.05 for cluster pair (NGC 5617 and Trumpler 22) and (NGC 3293 and NGC 3324), respectively.} 
\label{cc}
\end{center}
\end{figure}

To estimate the cluster reddening in the near-IR region, we used $(J-H)$ versus $(J-K)$ colour-colour diagrams
as shown in Fig. \ref{cc}. Stars plotted in this figure are the probable cluster members described in Sec. 3. The solid
line is the cluster's zero age main sequence (ZAMS) taken from Caldwell et al. (1993). The ZAMS shown by the dotted
line is displaced by the value of $E(J-H)$ and $E(J-K)$ for all clusters are given in Table \ref{para}.
In this figure, the solid line is theoretical isochrone taken from Marigo et al. (2017) of log(age)=8.25 and 7.05 for binary
clusters (NGC 5617 and Trumpler 22) and (NGC 3293 and NGC 3324), respectively. The ratio of $E(J-H)$ and $E(J-K)$ shows a good
agreement with the normal value 0.55 proposed by Cardelli et al. (1989). We have estimated the interstellar reddening,
$E(B-V)$ using the following relations (Fiorucci \& Munari, 2003):\\

~~~~~~~~~~~~~~~~~~~$E(J-H)=0.309\times E(B-V)$\\

~~~~~~~~~~~~~~~~~~~$E(J-K)=0.48\times E(B-V)$\\

Using the above relationships, we obtained the interstellar reddenings, $E(B-V)$ as, 0.55, 0.64, 0.23 and 0.45 for
the clusters NGC 5617, Trumpler 22, NGC 3293 and NGC 3324, respectively. Our derived value of $E(B-V)$ is similar
to Haug (1978) and slightly higher than KF91 for NGC 5617. Our $E(B-V)$ value for NGC 3324 is in good agreement with
the value obtained by Claria (1977).

\subsection{Age and distance to the clusters}

\begin{figure*}
\centering
\begin{center}
\hbox{
\includegraphics[width=6.5cm, height=6.5cm]{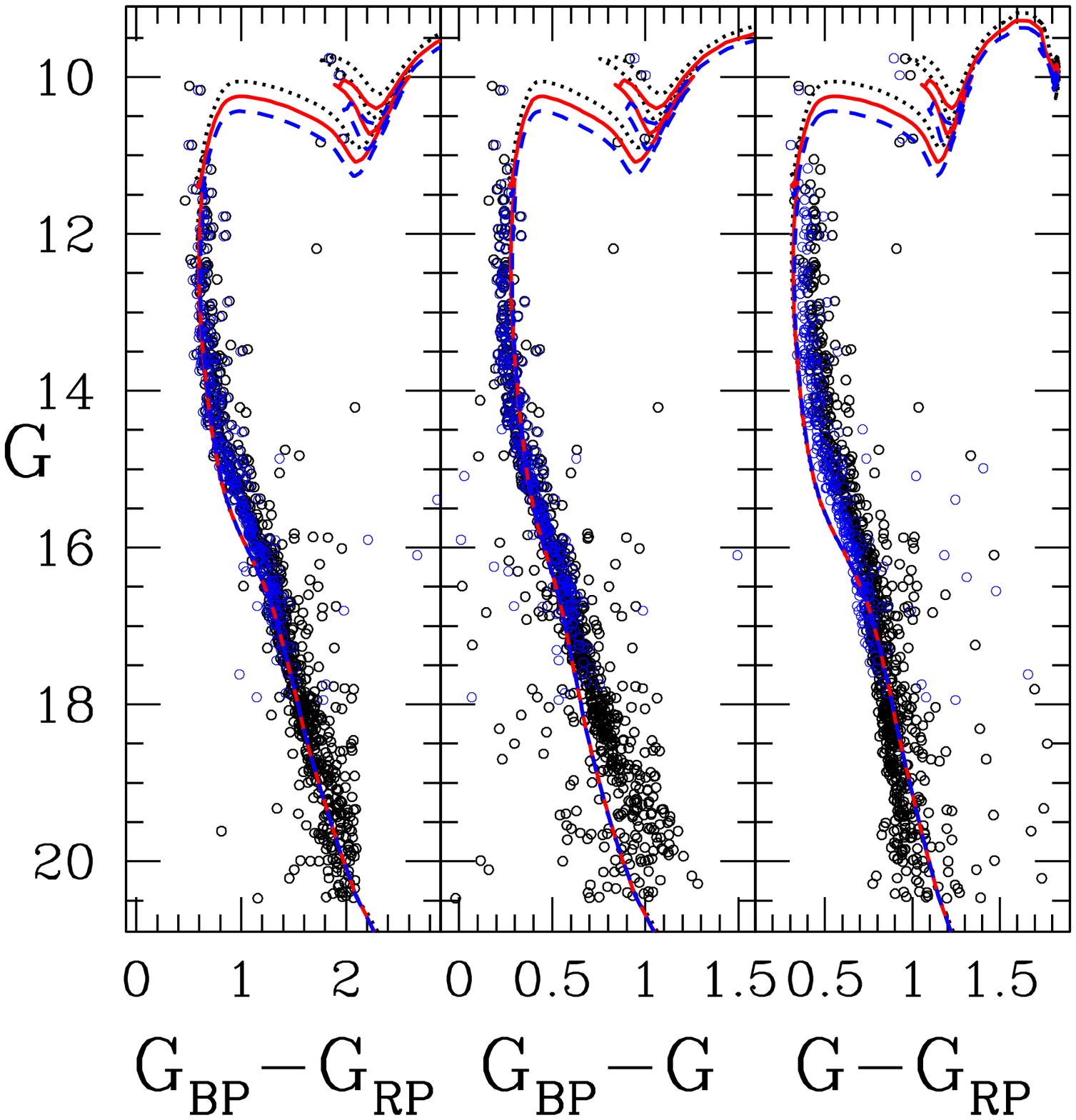}
\includegraphics[width=6.5cm, height=6.5cm]{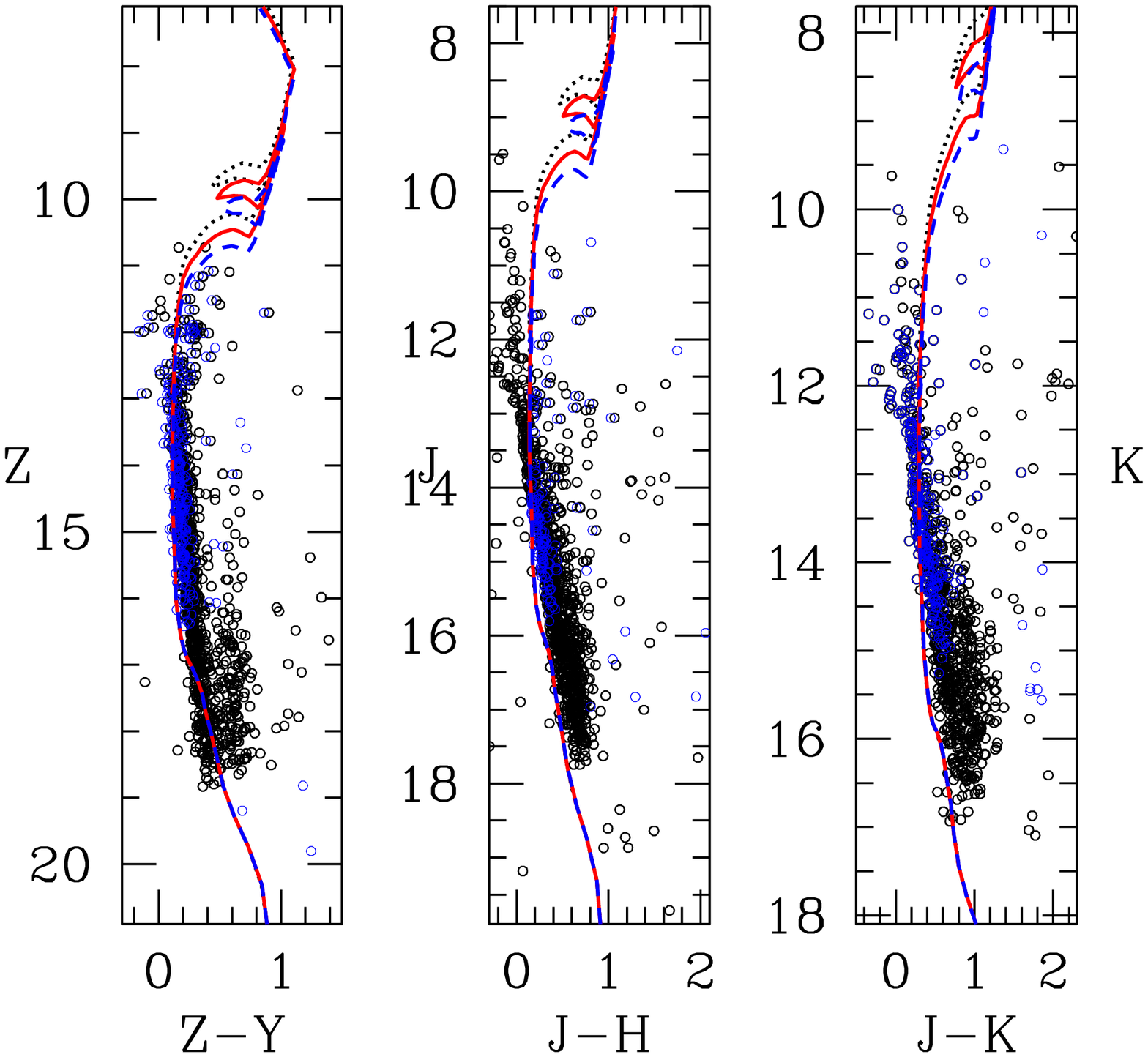}
}
\hbox{
\includegraphics[width=6.5cm, height=6.5cm]{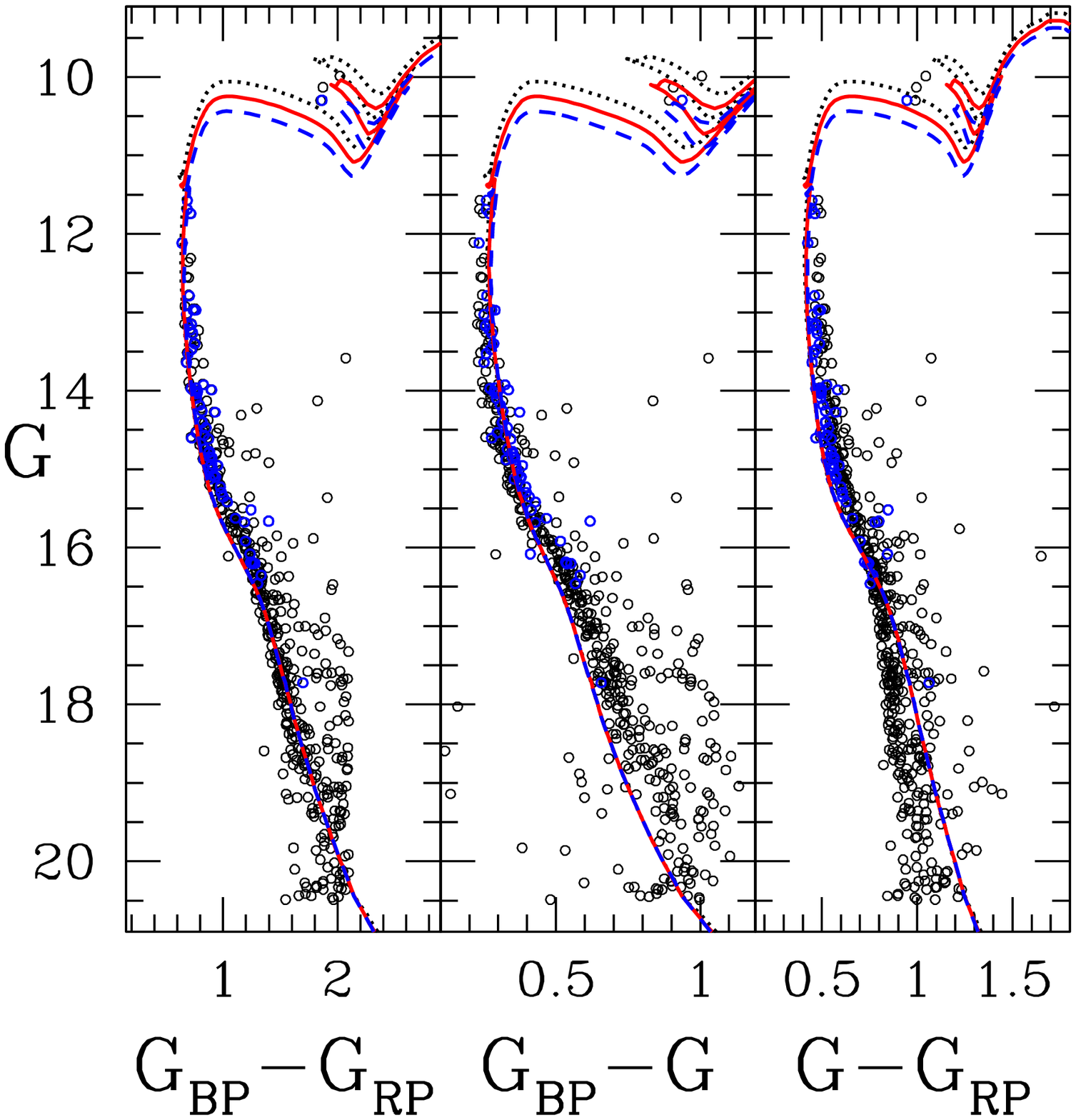}
\includegraphics[width=6.5cm, height=6.5cm]{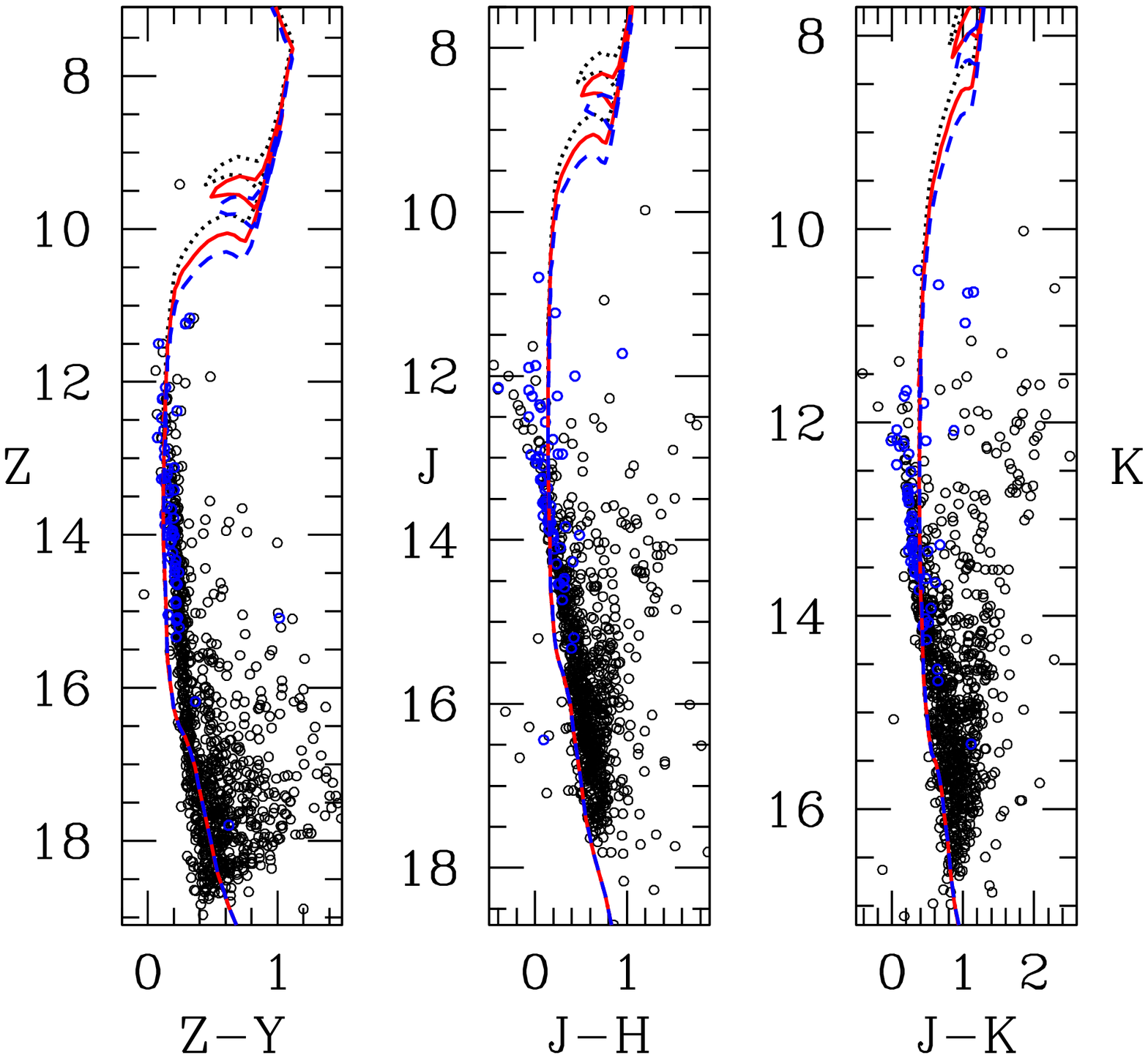}
}
\caption{The $G, (G_{BP}-G_{RP})$, $G, (G_{BP}-G)$, $G, (G-G_{RP})$, $Z, (Z-Y)$, $J, (J-H)$ and $K, (J-K)$ colour-magnitude
diagrams of open star cluster NGC 5617 (top panels) and Trumpler 22 (bottom panels). Black open circles show the most probable cluster members as selected from VPDs.
The curves represent the isochrones of (log(age)=7.90, 7.95 and 8.00) for both clusters. All these isochrones are taken from
Marigo et al. (2017) for solar metallicity. Blue dots are the matched stars with Cantat-Gaudin (2018) having membership
probability higher than $80\%$.} 
\label{dist_age}
\end{center}
\end{figure*}

\begin{figure*}
\centering
\begin{center}
\hbox{
\includegraphics[width=6.5cm, height=6.5cm]{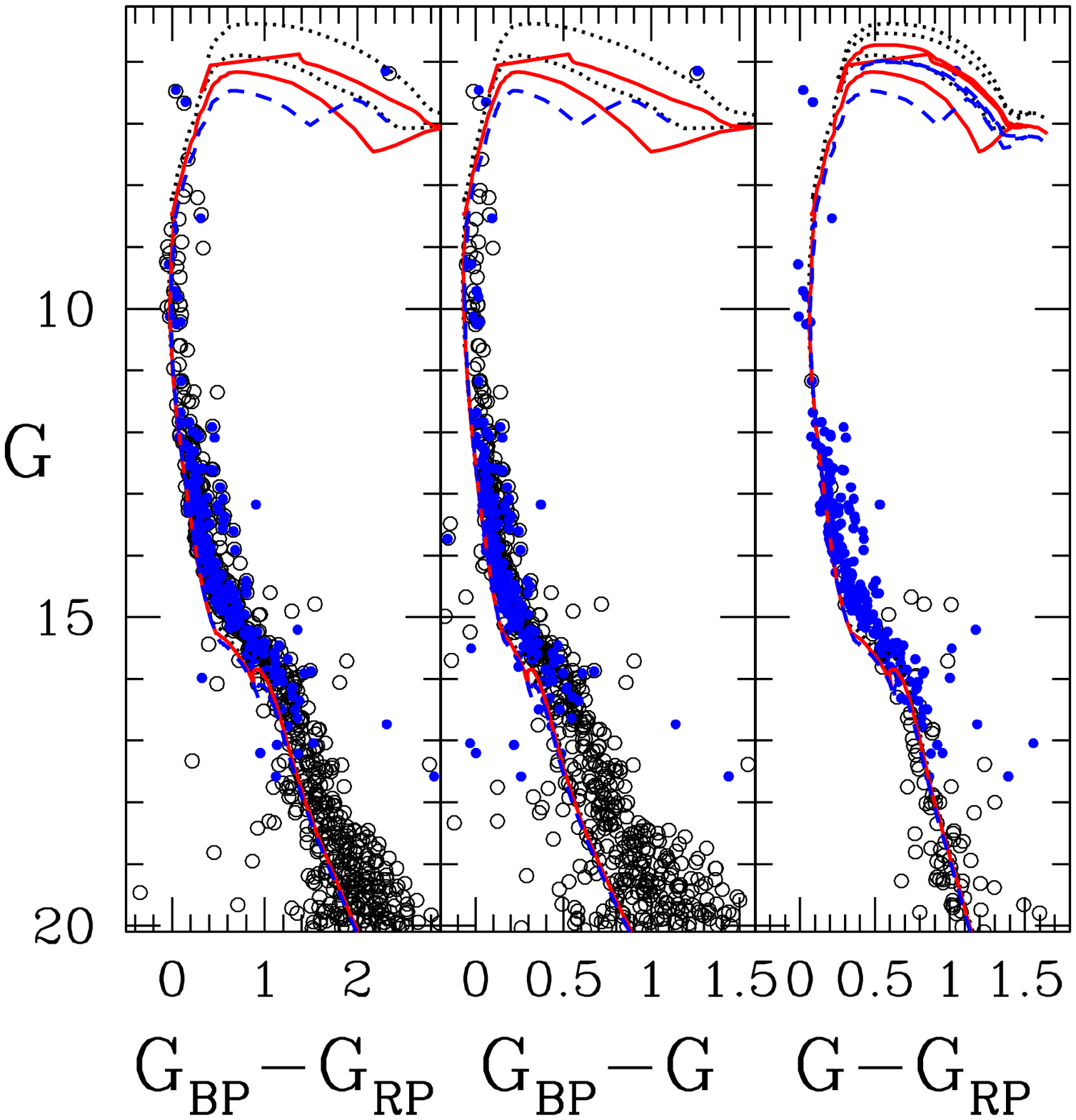}
\includegraphics[width=6.5cm, height=6.5cm]{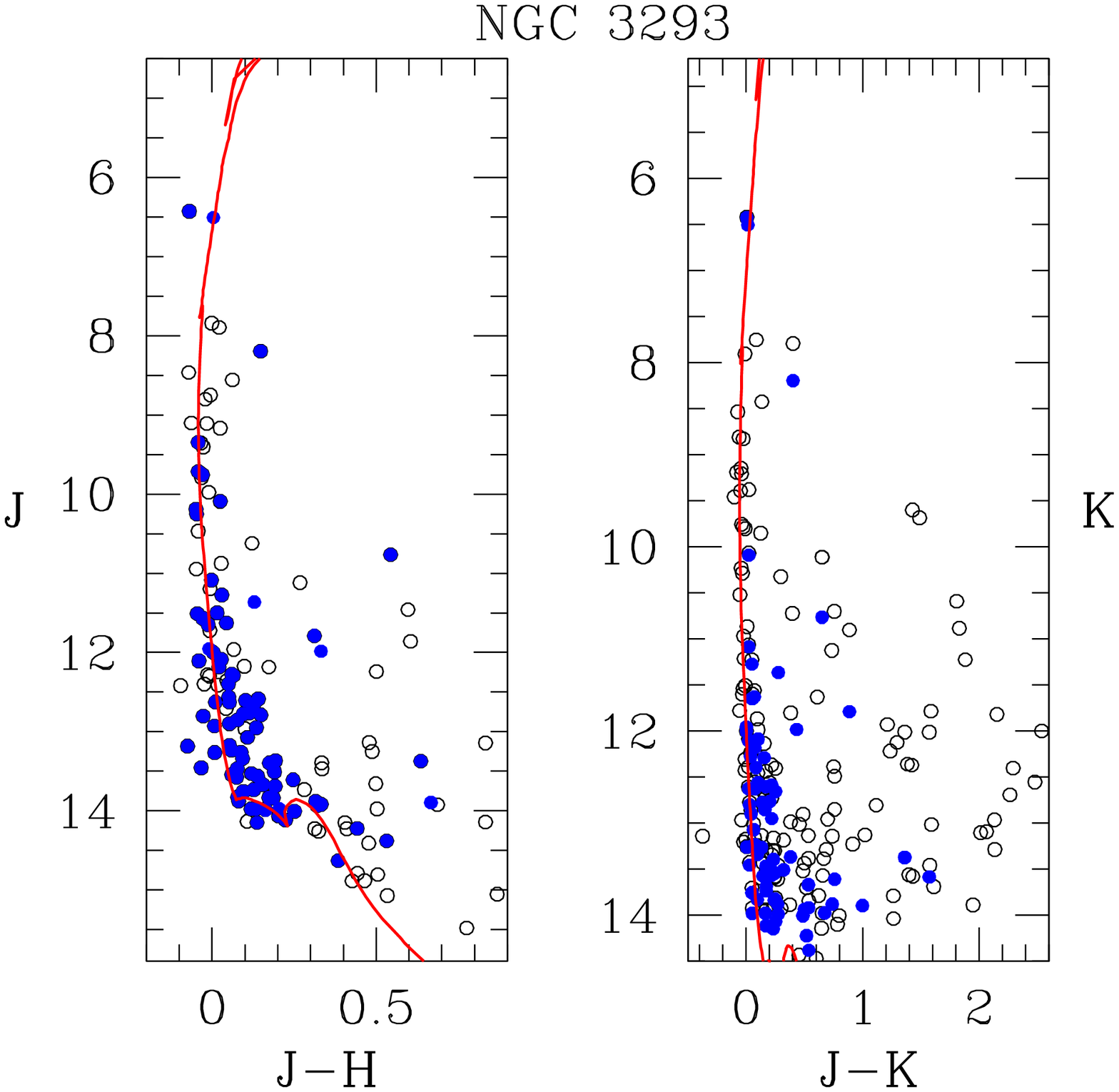}
}
\hbox{
\includegraphics[width=6.5cm, height=6.5cm]{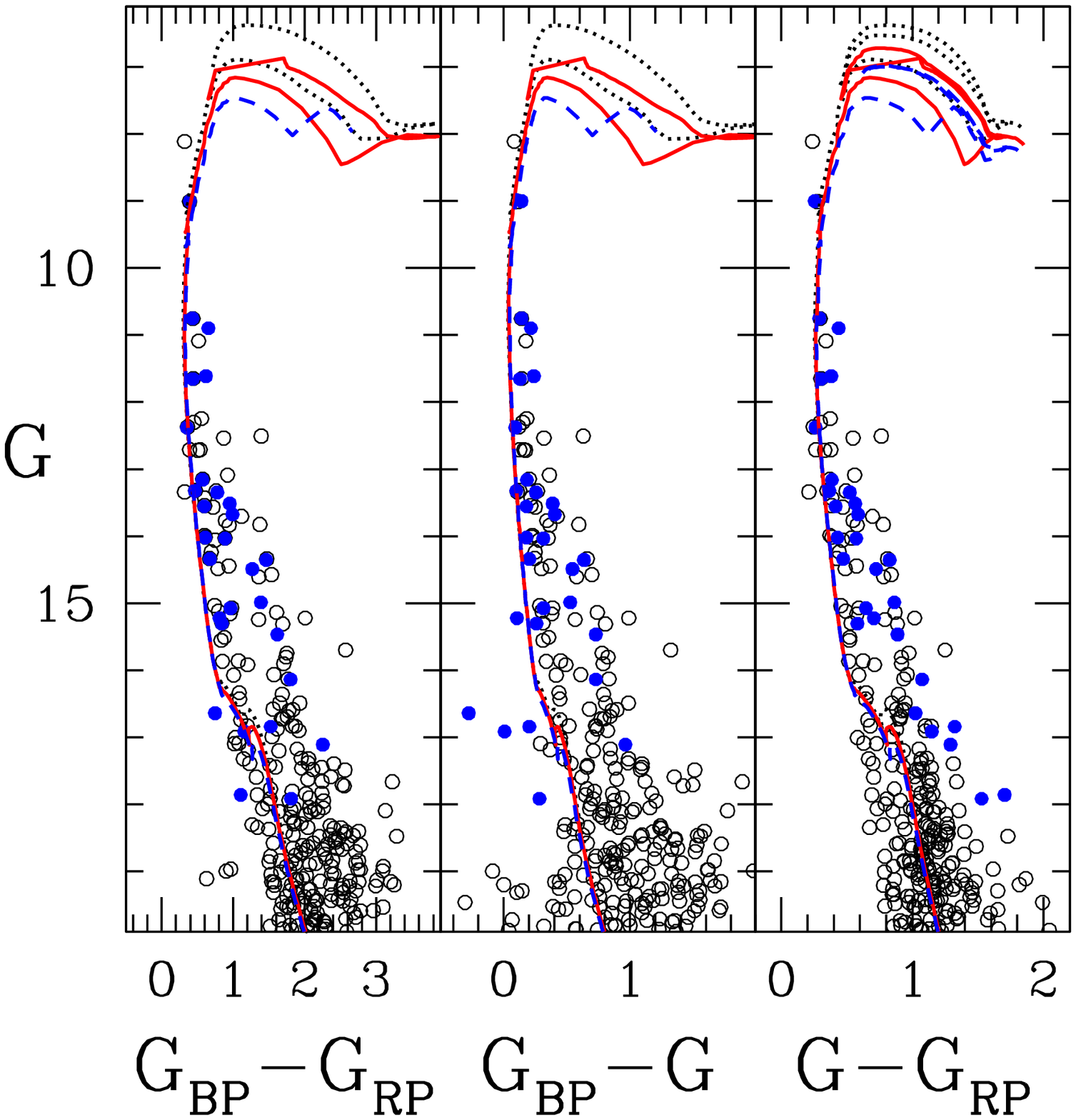}
\includegraphics[width=6.5cm, height=6.5cm]{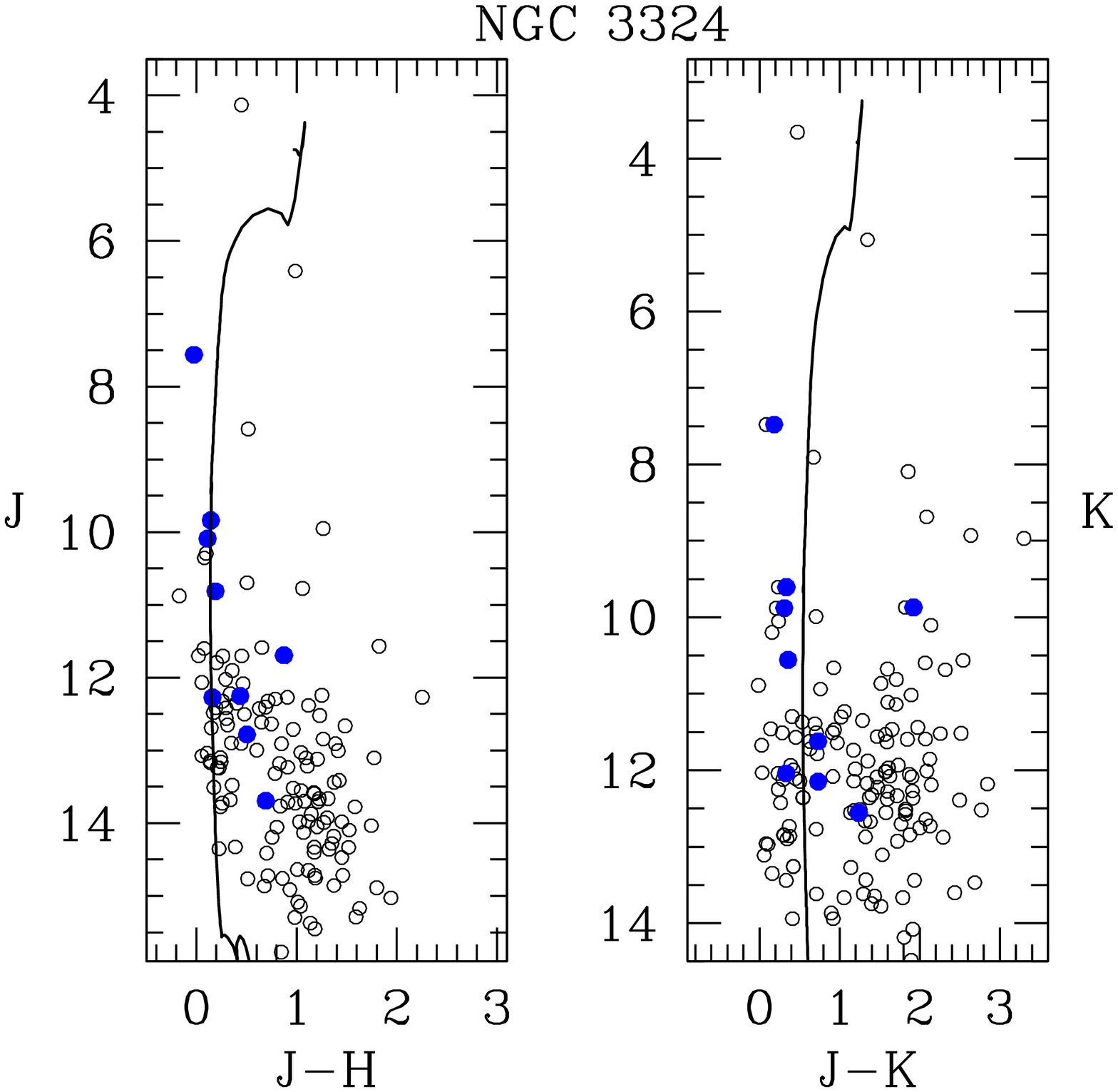}
}
\caption{The $G, (G_{BP}-G_{RP})$, $G, (G_{BP}-G)$, $G, (G-G_{RP})$, $J, (J-H)$ and $K, (J-K)$ colour-magnitude
diagrams of open star cluster NGC 3293 (top panels) and NGC 3324 (bottom panels). Black open circles are probable
cluster members as selected from VPDs. The curves are the isochrones of (log(age)=7.00, 7.05 and 7.10) in the CMDs of
the Gaia Bands. We used the middle age isochone of log(age)=7.05 in $J, (J-H)$ and $K, (J-K)$ CMDs. All these
isochrones are taken from Marigo et al. (2017) for solar metallicity. Blue dots are the matched stars with
Cantat-Gaudin (2018) having membership probability higher than $80\%$.}
\label{dist_age1}
\end{center}
\end{figure*}

To trace the Galactic structure and chemical evolution of the Galaxy using OCs, the distance, and age of OCs play the most
important role (Friel \& Janes 1993). We have estimated the mean value of $A_{G}$ for the studied clusters using the most probable members
from Gaia DR2. Our values are 1.58, 1.65, 0.86 and 0.75 for clusters NGC 5617, Trumpler 22, NGC 3293 and NGC 3324,
respectively. The main fundamental parameters (age, distance, and reddening) are estimated by fitting the solar metallicity
($Z=0.019$) isochrones of Marigo et al. (2017) to all the CMDs $(G, G_{BP}-G_{RP}), (G, G_{BP}-G), (G, G-G_{RP}) (Z, Z-Y),
(J, J-H) \& (K, J-K)$ as shown in Fig. \ref{dist_age} and Fig. \ref{dist_age1}. We have used only probable cluster members
in order to reduce the field star contamination in the cluster's area.

The galactocentric coordinates of the clusters $X$ (directed towards the galactic center in the Galactic disc), $Y$ (directed towards
the Galactic rotation) and distance from the galactic plane $Z$ (directed towards Galactic north pole) can be estimated using
clusters' distances, longitude, and latitude. The Galactocentric distance has been calculated by considering 8.5 kpc as the
distance of the Sun to the Galactic center. The estimated Galactocentric coordinates are listed in Table \ref{para}.

The estimation of the main fundamental parameters for the clusters are given below:

{\bf NGC 5617:} We fitted the theoretical isochrones of different ages (log(age)=7.90,7.95 and 8.00) in all the CMDs for the
cluster NGC 5617, shown in the upper panels of Fig. \ref{dist_age}. The best global fit is favorable for the middle isochrone with
log(age)=7.95 to the high mass cluster members. A satsifactory fitting of isochrones provides an age of $90\pm10$ Myr. The apparent
distance modulus ($(m-M)=13.70\pm0.4$ mag) provides a distance $2.5\pm0.30$ kpc from the Sun. The estimated distance is in good
agreement with the value of 3.0 kpc as given by Cantat-Gaudin  et al. (2018).\\

{\bf Trumpler 22:} In the CMDs of Trumpler 22, we have fitted exactly the similar age isochrones as shown in Fig. \ref{dist_age}. So, the age
of this object is the same as that of NGC 5617. The inferred apparent distance modulus $(m-M)=14.20\pm0.3$ mag provides a heliocentric distance
as $2.8\pm0.2$ kpc. This value of the distance is very close to the distance derived by Cantat-Gaudin et al. (2018). 

{\bf NGC 3293:} For cluster NGC 3293, we have fitted the theoretical isochrones of different ages (log(age)=7.00,7.05 and 7.10)
as shown in Fig. \ref{dist_age1}. Based on the best fitted middle isochrone of log(age)=7.05, we found the age of this object
as $12\pm2$ Myr. The inferred apparent distance modulus $(m-M)=12.90\pm0.2$ mag provides a heliocentric distance as $2.6\pm0.1$
kpc. This value of the distance is very close to the distance derived by Cantat-Gaudin et al. (2018). 

{\bf NGC 3324:} For this cluster also, the isochrones of the same age values as NGC 3293 were fitted (see Fig. \ref{dist_age1}). 
The inferred apparent distance modulus ($(m-M)=13.00\pm0.2$) mag provides a heliocentric distance of $2.8\pm0.2$ kpc. This value of the distance is very
close to the distance derived by Cantat-Gaudin et al. (2018).

\begin{figure}
\begin{center}
\hbox{
\includegraphics[width=4.5cm, height=4.5cm]{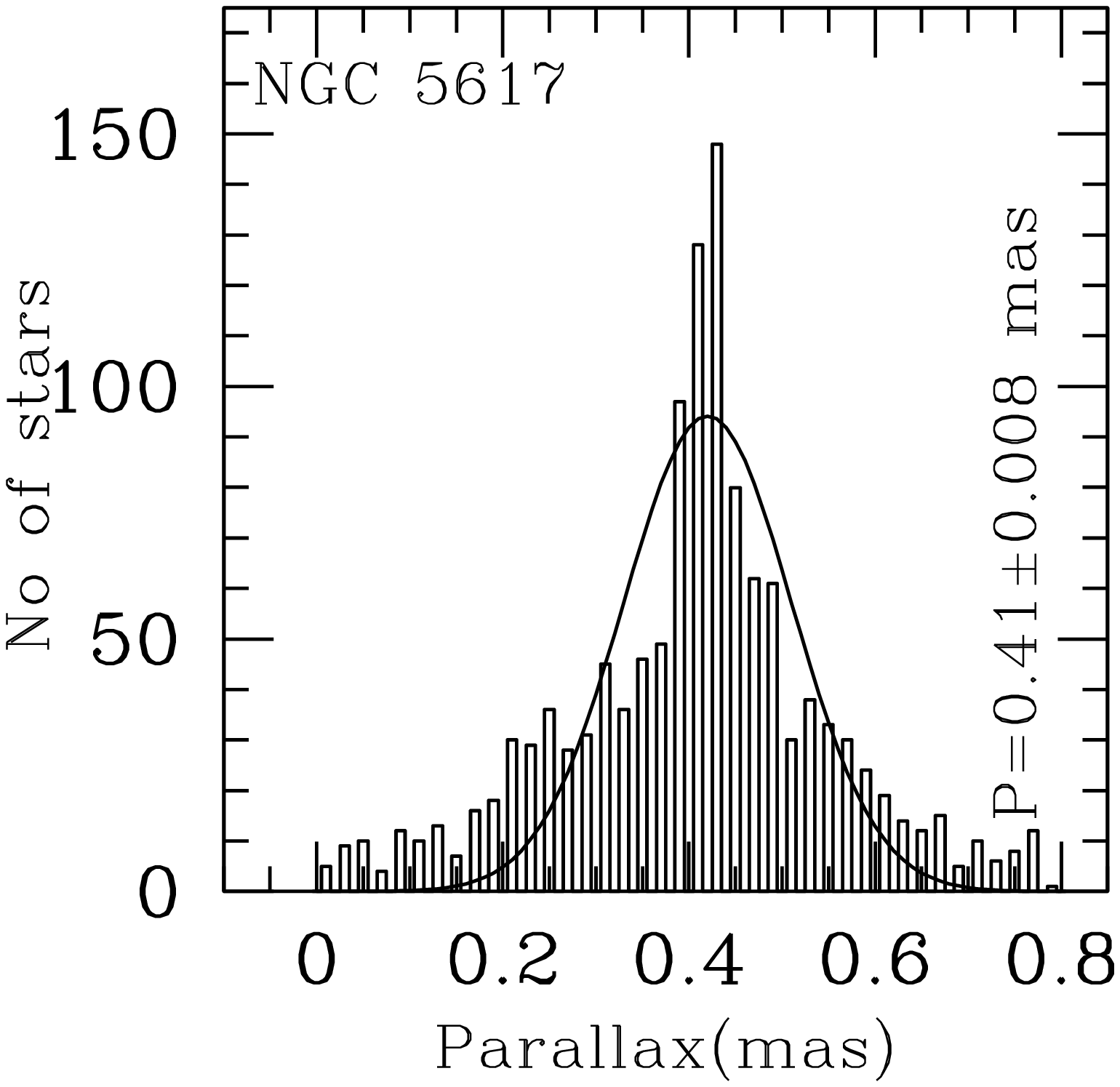}
\includegraphics[width=4.5cm, height=4.5cm]{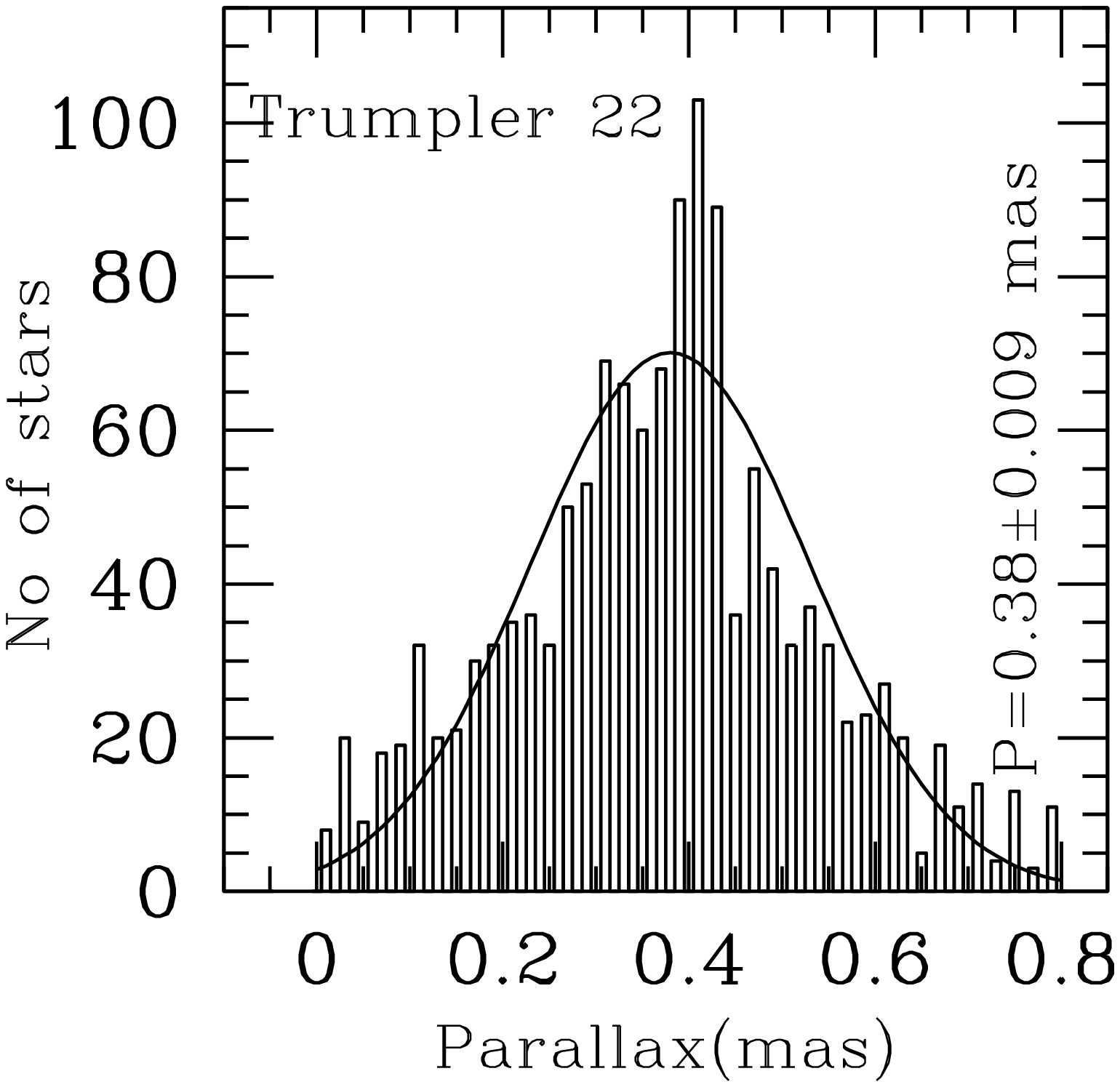}
}
\hbox{
\includegraphics[width=4.5cm, height=4.5cm]{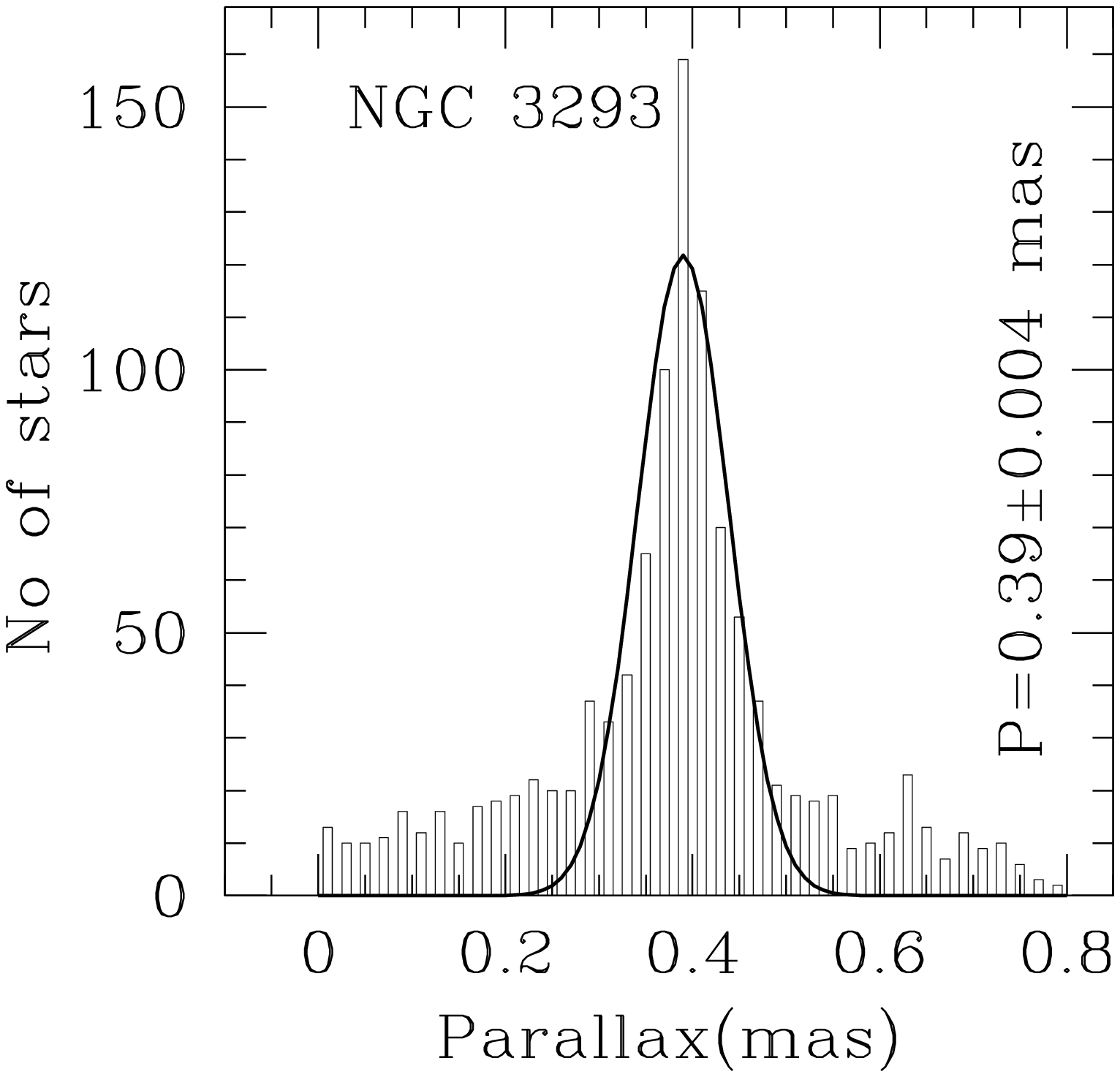}
\includegraphics[width=4.5cm, height=4.5cm]{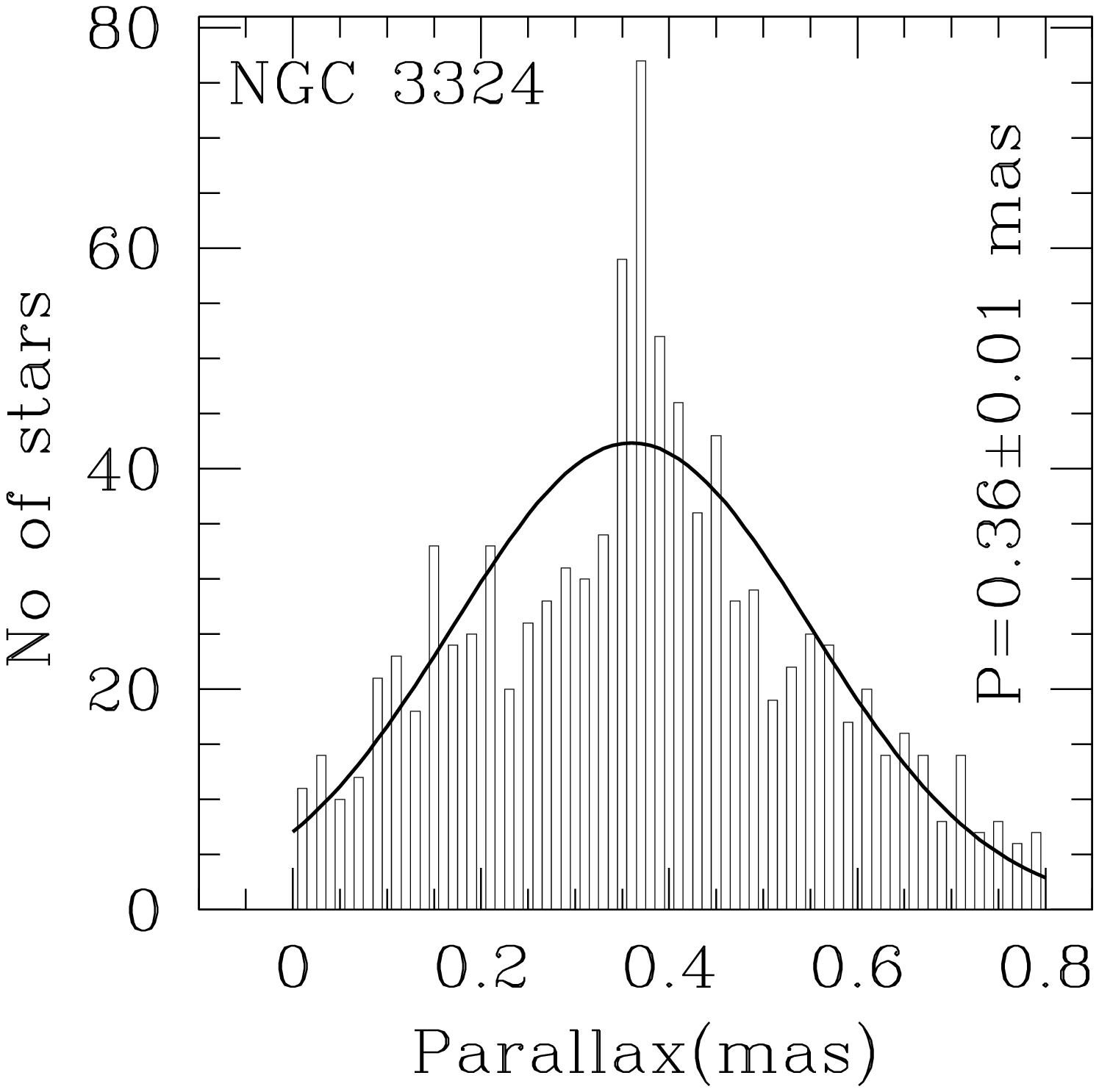}
}
\caption{Histogram for parallax estimation of the clusters NGC 5617, Trumpler 22, NGC 3293 and NGC 3324 using probable
cluster members based on clusters VPDs. The Gaussian function is fitted to the central bins provides a mean value of parallax. The
dashed line is the mean value of clusters parallax.} 
\label{pllax}
\end{center}
\end{figure}

We have used kinematical data from Gaia EDR3 to estimate the distances of clusters NGC 5617, Trumpler 22, NGC 3293 and NGC 3324. We
can estimate cluster distance using the mean parallax of probable members (Lauri et al. 2018). The Gaia~DR2 parallax has been
corrected for these clusters after adopting zero-point offset ($-$0.05 mas) as given by Riess et al. (2018). The histograms of parallax
using probable members in all clusters with 0.15 mas bins are shown in Fig. \ref{pllax}. The mean parallax is estimated as
$0.41\pm0.008$ mas, $0.38\pm0.009$ mas, $0.39\pm0.004$ mas and $0.36\pm0.01$ mas for the clusters NGC 5617,
Trumpler 22, NGC 3293 and NGC 3324, respectively and the corresponding distance values are $2.44\pm0.05$ kpc, $2.63\pm0.06$ kpc,
$2.56\pm0.02$ kpc and $2.77\pm0.07$ kpc. These obtained values of distance are reciprocal of cluster parallax. The mean
parallax for all clusters are very close to the parallax obtained by Cantat-Gaudin et al. (2018). We also determined the distance of the
clusters according to the method discussed by Bailer-Jones et al. (2018). In this way, our estimated values are $2.43\pm0.08$ kpc,
$2.64\pm0.07$ kpc, $2.59\pm0.1$ kpc and  $2.80\pm0.2$ kpc for clusters NGC 5617, Trumpler 22, NGC 3293 and NGC 3324, respectively.
The estimated values of cluster distance are also in good agreement with the results obtained using the isochrone fitting method as
described in the above paragraph.

\subsection{Young stellar object candidates}

A star can be considered a young stellar object candidate (YSO) if the free reddening parameter (Q) becomes less than
$-$0.05 mag (Buckner \& Froebrich (2013)). This Q value can be estimated for stars using the relationship
given by Buckner \& Froebrich (2013) as below,\\

~~~~~~~~~~~$Q=(J-H)-1.55\times(H-K)$,\\

\begin{figure}
\begin{center}
\hbox{
\includegraphics[width=4.5cm, height=4.5cm]{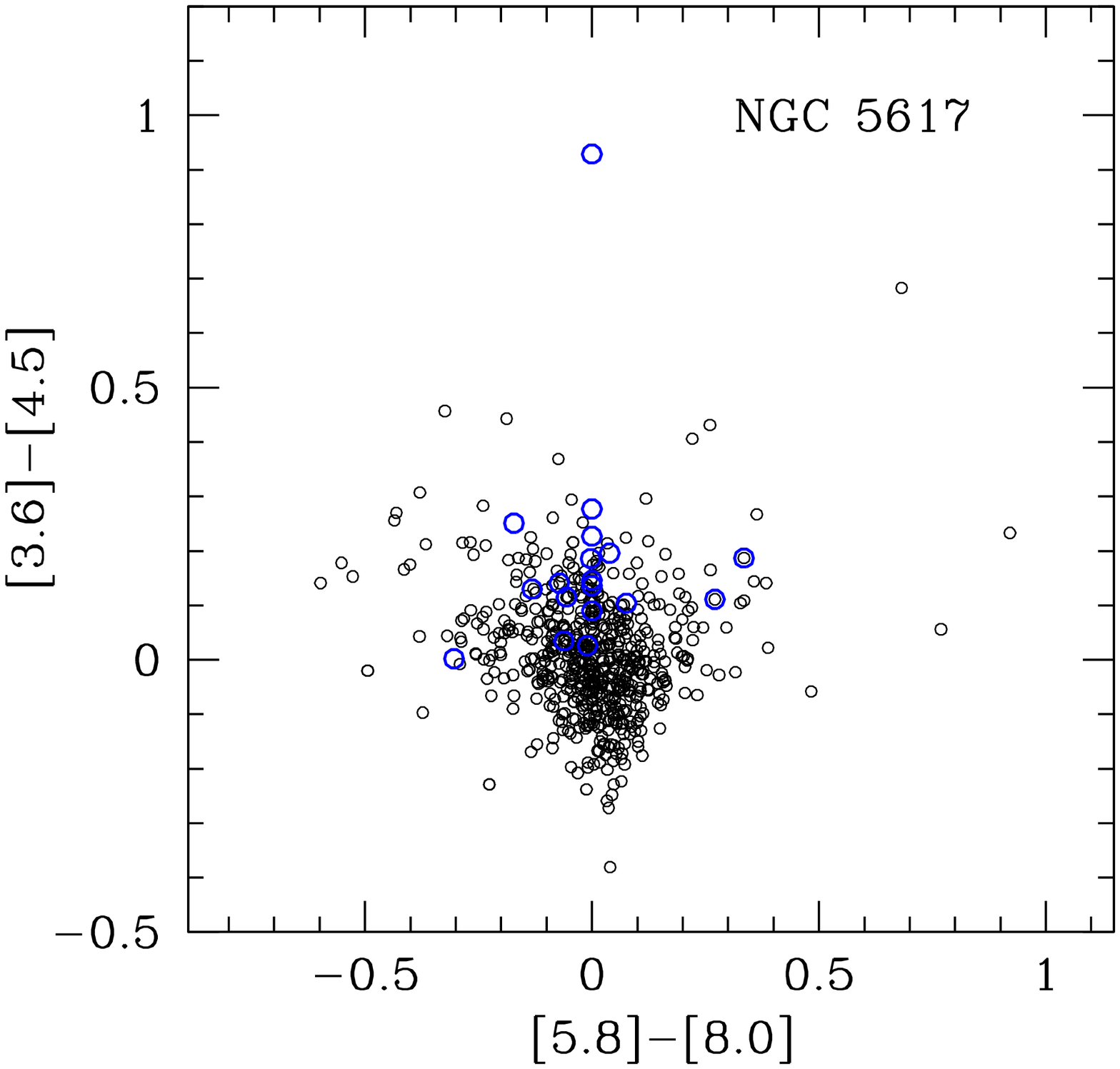}
\includegraphics[width=4.5cm, height=4.5cm]{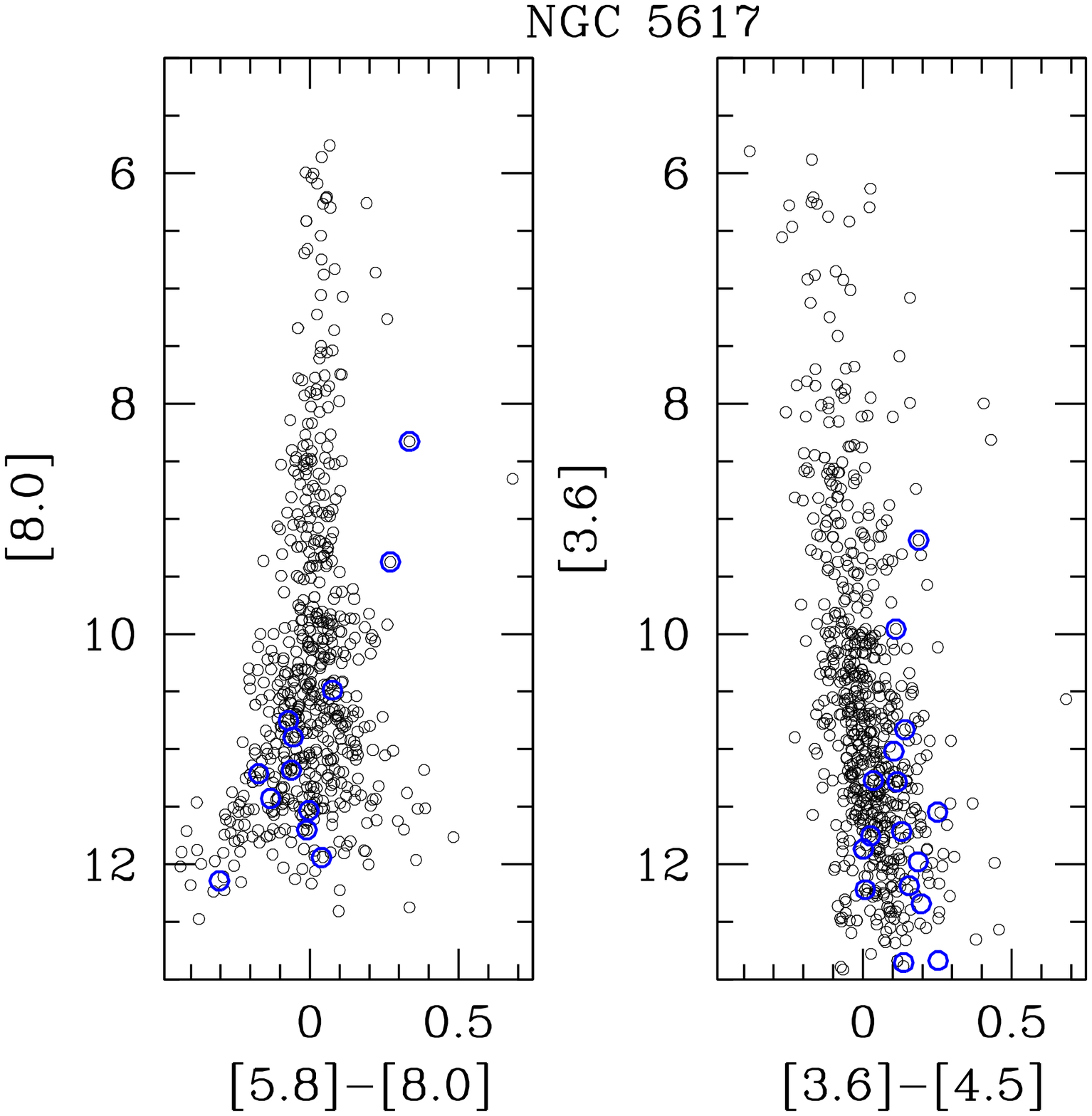}
}
\hbox{
\includegraphics[width=4.5cm, height=4.5cm]{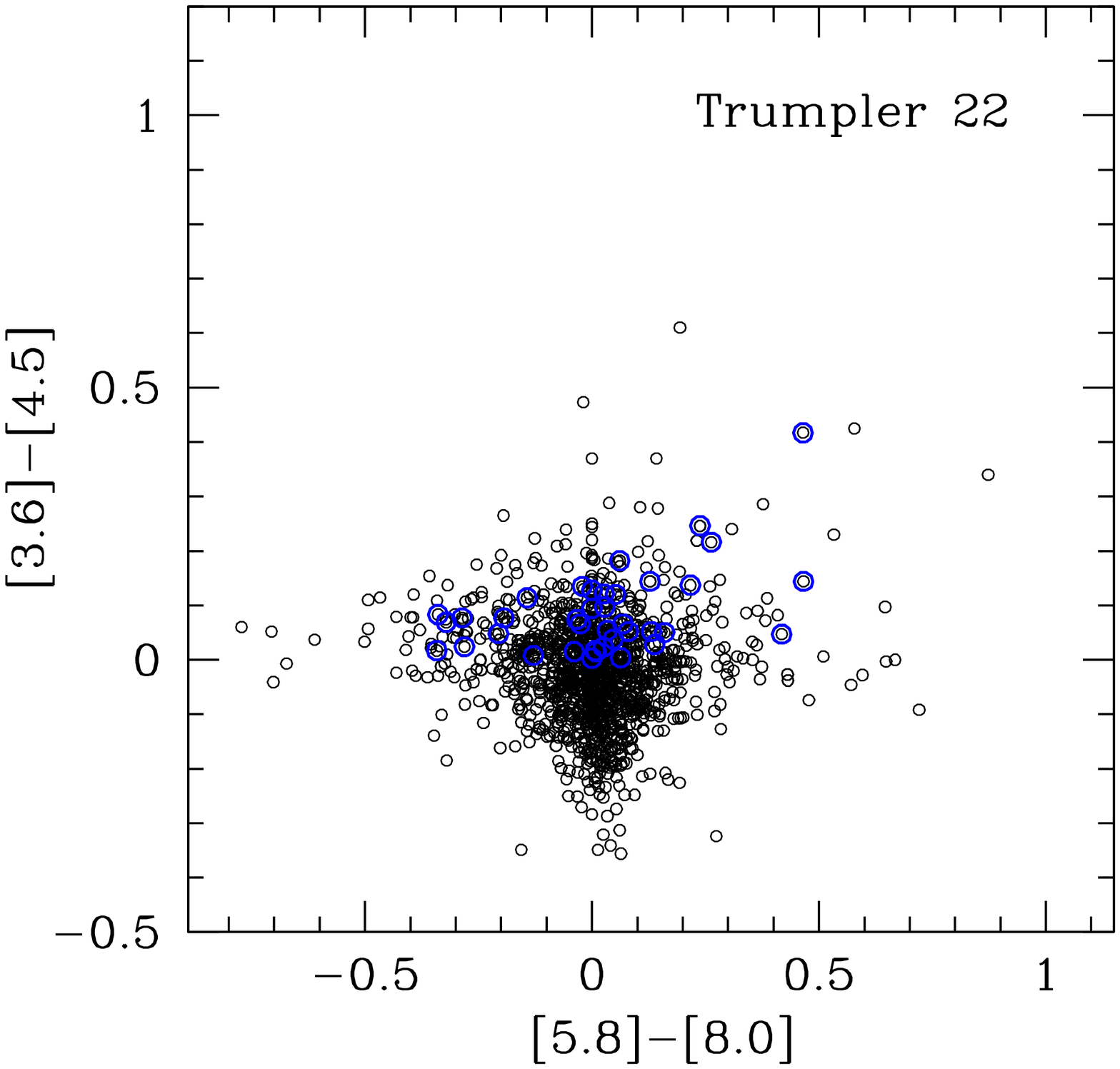}
\includegraphics[width=4.5cm, height=4.5cm]{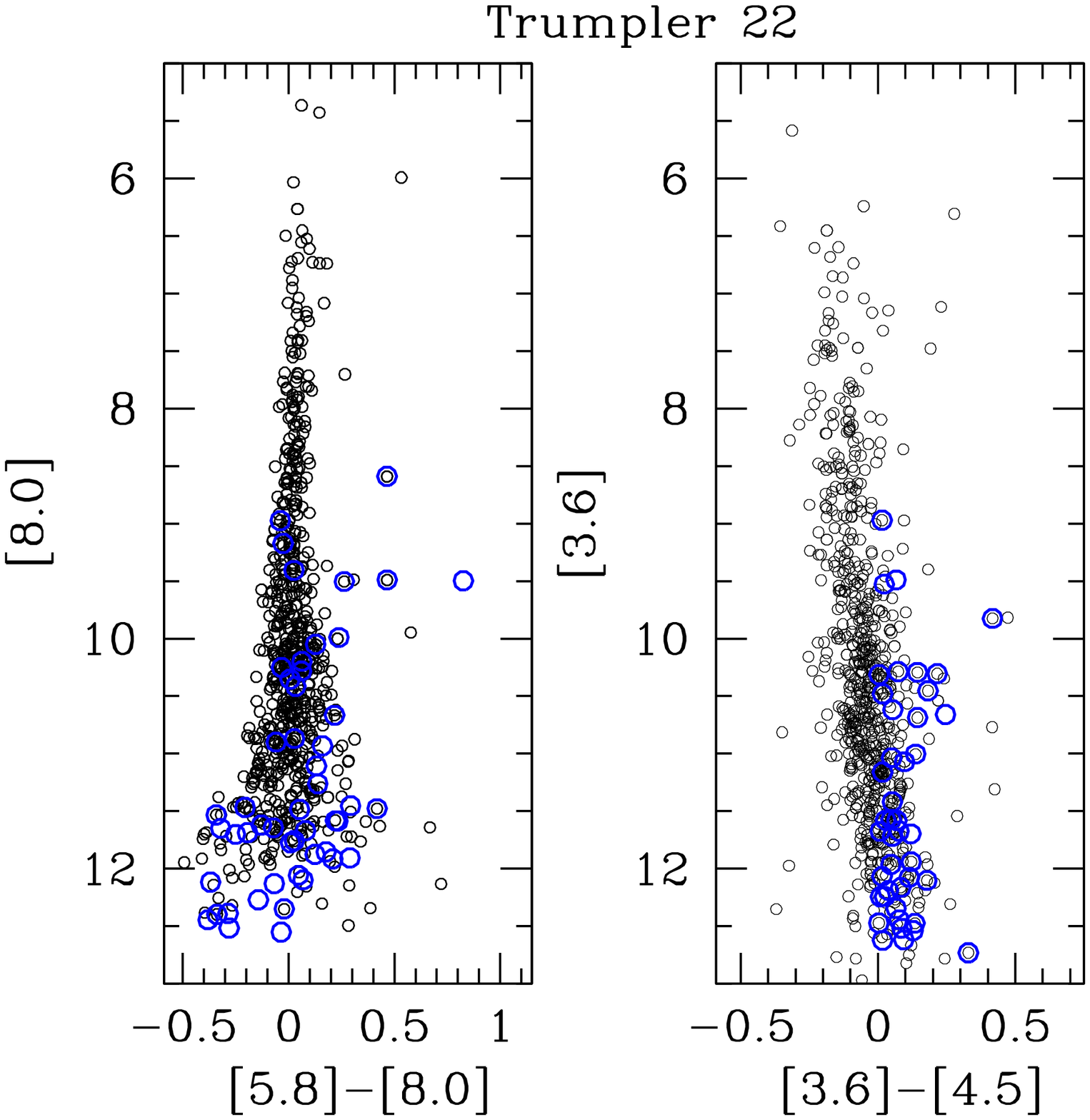}
}
\caption{(Top left panel) [5.8]-[8.0] vs [3.6]-[4.5] color-color diagram for NGC 5617. (Top right panel)
         [8.0],[5.8]-[8.0] and [3.6],[3.6]-[4.5] CMDs for NGC 5617. Same as Trumpler 22 in bottom panels. Blue dots
         are young stellar objects as identified towards the cluster region.} 
\label{yso}
\end{center}
\end{figure}

Here $J$, $H$, and $K$ are the VVV photometric magnitudes of stars. Using the above relation, we obtained a total of
18 and 44 members as YSOs towards the cluster region of NGC 5617 and Trumpler 22, respectively. These identified
YSOs have been plotted in Fig. \ref{yso} with blue dots in each panel.

\section{Dynamical study of the clusters}

\subsection{Luminosity function and Mass function}

\begin{figure}
\begin{center}
\hbox{
\includegraphics[width=4.5cm, height=4.5cm]{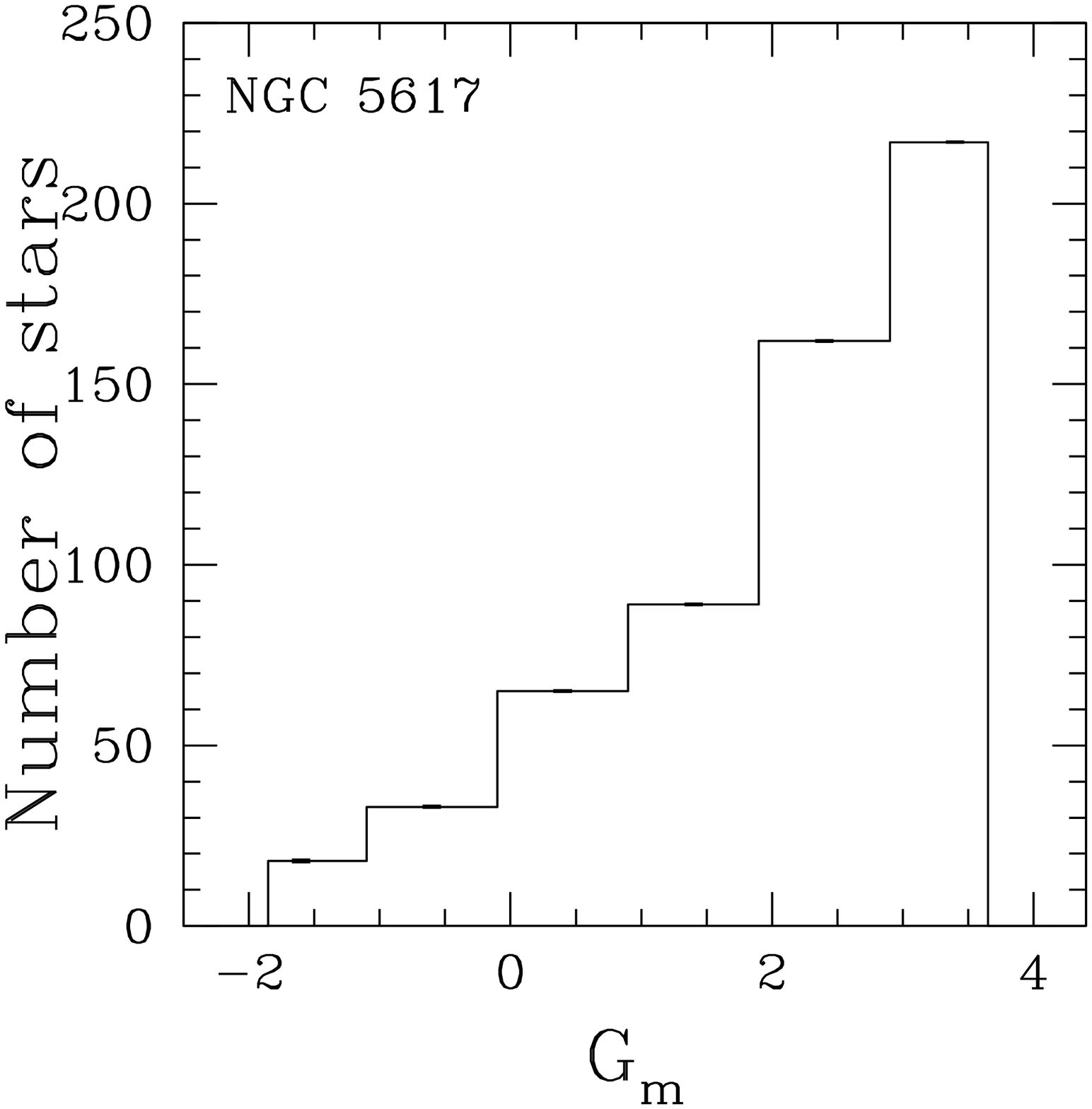}
\includegraphics[width=4.5cm, height=4.5cm]{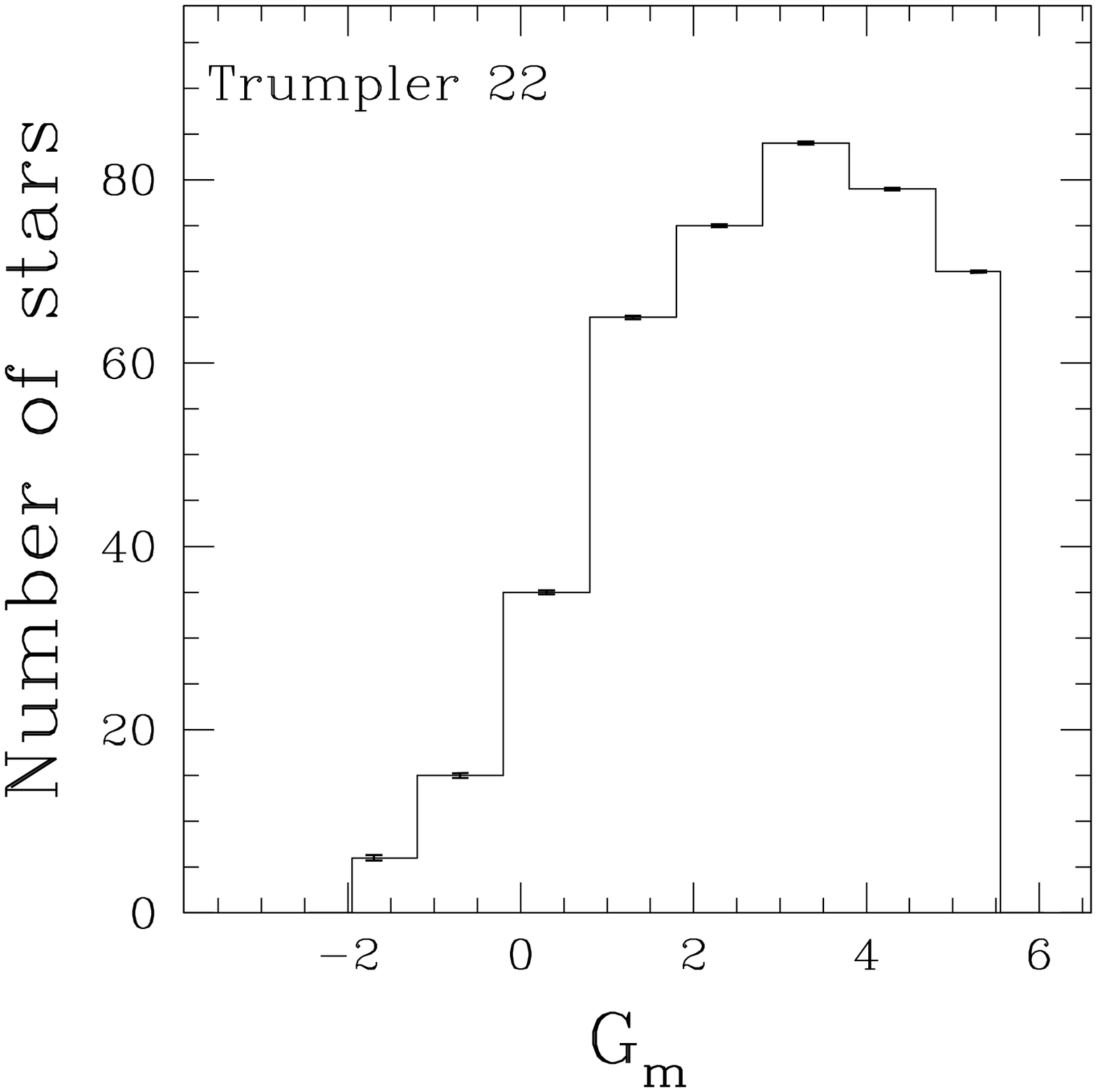}
}
\hbox{
\includegraphics[width=4.5cm, height=4.5cm]{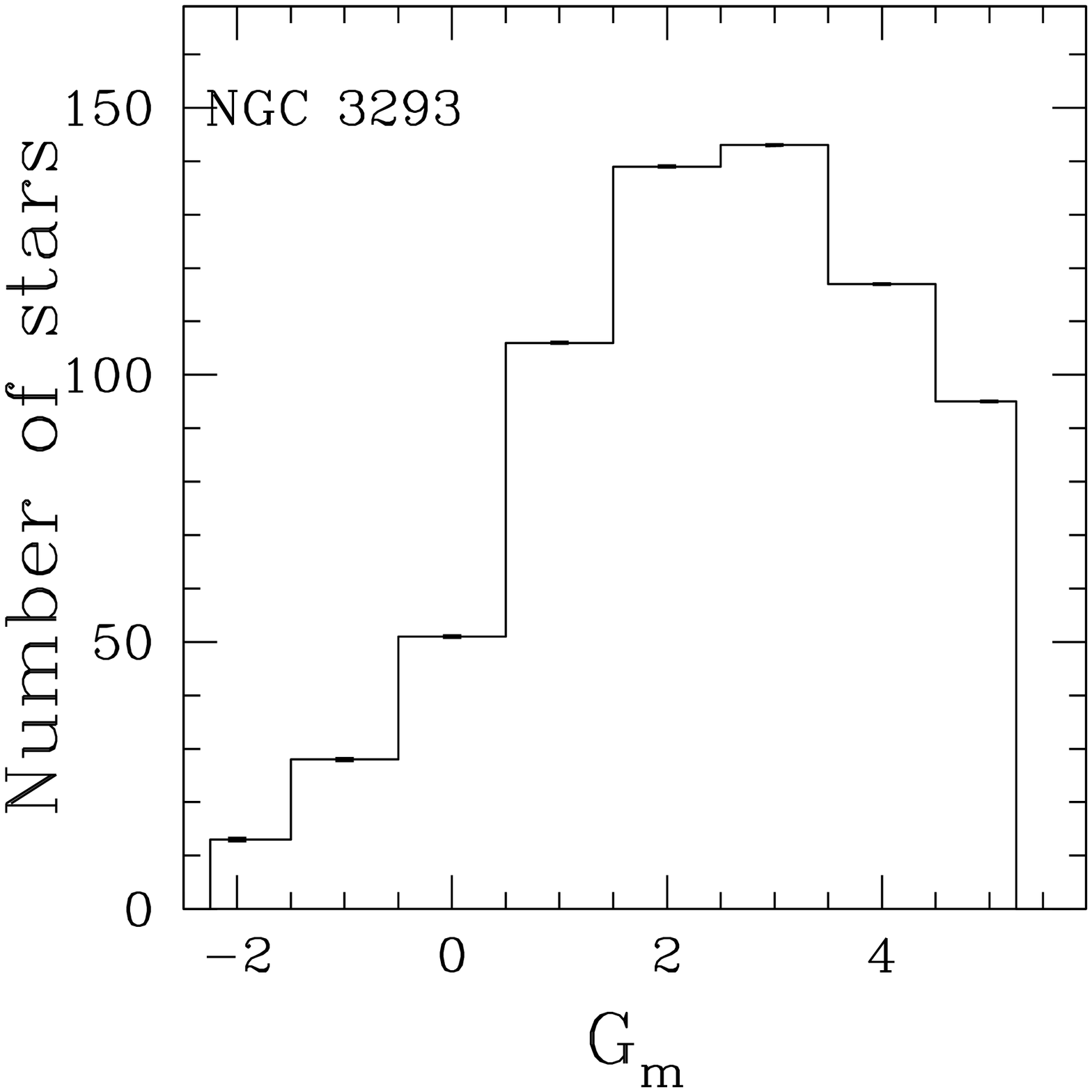}
\includegraphics[width=4.5cm, height=4.5cm]{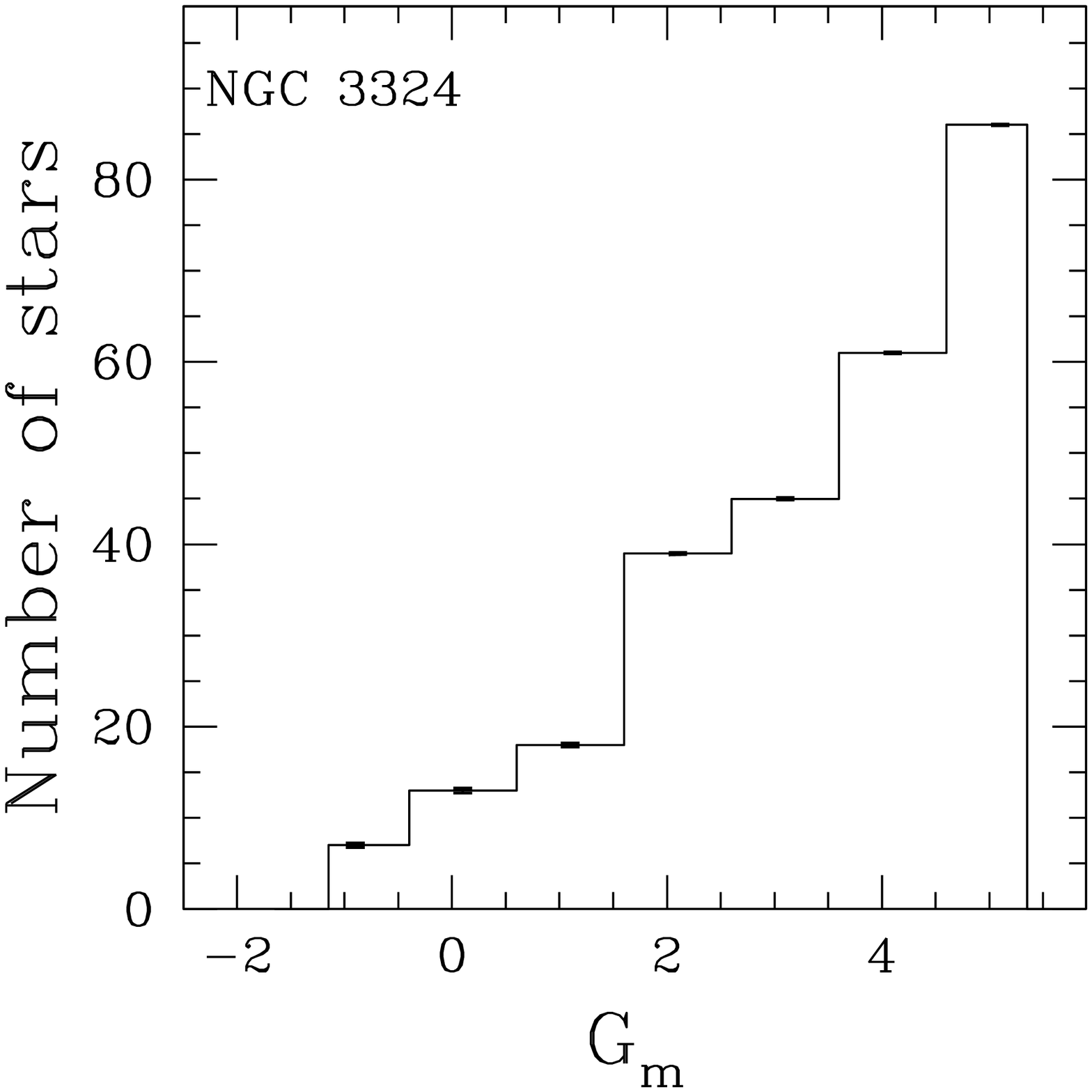}
}
\caption{Luminosity function of main sequence stars in the region of the clusters NGC 5617, Trumpler 22, NGC 3293 and NGC 3324}. 
\label{lf}
\end{center}
\end{figure}

Luminosity function (LF) and Mass function (MF) are primarily dependent on cluster membership and also connected with the
well known mass-luminosity relationship. To construct the LF for clusters NGC 5617, Trumpler 22, NGC 3293 and NGC 3324
we used $G$ versus $(G_{BP}-G_{RP})$ CMD. We converted the $G$ magnitudes of main sequence stars into the absolute magnitudes
using the distance modulus and reddening calculated in this paper for all clusters. A histogram is constructed with
1.0 mag intervals as shown in Fig. \ref{lf}. This figure exhibits that the LF continues to increase up to $M_{G}\sim$ 3.4, 2.0,
3.3 and 5.2 mag for clusters NGC 5617, Trumpler 22, NGC 3293 and NGC 3324, respectively.

\begin{figure}
\begin{center}
\hbox{
\includegraphics[width=4.3cm, height=4.3cm]{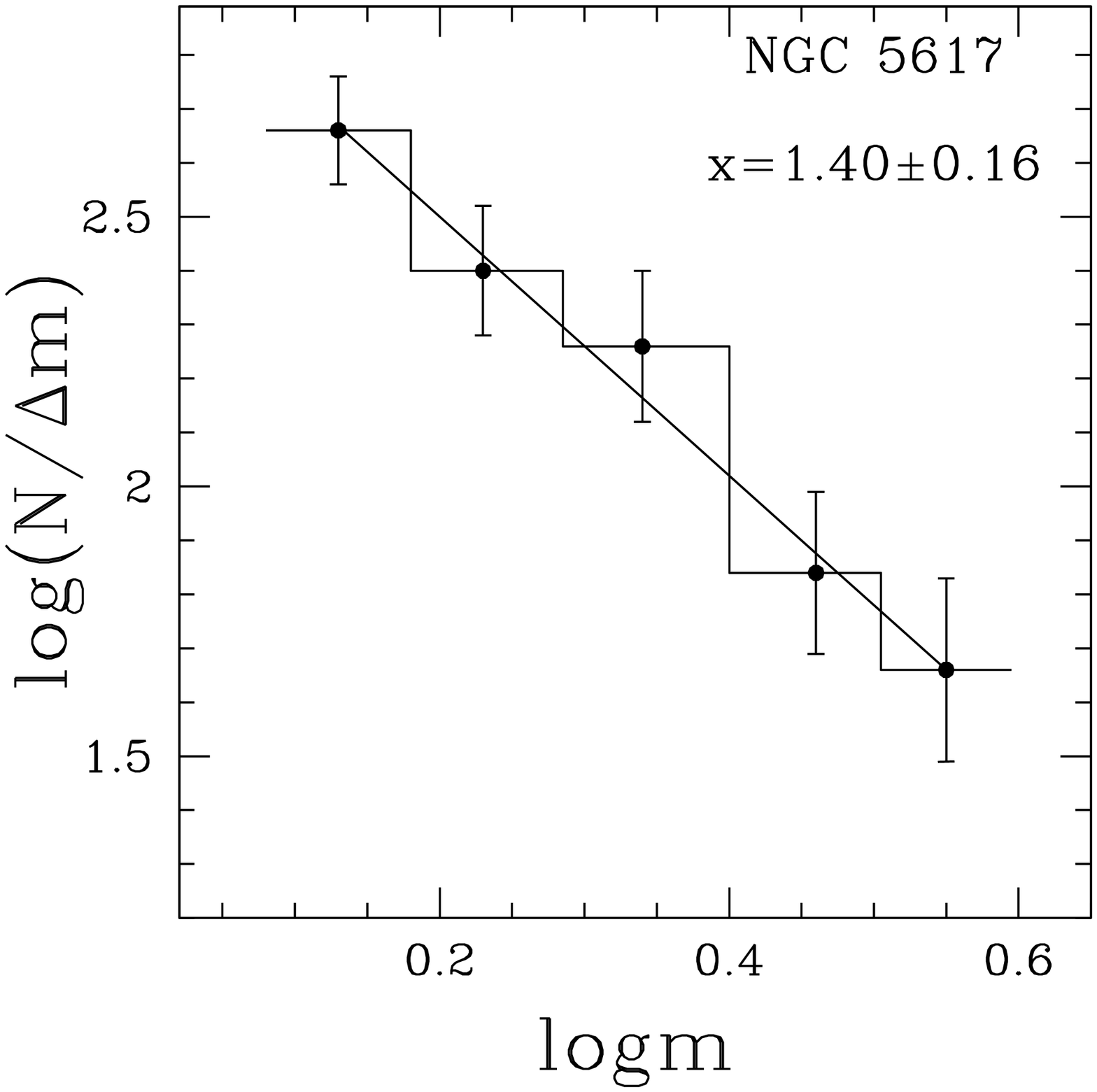}
\includegraphics[width=4.3cm, height=4.3cm]{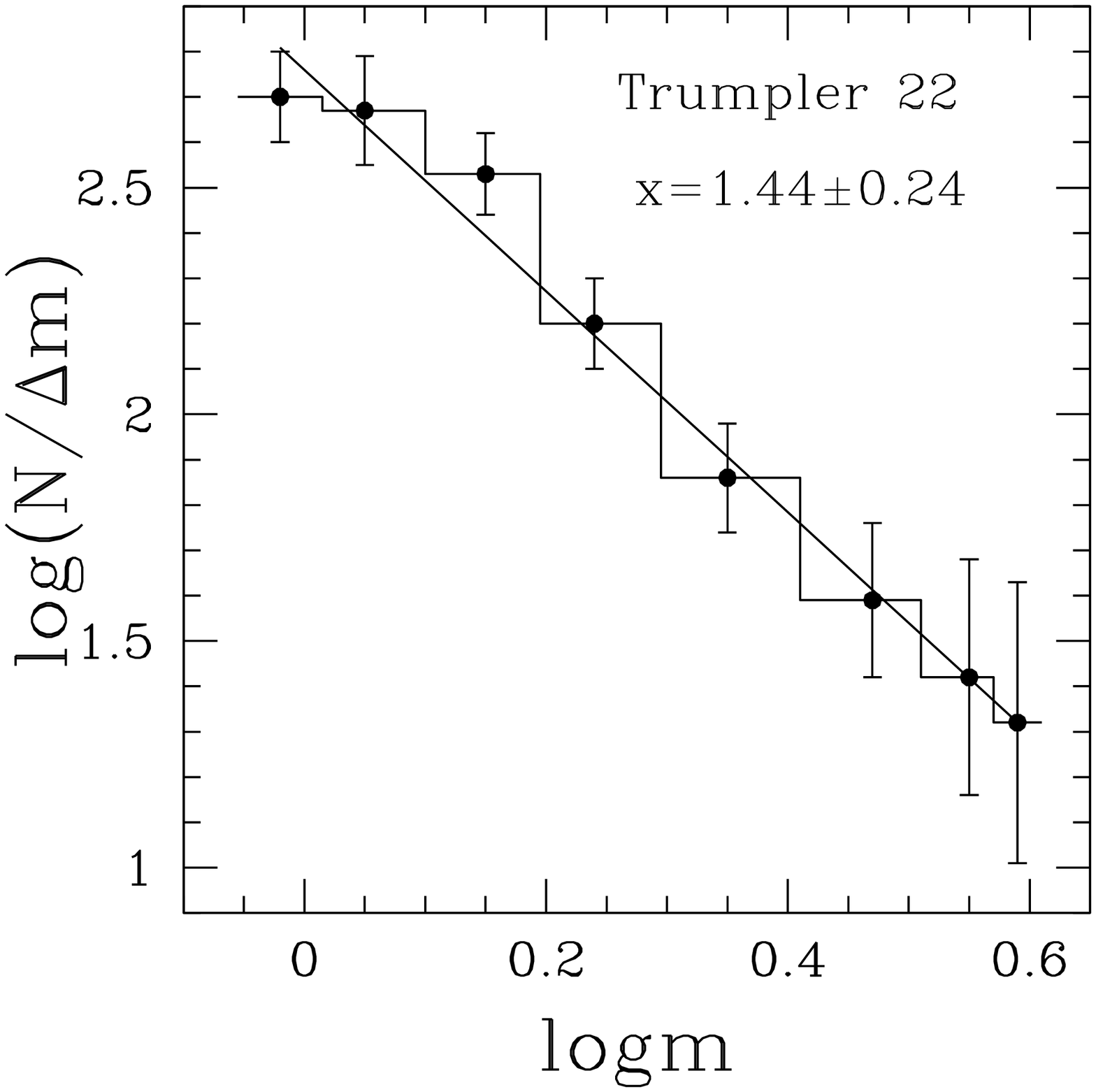}
}
\hbox{
\includegraphics[width=4.3cm, height=4.3cm]{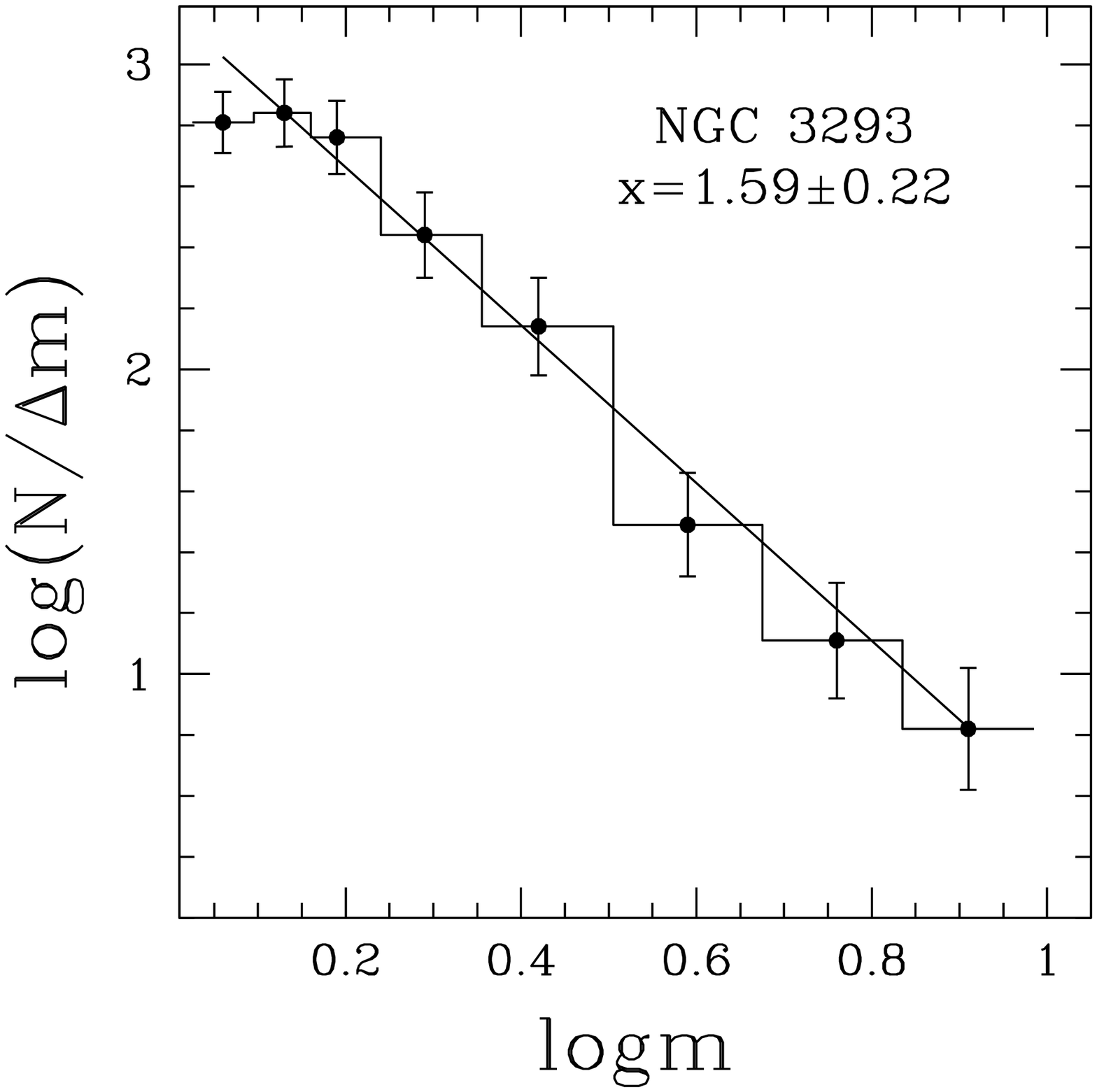}
\includegraphics[width=4.3cm, height=4.3cm]{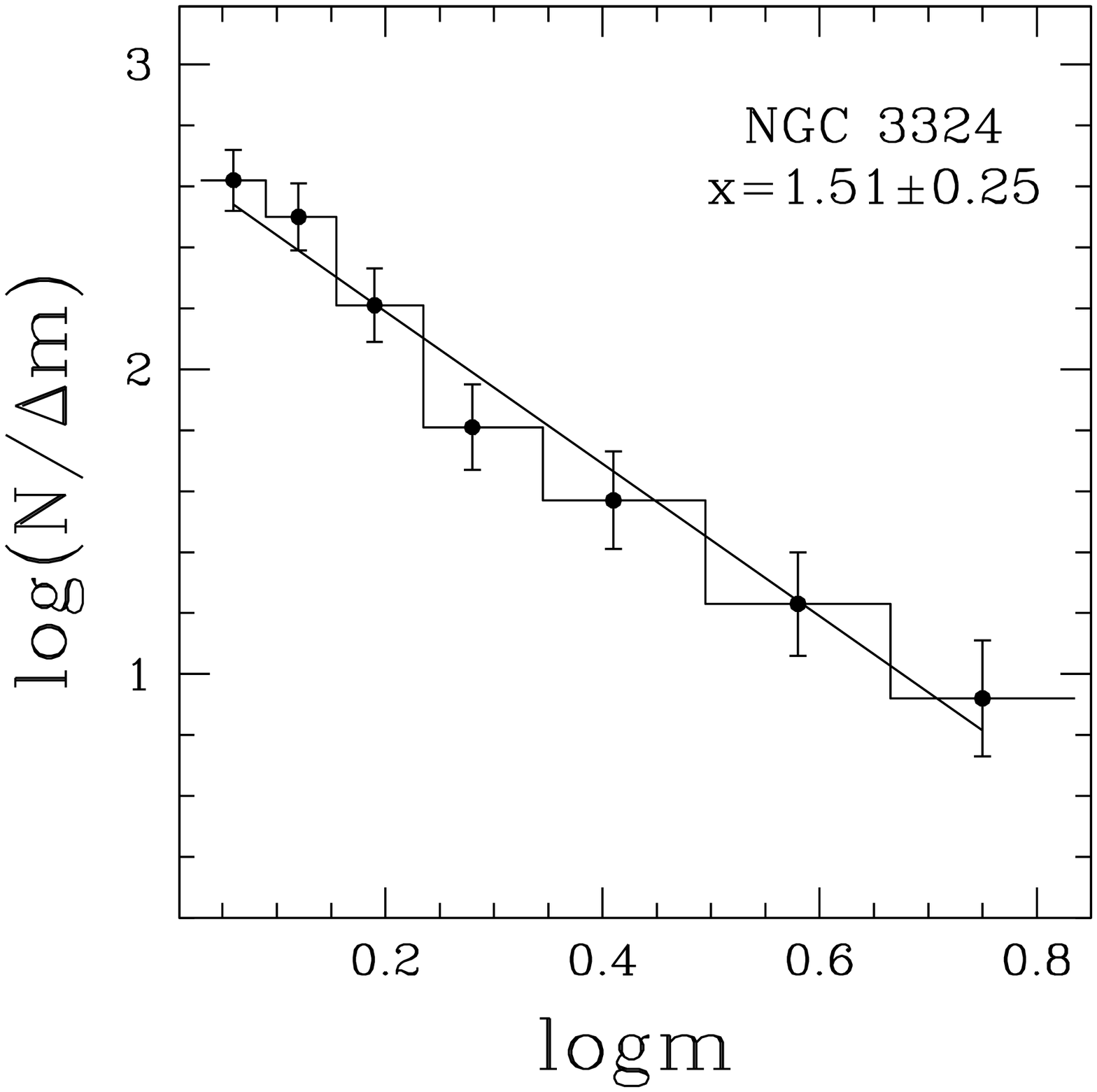}
}
\vspace{-0.5cm}
\caption{Mass function for the clusters under study derived using probable cluster members and
Marigo et al. (2017) isochrones. The error bars represent $\frac{1}{\sqrt{N}}$.} 
\label{mf}
\end{center}
\end{figure}

We have used the theoretical isochrones of Marigo et al. (2017) to convert the LF into MF. To understand the MF, we
have converted absolute mag bins to mass bins and the resulting mass function is shown in Fig. \ref{mf}. The MF slope can
be acquired by using a power-law given by,\\

$\log\frac{dN}{dM}=-(1+x)\log(M)$+constant\\

Where $dN$ is the probable cluster members in a mass bin $dM$ with central mass $M$ and $x$ is mass function slope.
Since Gaia data ($G$ mag) is not complete below G=19 mag (Arenou et al. 2018) then we took stars brighter than this limit,
which corresponds to stars more massive than 1 $M_{\odot}$.
The estimated values of the MF slopes are $x=1.40\pm0.16$, $1.44\pm0.24$, $1.59\pm0.22$ and $1.51\pm0.25$
for clusters NGC 5617, Trumpler 22, NGC 3293 and NGC 3324, respectively. These obtained values are satisfactory with the Salpeter's
initial mass function slope within error. The Total mass has been estimated for clusters using the derived mass function
slope. All MF related parameters in this section, like mass range, mass function slope, and the total mass estimated are
listed in Table \ref{massf_tab}.

\begin{table}
\centering
\caption{The main mass function parameters in clusters. 
}
\vspace{0.5cm} 
\begin{center}
\small
\begin{tabular}{lcccc}
\hline\hline
Object & Mass range & MF slope & Total mass & Mean mass \\
&      $M_{\odot}$ &  & $M_{\odot}$ & $M_{\odot}$
\\
NGC 5617  & $1.3-4.0$&$1.40\pm0.16$&$1230$&$2.10$ \\
Trumpler 22     & $1.0-4.1$&$1.44\pm0.24$&$755$&$1.76$ \\
NGC 3293     & $1.1-6.8$&$1.59\pm0.22$&$1457$&$2.10$ \\
NGC 3324     & $1.1-6.8$&$1.51\pm0.25$&$580$&$2.12$ \\
\hline
\end{tabular}
\label{massf_tab}
\end{center}
\end{table}

\subsection{Mass-segregation}

\begin{table}
\centering
\caption{Distribution of stars in different mass ranges along with the percentage of confidence level in mass-segregation effect for the
clusters.
}
\vspace{0.5cm}
\begin{center}
\small
\begin{tabular}{lcc}
\hline\hline
Object & Mass range & Confidence level\\
&      $M_{\odot}$ & $\%$ \\
\hline
NGC 5617  & $4.0-2.4, 2.4-1.2, 1.2-0.8$& 88 \\
Trumpler 22     & $4.1-2.6, 2.6-1.3, 1.3-0.8$& 75 \\
NGC 3293     & $6.9-2.2, 2.2-1.2, 1.2-0.8$& 77 \\
NGC 3324     & $6.9-2.1, 2.1-1.2, 1.2-0.8$& 70 \\
\hline
\end{tabular}
\label{masss_tab}
\end{center}
\end{table}

\begin{figure}
\begin{center}
\hbox{
\includegraphics[width=4.3cm, height=4.3cm]{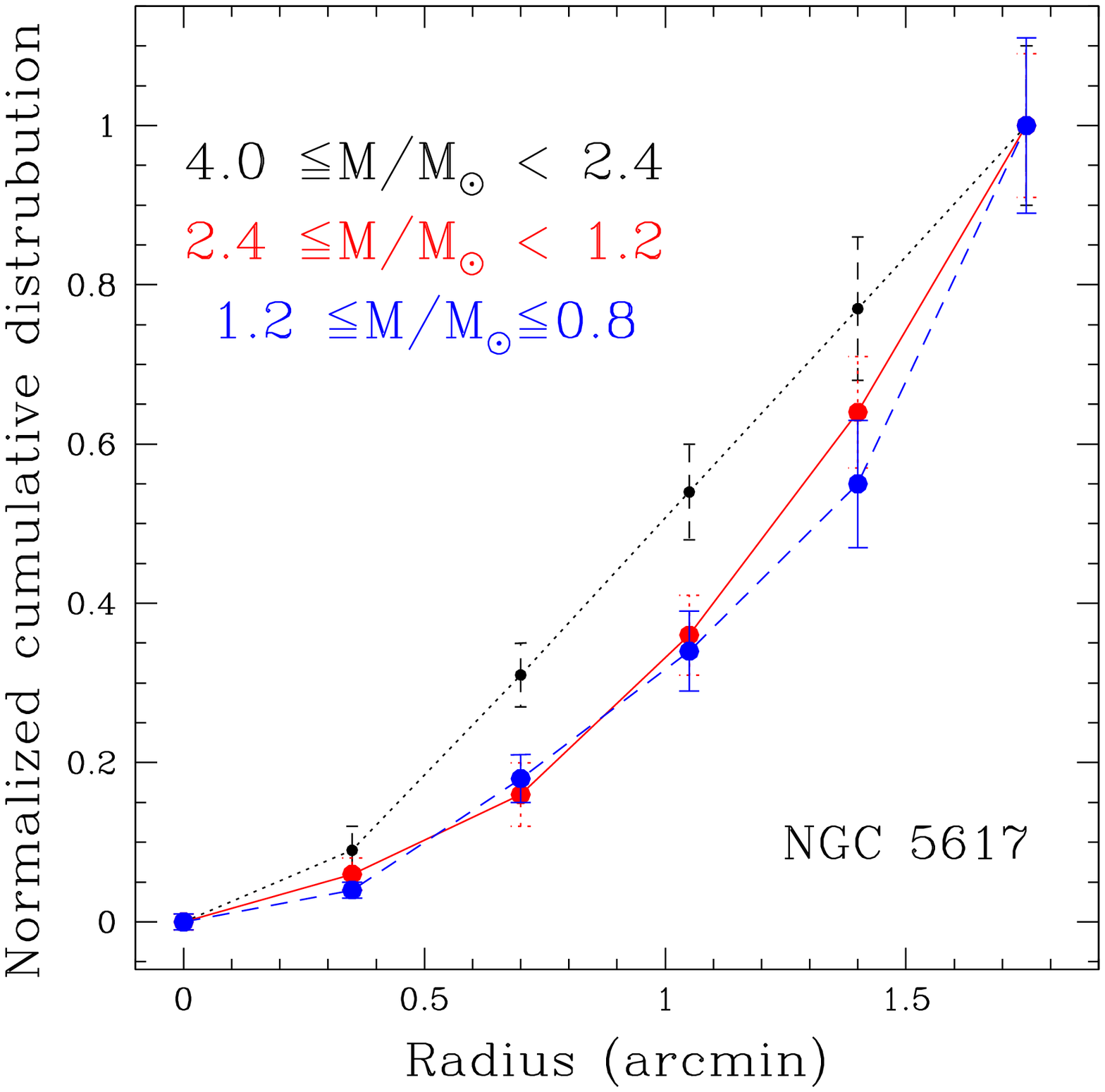}
\includegraphics[width=4.3cm, height=4.3cm]{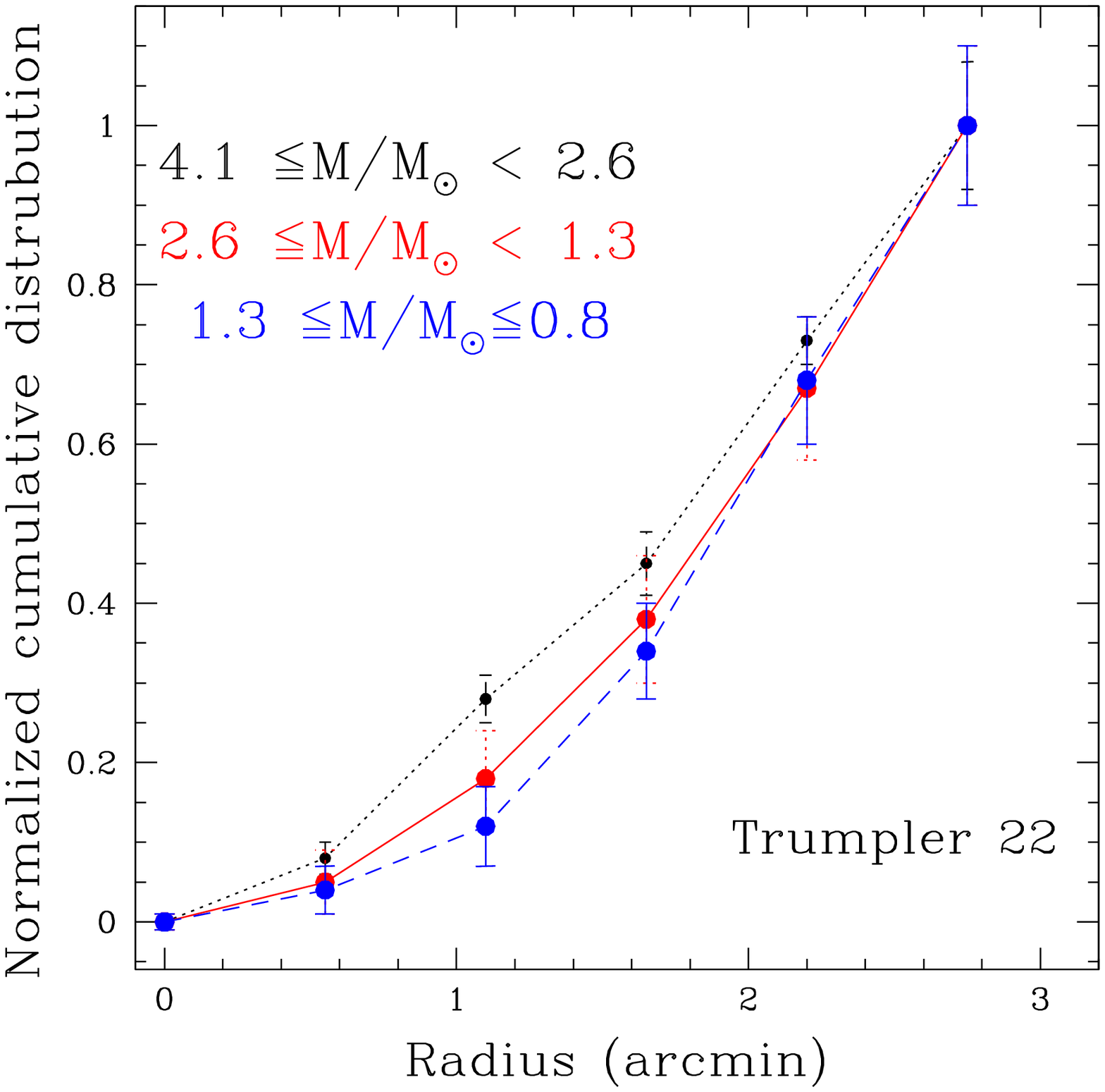}
}
\hbox{
\includegraphics[width=4.3cm, height=4.3cm]{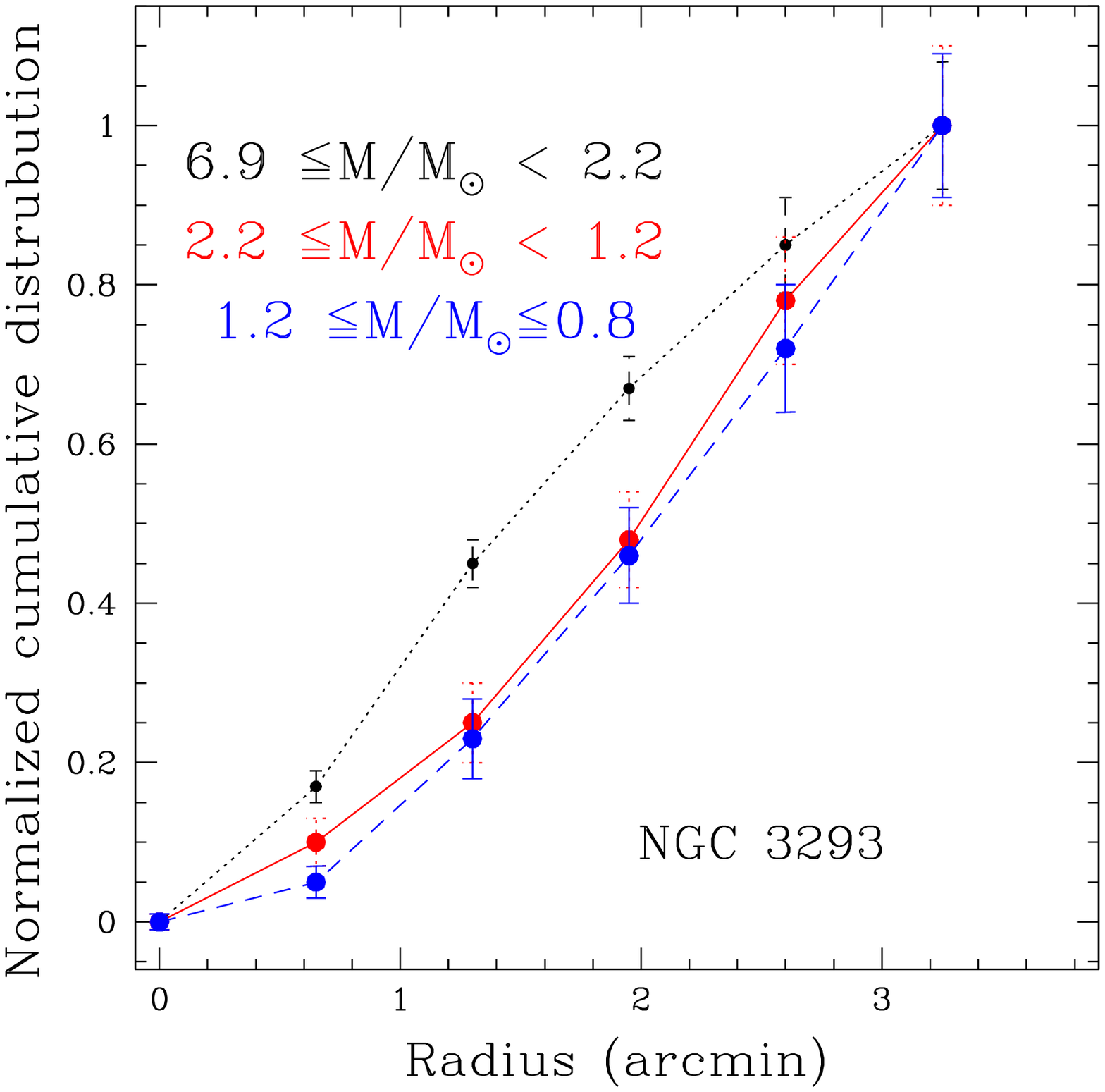}
\includegraphics[width=4.3cm, height=4.3cm]{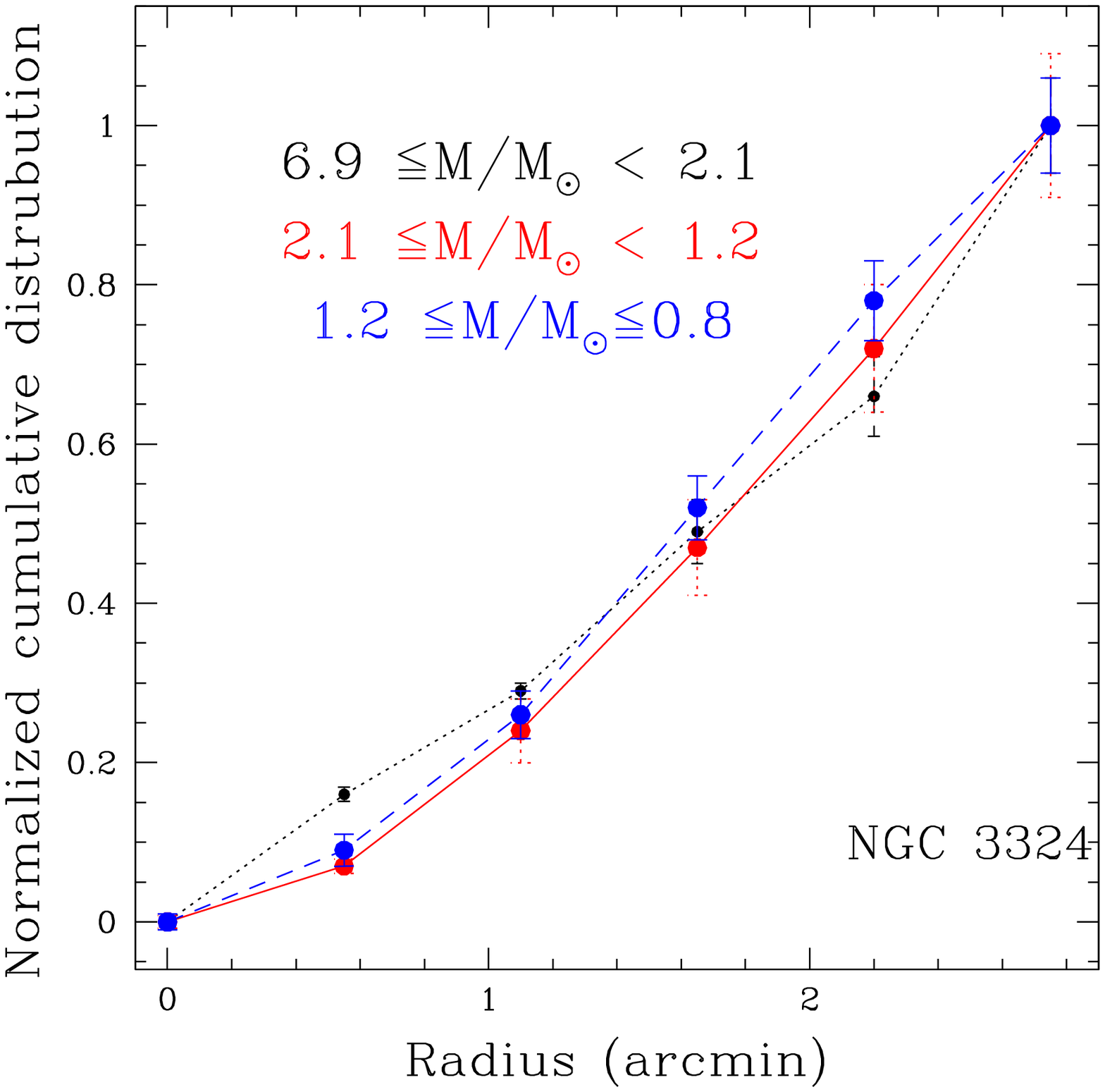}
}
\vspace{-0.5cm}
\caption{The cumulative radial distribution for NGC 5617, Trumpler 22, NGC 3293 and NGC 3324 in various mass range.} 
\label{mass_seg}
\end{center}
\end{figure}

In mass segregation, the higher mass stars gradually sink towards the cluster center and transfer their kinetic energy to
the more numerous lower-mass stellar component. Mass-segregation effect in clusters can be due to the dynamical evolution or
imprint of the star formation process or both. Considerable work has been done by many authors to check the mass segregation effect
in clusters (Hillenbrand \& Hartmann 1998; Lada \& Lada 1991; Brandl et al. 1996; Meylan 2000, Bisht et al. 2019, 2020). In this
study, we used only probable cluster members based on membership probability as described in section 3. To understand mass-
segregation cluster stars are divided into three different mass-ranges as shown in Table \ref{masss_tab} for clusters NGC 5617,
Trumpler 22, NGC 3293 and NGC 3324. The cumulative radial stellar distribution has been plotted for the main sequence
stars of all objects as shown in Fig. \ref{mass_seg}. This figure demonstrates the mass segregation effect in these
clusters based on the arrangement of massive and faint stars. We found the confidence level of mass-segregation as 88 $\%$,
75 $\%$, 77 $\%$ and 70 $\%$ for the clusters NGC 5617, Trumpler 22, NGC 3293 and NGC 3324, respectively based on
Kolmogrov-Smirnov test.

In the lifetime of star clusters, encounters between its member stars gradually lead to an increased degree of energy
equipartition throughout the clusters. The time scale on which a cluster will lose all traces of its initial conditions
is well represented by its relaxation time $T_{R}$, which is given by\\

~~~~~~~~~$T_{R}=\frac{8.9\times10^5\sqrt{N}\times{R_{h}}^{3/2}}{\sqrt{m}\times log(0.4N)}$

In the above formula, $N$ denotes the cluster members, $R_{h}$ is the radius within which half of the cluster mass is
accommodated and $m$ is the mean mass of the cluster stars (Spitzer \& Hart 1971).

The value of $R_{h}$ can be estimated based on the transformation equation given in Larsen (2006),\\

~~~~~~~~~~~~~$R_{h}=0.547\times R_{c}\times(\frac{R_{t}}{R_{c}})^{0.486}$\\

where $R_{c}$ is core radius while $R_{t}$ is tidal radius. We obtaind the value of half light radius as
1.67, 2.50, 2.52 and 2.28 pc for clusters NGC 5617, Trumpler 22, NGC 3293 and NGC 3324, respectively.

We estimated the value of $T_{R}$ as 13.5, 24.5, 26 and 17 Myr for NGC 5617, Trumpler 22, NGC 3293 and NGC 3324, respectively.
The dynamical evolution parameter ($\tau=\frac{age}{T_{E}}$) are found to be greater than 1 for clusters  NGC 5617 and
Trumpler 22, which concludes that these objects are dynamically relaxed. The value of $\tau$ is less than 1 for clusters
NGC 3293 and NGC 3324. Hence our study demonstrates that the binary clusters NGC 3293 and NGC 3324 are not dynamically relaxed.


\subsection{Dissociation time of clusters}

The compact N-body stellar groups have to be eventually demolished tidally, either by the Galactic field or by the nearby
transit of any giant molecular clouds (Converse \& Stahler 2011). We have estimated the dissociation time of both clusters
using the relationship given by Binney \& Tremaine (2008)\\

~~~~~~~~~~~~~~~$t_{dis}=250(\frac{M}{300})^{1/2}\times(\frac{R_{h}}{2})^{-3/2}$\\

$M$ and $R_{h}$ are the total cluster mass and half mass radius of the clusters. We have estimated dissociation time
as 664, 282, 390 and 285 Myr for clusters NGC 5617, Trumpler 22, NGC 3293 and NGC 3324, respectively.
These estimated values of dissociation time for all objects are very high than their relaxation times.
These obtained values of dissociation time indicate that NGC 5617, Trumpler 22, NGC 3293 and NGC 3324
should turn apart after the death of their bright members throughout the quick expansion of these objects.

\subsection{Tidal Radius of the clusters}

Tidal interactions are crucial to understand the initial structure and dynamical evolution of the clusters (Chumak et al. 2010).
Tidal radius is the distance from the cluster center where gravitational acceleration caused by the cluster becomes equal
to the tidal acceleration due to parent Galaxy (von Hoerner 1957). The Galactic mass $M_{G}$ inside a Galactocentric
radius $R_{G}$ is given by (Genzel \& Townes, 1987),\\

~~~~~~~~~~~~~~~$M_{G}=2\times10^{8} M_{\odot} (\frac{R_{G}} {30 pc})^{1.2}$\\

Estimated values of Galactic mass inside the Galactocentric radius (see Sec. 4.5) are found as $2.4\times10^{11} M_{\odot}$ and
$1.6\times10^{11} M_{\odot}$ for cluster pairs (NGC 5617 and Trumpler 22) and (NGC 3293 and NGC 3324), respectively.
Kim et al. (2000) has introduced the formula for tidal radius $R_{t}$ of clusters as, \\

~~~~~~~~~~~~~~~$R_{t}=(\frac{M} {2M_{G}})^{1/3}\times R_{G}$\\

where $R_{t}$ and  $M$ are the tidal radius and the total mass of the clusters. The estimated values of the tidal radius
are 15.06, 12.94, 13.36 and 9.98 pc for NGC 5617, Trumpler 22, NGC 3292 and NGC 3324, respectively.

\section{Binarity of the clusters}

Using the photometry and high resolution spectroscopy, de Silva et al. (2018) studied the physical connection 
between NGC 5617 and Trumpler 22. Based on the age and chemical composition they concluded that these clusters
are a primordial binary cluster pair in the Milky Way. To check the physical connection we estimated the separation
between these two binary clusters using the relation given by de la Fuente Marcos \& le la Fuente Marcos (2010),\\

~~~~~~~~~~~~$P_{orb}(Myr)=94(\frac{S_{0}}{1+e_{0}})^{3/2}\times\frac{1}{\sqrt{M_{1}+M_{2}}}$\\

where $P_{orb}$ is orbital time period, $e_{0}$ is eccentricity, $M_{1}$ and $M_{2}$ are the total masses of
clusters and $S_{0}$ is the separation between the binary clusters. Using the above relation, we obtained
the separation as $\sim$ 18 pc and $\sim$ 19 pc for clusters pairs (NGC 5617 and Trumpler 22) and (NGC 3293 and NGC 3324).
The small value of separation indicates that these objects are bound.

We analyzed the orbits of both the cluster pairs. All clusters are orbiting in a circular orbit. Their close
values of orbital parameters indicate that they are moving together. Their distance and age also indicate that they have
formed in a similar time scale. Therefore, based on the motion we can conclude that these clusters are a
cluster pairs of our Galaxy.

\section{Conclusions}
\label{con}

\begin{table*}
\centering
\caption{Various fundamental parameters of the clusters NGC 5617, Trumpler 22, NGC 3293 and NGC 3324.}
\vspace{0.5cm}
\begin{center}
\begin{tabular}{ccccc}
\hline\hline
Parameter &  NGC 5617 &  Trumpler 22 & NGC 3293 & NGC 3324 \\
\hline\hline
\\
RA(deg)                                    & $217.45\pm0.008$    & $217.82\pm0.006$  & $158.95\pm0.007$  & $159.35\pm0.01$  \\
DEC(deg)                                   & $-60.72\pm0.002$    & $-61.16\pm0.004$ & $-58.23\pm0.004$ & $-58.62\pm0.003$   \\
Radius(arcmin)                             & 3.5                & 5.5               & 6.5              &  5.5               \\
Radius(parsec)                             & 2.6                & 4.3               & 5.0              &  4.5               \\
$\mu_{\alpha}cos\delta$($mas~ yr^{-1}$)    & $-5.66\pm0.01$     & $-5.13\pm0.01$   & $-7.65\pm0.01$  & $-7.06\pm0.01$     \\
$\mu_{\delta}$($mas~ yr^{-1}$)             &  $-3.19\pm0.01$    &  $-2.70\pm0.01$   &  $3.36\pm0.009$  &  $ 2.85\pm0.01$    \\
Parallax(mas)                              &  $0.41\pm0.008$        &  $0.38\pm0.009$         &  $0.39\pm0.004$ &  $0.36\pm0.1$         \\
Age(Myr)                                   &  $90\pm10$        &  $90\pm10$         &  $12\pm3$       &  $ 12\pm3$         \\
Metal abundance                            &  $0.019$            &  $0.019$         &  $0.019$         &  $0.019$              \\
E(J-H) (mag)                               &  $0.17\pm0.03$     &  $0.20\pm0.03$    &  $0.07\pm0.02$   &  $0.14\pm0.03$     \\
E(J-K) (mag)                               &  $0.32\pm0.05$     &  $0.39\pm0.06$    &  $0.15\pm0.05$   &  $0.31\pm0.07$     \\
E(B-V) (mag)                               &  $0.55\pm0.05$     &  $0.64\pm0.05$    &  $0.23\pm0.03$   &  $0.45\pm0.05$      \\
$R_{V}$                                    &  $\sim 3.1$        &  $\sim 3.1$       &  $\sim  4$             &  $\sim 2$                \\
Distance modulus (mag)                     &  $13.70\pm0.40$    &  $14.20\pm0.30$   &  $12.90\pm0.20$  &  $13.00\pm0.20$     \\
Distance (Kpc)                             &  $2.43\pm0.08$     &  $2.64\pm0.07$    &  $2.59\pm0.10$   &  $2.80\pm0.2$      \\
$X$(Kpc)                                   &  -2.30              &  -2.60           &   0.70           &    0.78                \\
$Y$(Kpc)                                   &   7.85             &    7.83           &  -2.50          &   -2.71             \\
$Z$(Kpc)                                   &  -0.004              &  -0.020         &   0.003          &   -0.007                \\
$R_{GC}$(Kpc)                                   &  $10.90\pm0.5$ &  $11.20\pm0.8$  &   $7.9\pm0.4$   &   $8.0\pm0.3$          \\
Total Luminosity(mag)                      &   $\sim 3.4$       &   $\sim 3.3$      &   $\sim 3.3$     &   $\sim 5.2$        \\
Cluster members                            &   584              &   429             &   692           &    273               \\
MF slope                                  &   $1.40\pm0.16$    &   $1.44\pm0.24$   &   $1.59\pm0.22$  &   $1.51\pm0.25$     \\
Total mass ($M_{\odot}$                    &   $ \sim 1230 $      &   $\sim 755 $     &   $ \sim 1457 $    &   $\sim 580 $       \\
Average mass($M_{\odot}$)                  &   $2.10$           &   $1.76$          &   $2.10$         &   $2.12$            \\
Relaxation time(Myr)                       &   13.5              &   24.5            &      26            &    17                \\
Dynamical evolution parameter ($\tau$)     &   $ \sim 6.5$         &   $ \sim 3.7$         &   $\sim 0.46$       &   $\sim 0.7$         \\
\hline
\end{tabular}
\label{para}
\end{center}
\end{table*}

One of the outcomes of this study is the estimation of membership probability of stars in the field of
the two binary clusters (NGC 5617 and Trumpler 22) and (NGC 3293 and NGC 3324). We have estimated all
the fundamental parameters of the clusters as shown in Table \ref{para}.
The main conclusions of the present study are as follows: 

\begin{enumerate}

\item The new center coordinates are derived for all clusters and are listed in Table \ref{para}.



\item  Colour-colour diagrams have been constructed after combining Gaia~ EDR3, VVV, VPHAS, APASS, and WISE database. The diagrams
       show that the interstellar extinction law is normal towards the cluster's area of NGC 5617 and Trumpler 22. We found
       extinction law is abnormal for clusters NGC 3293 and NGC 3324.

\item  The distance estimation from parallax are well supported by the values estimated using the isochrone fitting method to the
       clusters CMDs. We obtained a similar age of cluster pairs by comparing with the theoretical isochrones of solar metallicity
       taken from Marigo et al. (2017).
       

\item  We obtained 18 and 44 stars towards the cluster region of NGC 5617 and Trumpler 22 as the YSOs reddening parameter (Q) method.

\item We determined LFs and MFs of both objects by considering the members as selected on the basis of membership probability of
      stars. The MF slopes are in fair agreement with the Salpeter (1955) value for the clusters under study. 



\item The presence of mass-segregation is examined for these clusters using probable cluster members. We found that the massive
      stars are concentrated towards the inner regions of the clusters in comparison to faint stars. The confidence level
      of mass segregation is found as 88 $\%$, 75 $\%$, 77 $\%$ and 70 $\%$ for NGC 5617, Trumpler 22,
      NGC 3293 and NGC 3324, respectively on the basis of the K-S test. Our study indicates that NGC 5617 and Trumpler 22 are
     dynamically relaxed while NGC 3293 and NGC 3324 are not relaxed.

\item Galactic orbits and orbital parameters are estimated using Galactic potential models. We found that these objects are 
      orbiting in a boxy pattern in circular orbit. The different orbital parameters are listed in Table \ref{inp} and 
      \ref{orpara} for the clusters under study. Present analysis shows that clusters (NGC 5617 and Trumpler 22) and 
      (NGC 3293 and NGC 3324) are physically connected and are cluster pairs of Milky Way.

\end{enumerate}

\section{ACKNOWLEDGEMENTS}

The authors are thankful to the anonymous referee for useful comments, which improved the contents of the paper
significantly. This work has been financially supported by the Natural Science Foundation of China (NSFC-11590782, NSFC-11421303).
Devesh P. Sariya and Ing-Guey Jiang are supported by the grant from the Ministry of Science and Technology (MOST),
Taiwan. The grant numbers are MOST 105-2119-M-007 -029 -MY3 and MOST 106-2112-M-007 -006 -MY3. This work has made use
of data from the European Space Agency (ESA) mission Gaia (https://www.cosmos.esa.int/gaia), processed by the Gaia Data
Processing and Analysis Consortium (DPAC, https://www.cosmos.esa.int/web/gaia/dpac/consortium). Funding for the DPAC has
been provided by national institutions, in particular the institutions participating in the Gaia Multilateral Agreement.
In addition to this, It is worth to mention that, this work has been done by using WEBDA.

\section{Data Availability}

We have used the different data sets for the analysis of two pairs of binary clusters (NGC 5617 and Trumpler 22) and (NGC 3293 and NGC 3324),
which are publicly available as following,

\begin{itemize}

\item GAIA EDR3\\
https://vizier.u-strasbg.fr/viz-bin/VizieR-3?-source=I/350/gaiaedr3\\

\item WISE\\
https://vizier.u-strasbg.fr/viz-bin/VizieR-3?-source=II/311/wise\\

\item VVV\\
http://vizier.u-strasbg.fr/viz-bin/VizieR?-source=II/348\\

\item APASS\\
http://vizier.u-strasbg.fr/viz-bin/VizieR?-source=II
\\
\item GLIMPSE\\
https://vizier.u-strasbg.fr/viz-bin/VizieR?-source=
\\
\item VPHAS\\
http://vizier.u-strasbg.fr/viz-bin/VizieR?-source=II

\end{itemize}

\end{document}